\documentclass[aps,prd,superscriptaddress,nofootinbib,preprintnumbers,12pt]{revtex4-1}
\usepackage{amsmath}
\usepackage{graphicx}
\usepackage{subfigure}
\usepackage{amssymb}
\usepackage{xcolor}
\usepackage{multirow}
\usepackage{cancel}
\usepackage{color}
\usepackage{ulem}
\usepackage{listings}
\usepackage{xcolor}
\usepackage{float}
\usepackage{slashed}
\usepackage{wrapfig}
\usepackage{mathtools}
\usepackage[colorlinks,citecolor=blue]{hyperref}
\usepackage[T1]{fontenc}  
\usepackage[utf8]{inputenc}  
\usepackage{times}
\numberwithin{equation}{section}

\begin{document}
	
\title{Low-Scale Leptogenesis from\\Resonant Thermal Lepton Flavour Coherences}

\author{Shao-Ping Li}
\email{lisp@het.phys.sci.osaka-u.ac.jp}
\affiliation{Department of Physics, The University of Osaka, Toyonaka, Osaka 560-0043, Japan}

\affiliation{Marietta Blau Institute for Particle Physics, Austrian Academy of Sciences, Dominikanerbastei 16, A-1010 Vienna, Austria}

\author{Apostolos Pilaftsis}
\email{apostolos.pilaftsis@manchester.ac.uk}
\affiliation{Department
 of Physics and Astronomy, University of Manchester, Manchester, M13 9PL, United Kingdom}

\begin{abstract}

\bigskip
\noindent
Resonant heavy-neutrino mixing  and sterile neutrino oscillations are two prominent mechanisms to realize low-scale leptogenesis, with singlet neutrino masses below 
TeV~energies\- that could be probed in current and future laboratory experiments. In their minimal settings\-, both mechanisms require a significant degree of degeneracy in the singlet neutrino masses to compensate for the suppression that results from the small neutrino Yukawa couplings.
After further developing the flavour-covariant Kadanoff-Baym formalism, we study in detail a novel dominant mechanism for low-scale leptogenesis which becomes greatly enhanced by resonant thermal lepton-flavour coherences at the two-loop level. This mechanism works successfully for both Dirac and Majorana singlet neutrinos, and it does not rely on whether these singlet neutrinos are quasi-degenerate or not. In~particular, it implies that successful low-scale leptogenesis in the type-I seesaw framework can be naturally realised with heavy neutrino masses that could be as~low as~GeV.

\end{abstract}

\maketitle

\tableofcontents

\clearpage

\section{Introduction}
\label{sec:intro}

Right-handed neutrinos offer a minimal and promising extension of the Standard Model~(SM), through the possible addition of lepton-number ($L$) violating Majorana masses,  which allow to address several fundamental questions in modern particle physics and cosmology. In particular, these singlet neutrinos can naturally explain the smallness in mass  of the active SM neutrinos by virtue of the seesaw mechanism~\cite{Minkowski:1977sc,Mohapatra:1979ia,Yanagida:1980xy,Schechter:1980gr,Schechter:1981cv}, account via leptogenesis~\cite{Fukugita:1986hr,Luty:1992un} for the matter-antimatter or baryon asymmetry in our Universe (BAU), and significantly contribute to the dark matter (DM)  relic abundance~\cite{Dodelson:1993je,Shi:1998km,Asaka:2005pn,Asaka:2005an}. 
    
In its original formulation~\cite{Fukugita:1986hr}, leptogenesis was realized by ultra-heavy Majorana neutrinos with masses of the order $10^{14}$~GeV close to the scale of SO(10) Grand Unification Theory~(GUT). The~enormity of this GUT scale, which is many orders of magnitude higher than the electroweak scale, renders any direct exploration of the ultra-heavy Majorana sector practically impossible. Therefore, considerable efforts were made to 
lower the scale of leptogenesis below TeV energies so that the heavy neutrino sector responsible for the BAU could be probed, to a good extent, at high-energy colliders and in other low-energy experiments at the intensity frontier.
One such low-scale scenario for successful leptogenesis is
Resonant Leptogenesis (RL)~\cite{Pilaftsis:1997jf,Pilaftsis:2003gt,Pilaftsis:2005rv}, where lepton-number asymmetries due to heavy-neutrino self-energy contributions~\cite{Flanz:1994yx,Covi:1996wh} can be maximally enhanced up to order one~\cite{Pilaftsis:1997dr}, thus avoiding the lower bound of $10^9$~GeV in the heavy neutrino masses~\cite{Davidson:2002qv}. Interestingly enough,
heavy-neutrino flavour effects in RL provide new possibilities for modelling scenarios with observable signatures of charged-lepton flavour violation~\cite{Pilaftsis:2004xx,Pilaftsis:2005rv}.   

Another interesting low-scale leptogenesis scenario uses the ARS mechanism~\cite{Akhmedov:1998qx,Asaka:2005pn}, where the required lepton asymmetry is created by coherent sterile neutrino oscillations\footnote{Note that the terms: right-handed neutrino, heavy neutrino, singlet neutrino, and sterile neutrino, will be inter\-changeably used in this paper.}. In the ARS scenario, the singlet neutrinos can be as light as GeV
and hence be accessible at beam-dump experiments, whilst electroweak-scale heavy neutrinos may be observed  by displaced vertex searches at high-energy colliders. In addition, low-scale leptogenesis benefits from the fact that it does not require high reheating temperatures, which may become a limiting factor for some cosmological models with heavy stable relics, such as gravitinos.

In realizing low-scale leptogenesis with two singlet neutrino flavours, a high degree of degeneracy in the neutrino masses is typically needed. In this setting, low-scale leptogenesis scenarios generally need feeble Yukawa couplings to provide the out-of-equilibrium condition in producing a net baryon  asymmetry via the electroweak sphaleron processes~\cite{Kuzmin:1985mm}. Then, as mentioned above, strongly degenerate singlet neutrino masses can significantly enhance the lepton asymmetry to obtain viable leptogenesis. Nevertheless, such mass degeneracy may not be necessary in some special situations, e.g., in flavoured leptogenesis with temperatures above $10^5$~GeV~\cite{Drewes:2012ma,Canetti:2014dka}, and in leptogenesis with contributions from more than two singlet neutrino flavours, as more degrees of freedom are involved~\cite{Drewes:2016gmt,Abada:2018oly}.

\begin{figure}[t]
	\centering
\includegraphics[scale=0.62]{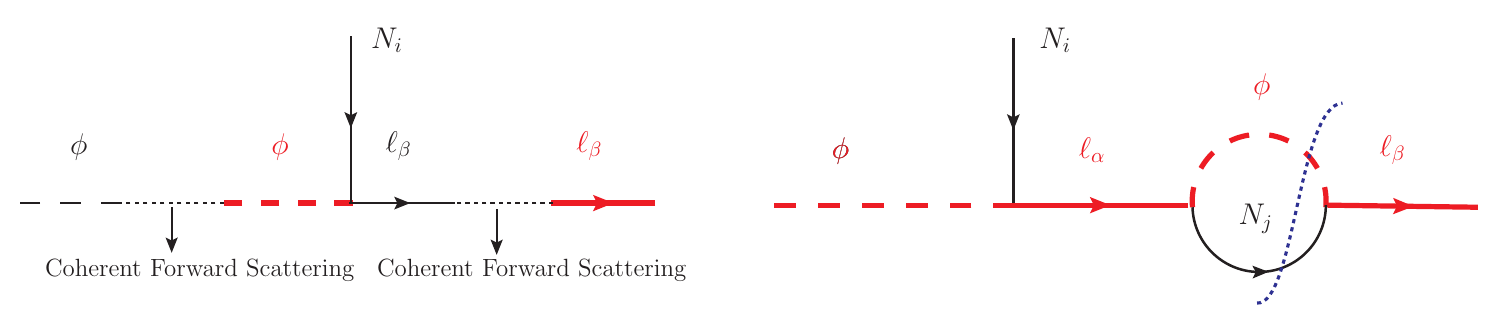} 
	\caption{\label{fig:Higgsdecay} Tree-level and one-loop diagrams for Higgs decay to leptons and relativistic singlet neutrinos.
    The free Higgs undergoes coherent forward scattering with the background plasma, acquiring a thermal mass  to trigger the decay before gauge symmetry breaking. The produced leptons also undergo coherent forward scattering, enhancing the absorptive part of the one-loop amplitude induced by the blue-dashed thermal cut line. The red, thick lines for lepton and Higgs doublets reflect thermal corrections after fast coherent forward scattering.}
\end{figure}

In this paper, we will explicitly demonstrate how scenarios  with  two singlet neutrino flavours in the GeV scale can realise low-scale leptogenesis, without relying on quasi-degenerate heavy neutrino masses or on flavoured leptogenesis. The channel under consideration appears in SM Higgs decay to leptons and light sterile neutrinos, where the thermal mass effect of the SM Higgs triggers the decay before the sphaleron decoupling. In addition, this thermally induced Higgs decay can also create a CP-violating source from higher-order quantum-statistics corrections. A graphic depiction is given in Fig.~\ref{fig:Higgsdecay}. The CP-violating source exhibits a large, finite-temperature induced resonant enhancement in the quasi-thermal SM leptons rather than in the nonthermal sterile neutrinos. Hence, we call this scenario of leptogenesis from resonant lepton-flavour coherences\footnote{In Section~\ref{sec:mix-osc-cor}, we will provide qualitative understanding of the concepts and effects from flavour mixing and oscillations, where lepton-flavour off-diagonal correlations correspond to the quantum-mechnical coherences. },  or in short, Thermal Resonant Leptogenesis (TRL). TRL works without the need to introduce  extra fields  beyond the  type-I seesaw framework.  Unlike the standard RL, the strength of the resonant enhancement from lepton-flavour coherences is almost fully determined by the SM sector, within the context of perturbative thermal Quantum Field Theory (QFT).

To consolidate our findings, we use two equivalent methods based on non-equilibrium QFT, in terms of the Schwinger–Keldysh Closed-Time-Path (CTP) formalism~\cite{Chou:1984es,Calzetta:1986cq}, which enable us to reliably calculate the CP-violating source from: (i)~flavour-covariant Kadanoff-Baym (KB) formalism~\cite{Prokopec:2003pj,Prokopec:2004ic,Beneke:2010dz,BhupalDev:2014pfm} of quasi-thermal leptons and (ii)~two-loop self-energy graphs of sterile neutrinos. Building upon the first method  allows one to understand the CP-violating effect as the interplay among lepton-flavour oscillations, lepton-flavour mixing, and the out-of-equilibrium neutrino Yukawa interactions, while the diagrammatic method can clearly display the origin of the resonant enhancement from thermal corrections to lepton propagators. 

To estimate the contribution to TRL from GeV-scale right-handed neutrinos, we provide a simple but rather instructive numerical analysis. We also make a comparison with RL from quasi-degenerate neutrinos, identifying the regime where TRL  dominates low-scale lepto\-genesis. A~dedi\-cated numerical analysis, which accounts for the active neutrino masses and mixing and identifies the observational regions of vital parameter space in laboratory experiments, is presented in a companion work~\cite{Li:2026rqg}.

The remainder of the paper is organized as follows. In Section~\ref{sec:sota}, we present a short revisit to type-I seesaw framework,  discuss the state-of-the-art studies of finite-temperature effects in realizing leptogenesis, and outline the novelties of the current  work.  Subsequently, in Section~\ref{sec:FC-leptons}, we use the flavour-covariant KB formalism and the two-loop diagrammatic method to calculate the lepton-number asymmetry generated in the CP-violating Higgs decay, and provide complementary and qualitative understanding of the effects from lepton-flavour mixing and oscillations that yield the resonant flavour coherences.  Section~\ref{sec:num}  is devoted to a simple estimate of the final BAU, showing how TRL can successfully explain the BAU problem. Finally, we draw our conclusions in Section~\ref{sec:con}. To facilitate understanding of the two methods used in this work, we provide detailed technical discussions and derivations in the Appendices~\ref{app:full-KB}-\ref{app:collision-rates}, where we clarify the conventions and approximations adopted in low-scale type-I leptogenesis\footnote{We will call it low-scale type-I leptogenesis for short.} and present possible generalizations and extensions. Appendix~\ref{app:vertex} shows that the vertex corrections to the TRL mechanism are sub-dominant, and Appendix~\ref{app:RFL-N} is devoted to  details of the CTP calculations for RL that is induced by quasi-degenerate neutrinos, which forms the basis for the comparison made in Section~\ref{sec:2loo-diagram}.

\section{State of the art of finite temperature effects} \label{sec:sota}

In the type-I seesaw framework, the Yukawa interactions relevant for leptogenesis is given by
\begin{align}\label{eq:lag}
    \mathcal{L}=-y_{\alpha\beta}\bar\ell_\alpha \phi P_R e_{\beta}- y'_{\alpha i} \bar \ell_\alpha\tilde{\phi}P_R N_i-\frac{1}{2}M_{ij}\overline{N_i^c} P_R N_j+\rm h.c.\,,
\end{align}
where $\ell, \phi, e, N$ denote the lepton doublet, Higgs doublet, charged-lepton singlet, and right-handed neutrinos, respectively, with $\tilde{\phi}=i \sigma_2 \phi$,  $\phi=(0, (v+h)/\sqrt{2})^T$ in the unitary gauge, and $v\approx 246$~GeV the electroweak vacuum expectation value. We will work in the basis where the charged-lepton Yukawa matrix and the neutrino mass matrix  are diagonal, i.e., $y_{\alpha\beta}\equiv y_{\alpha}\delta_{\alpha\beta}, M_{ij}\equiv M_i \delta_{ij}$. For clarity, we  will use throughout Greek indices ($\alpha,\beta,\gamma$) for charged-lepton flavours and Latin indices ($i,j,k$)  for neutrino flavours. 

For light sterile neutrinos, if their masses are much smaller than the temperatures during which leptogenesis  is mostly active,  they can populate in the relativistic regime where the helicity acts as an approximately conserved quantum number to distinguish sterile neutrinos and their antiparticle states. Then, one can define a generalized lepton number~\cite{Akhmedov:1998qx,Asaka:2005pn,Abada:2018oly}
	\begin{align}
		L_{\rm tot}=L_{\rm SM}+L_N\,,
	\end{align}
where $L_{\rm SM}$ denotes the lepton number in the SM sector and $L_N$ the quantum number of  $N$-helicity. The $L_{\rm tot}$-violating (or lepton-number violating) processes will then scale with the sterile neutrino masses, which is suppressed by $M^2/T^2\ll 1$. In this work, we will consider such relativistic Majorana neutrinos, and illustrate a leptogenesis mechanism that is induced by $L_{\rm tot}$-conserving processes. Since these processes do not require the Majorana mass insertion or the neutrino chirality flip, the conclusions will also be applicable to Dirac neutrinos, where the corresponding $L_{\rm tot}$ is conserved due to global $U(1)$ symmetry. 

At finite temperatures, the background plasma  effects open up more quantum-statistical channels to create the CP asymmetry~\cite{Beneke:2010wd,Garbrecht:2010sz,Garny:2010nj,Garbrecht:2012pq,Frossard:2012pc,Hamada:2016oft,Li:2020ner,Li:2021tlv}.  In the  type-I seesaw framework, the impact of  thermal mass correction to the SM Higgs has been identified in leptogenesis~\cite{Pilaftsis:1997jf,Giudice:2003jh}, and was later applied to realize leptogenesis from thermally induced CP violation~\cite{Hambye:2016sby,Hambye:2017elz}, which is kinematically forbidden in the zero temperature limit. Such a kind of forbidden leptogenesis considered the $L_{\rm tot}$-violating effect from GeV-scale sterile neutrinos, and like RL, still requires a high-degree degeneracy of sterile neutrino masses. Different from  the ARS mechanism~\cite{Akhmedov:1998qx}, the asymmetries of leptons and sterile  neutrinos in this kind of forbidden leptogenesis are generated simultaneously from Higgs decay. It is also different from the traditional RL, where the CP-violating process is realized in  neutrino decay rather than Higgs decay.  For all the aforementioned leptogenesis, there is a common basis for the generation of CP asymmetry: the vacuum or thermal absorptive part of  quantum corrections arises from sterile neutrino mixing.

Thermal corrections to the SM leptons provide another CP-violating source in leptogenesis, which is a combined effect of the CP-violating phase from neutrino Yukawa couplings and the absorptive part from lepton self-energy diagrams. This effect, which is $L_{\rm tot}$-conserving but lepton-flavour dependent, is caused by off-diagonal flavour correlations of thermal lepton doublets, and was noticed in high-scale lepton-flavoured leptogenesis~\cite{Beneke:2010dz} and later detailed in \cite{Garbrecht:2010sz,Garbrecht:2012pq} where  a resonant enhancement factor as maximal as $10^5$  can appear. A large thermal correction to leptons was also considered in \cite{Hamada:2016oft,Hamada:2018epb} at inflationary temperatures, where lepton mixing appears through the thermal mass matrix and assists in generating the CP-violating source. In this case, however, a resonant  enhancement is not expected due to large  lepton thermal masses. 
For low-scale leptogenesis, on the other hand, the $L_{\rm tot}$-conserving effect was found to have a  resonant enhancement factor as maximal as $10^8$~\cite{Kanemura:2024dqv,Kanemura:2024fbw}, which can help to realize leptogenesis that is not attainable if one follows the traditional mechanism.  As will be discussed in this paper, the reason that leads to the  different enhancement factors comes from  the identification of  damping effects and higher-order source terms.

While the development of low-scale leptogenesis from singlet neutrino mixing has brought us   a  clearer  picture of the physical process in recent years, and the behaviour of RL  from sterile neutrino mixing and the ARS mechanism from sterile neutrino oscillations can also be united by a common density-matrix formalism~\cite{Klaric:2020phc}, it is currently not clear yet about the physics of low-scale leptogenesis that gets significant enhancement from quasi-thermal leptons. In particular,  the kinetic equations are more complicated, technically built beyond the simple treatment of Boltzmann equations, and the CP-violating source necessary for the BAU explanation is a purely thermal effect.  These kinetic equations should include consistently the finite-temperature effects,  flavour mixing, and potentially coherent correlation effects that may lead to flavour oscillations. The  construction, derivation, and working regimes warrant comprehensive details and intuitive understanding. Besides, all these effects may not be fully captured in the vacuum density-matrix formalism~\cite{Sigl:1993ctk}. In this work, we aim to provide these details for low-scale leptogenesis in the type-I  seesaw framework, though the formulae are readily generalisable beyond the type-I  seesaw. The novelties and highlights in this paper will include the following points: 
\begin{itemize}
    \item We build the flavour-covariant KB equations for quasi-thermal leptons in the low-scale regime, and find that independently of the mass degeneracy in sterile neutrinos, the low-scale type-I leptogenesis can be realized via  thermally induced Higgs decay to GeV-scale sterile neutrinos. Therefore,  TRL contributes as a third channel beyond the ARS mechanism and the traditional RL. 
    \item We present two methods to evaluate the CP-violating source. The equivalence between the two methods implies that lepton-flavour coherences (induced by lepton-doublet mixing and oscillations) and the resonant behaviour of thermal leptons can be united into the flavour-covariant KB equations to describe purely thermal-induced leptogenesis (vacuum-forbidden leptogenesis). This may be regarded as a finite-temperature generalization analogous to the unity found in \cite{Klaric:2020phc} based on the density-matrix formalism.
    \item  We demonstrate that two-loop source terms are important to maintain the  lepton-flavour correlations (coherences) and hence help to boost low-scale leptogenesis, which were not identified in former studies~\cite{Beneke:2010dz,Garbrecht:2010sz,Garbrecht:2012pq}. The two equivalent technical methods presented in this work also provide another check for the importance of the two-loop source terms.
    \item We present detailed calculations of the CP- and $L_{\rm tot}$-violating Higgs decay in the CTP formalism, providing another understanding complementary to the density-matrix formalism~\cite{Hambye:2017elz}.  Following this formula,  one can identify the regime in which low-scale leptogenesis is dominated by TRL or by the traditional RL.
\end{itemize}

\section{Flavour-covariant Kadanoff-Baym equations for leptons}\label{sec:FC-leptons}
We start in this section the construction towards the kinetic evolution of lepton asymmetry  following the CTP formalism. As the first method to derive the CP-violating source, we present  in Section~\ref{sec:0thKB} the leading-order flavour-covariant KB equations for leptons, with the technical and fundamental details documented in Appendixes~\ref{app:full-KB}-\ref{app:constraint-KB}.  Then,  in Section~\ref{sec:lepton-asymmetry}, we calculate the off-diagonal correlations  for lepton flavours, and substitute the solution  to determine the flavour-diagonal entries. More comprehensive discussions are added in Appendix~\ref{app:collision-rate-1} on the damping and source terms for the off-diagonal correlations and Appendix~\ref{app:collision-rate-2} for the diagonal entries, assisting the understanding of  Section~\ref{sec:lepton-asymmetry}.  After that, we provide a two-loop diagrammatic perspective as the second method in Section~\ref{sec:2loo-diagram} to calculate the CP-violating source, where we confirm the equivalence between the two methods under the total lepton-number conservation.  The technical details on which Section~\ref{sec:2loo-diagram} is based is presented in Appendix~\ref{app:RFL-N}. Finally, we discuss in Section~\ref{sec:mix-osc-cor} the interplay between  flavour mixing and flavour oscillations that give rise to TRL.

\subsection{Leading-order Kadanoff-Baym equations}\label{sec:0thKB}

The starting point in this section is the truncated lowest order of the KB equations derived from the Schwinger-Dyson equations in flavour-covariant formulation, where we present the detailed derivation in Appendix~\ref{app:full-KB} for consistency check and for visibility of approximations we apply in leptogenesis. 

The leading-order KB  kinetic and constraint equations for quasi-thermal massless leptons are obtained by taking $\sin\Diamond=0$ in \eqref{eq:KB-H}  and \eqref{eq:KB-antiH},  yielding 
\begin{align}i\partial_t[\gamma^0iS_{<(>)}]_{\alpha\beta}&-\left[{\bf k}\cdot{\boldsymbol{ \gamma}}\gamma^0+\text{Re}\Sigma_{R}\gamma^0,\gamma^0iS_{<(>)}\right]_{\alpha\beta}-\left[i\Sigma_{<(>)}\gamma^0,\gamma^0 \text{Re}S_{R}\right]_{\alpha\beta}
    \nonumber \\[0.2cm]
  &=-\frac{i}{2} \left(\left\{\gamma^0 iS_<,i\Sigma_>\gamma^0\right\}_{\alpha\beta}-\left\{\gamma^0iS_>,i\Sigma_<\gamma^0\right\}_{\alpha\beta}\right),\label{eq:S>-sim}
    \\[0.2cm]
  2k_0[\gamma^0iS_{<(>)}]_{\alpha\beta}&-\left\{{\bf k}\cdot{ \boldsymbol{ \gamma}}\gamma^0+\text{Re}\Sigma_{ R}\gamma^0,\gamma^0iS_{<(>)}\right\}_{\alpha\beta}-\{i\Sigma_{<(>)}\gamma^0,\gamma^0 \text{Re}S_R\}_{\alpha\beta}
    \nonumber \\[0.2cm]
   & =\frac{i}{2}\left(\left[\gamma^0 iS_<,i\Sigma_>\gamma^0\right]_{\alpha\beta}-\left[\gamma^0iS_>,i\Sigma_<\gamma^0\right]_{\alpha\beta}\right),  \label{eq:S<-sim}
\end{align}
where $\partial_t\equiv \partial/\partial t$, $\{A,B\}, [A,B]$ are the anti-commutation and commutation between $A$ and $B$, and $\boldsymbol{ \gamma}, \gamma^0$ are the Dirac matrices.   $\Sigma_{<,>}$ are the Wightman self-energy amplitudes, and $\text{Re}S_R, \text{Re}\Sigma_R$ denote the real part of retarded propagator and retarded self-energy, respectively.  $S_{<,>}$ denote the full lepton Wightman propagators, which, following Appendix~\ref{app:lepton-Wightman}, can be  written as 
\begin{align}\label{eq:S<-approx}
     [iS_<]_{\alpha\beta}&=-2\pi P_L\slashed{k} \delta(k^2-2\delta _{\alpha \gamma}\tilde{m}_{\gamma}^2)\left[\theta(k_0)f_{\gamma\beta}(k_0,t)-\theta(-k_0) \left(\delta_{\gamma\beta}-\bar f_{\gamma\beta}(-k_0,t)\right)\right],
\\[0.2cm]
       \left[iS_{>}\right]_{\alpha \beta}&=-2\pi P_L\slashed{k}\delta(k^2-2\delta _{\alpha \gamma}\tilde{m}_{\gamma}^2)\left[-\theta(k_0)\left(\delta_{\gamma\beta}-f_{\gamma\beta}(k_0,t)\right)+\theta(-k_0)\bar f_{\gamma\beta}(-k_0,t)\right].\label{eq:S>-approx}
\end{align}
Here, $\alpha, \beta, \gamma$ are flavour indices with the repeated ones being summed,  $\sqrt{2}\tilde{m}_\alpha$ denotes the asymptotic thermal  mass of lepton flavour $\alpha$~\cite{Weldon:1982bn,Drewes:2013iaa,Li:2023ewv}, and  the chirality operator $P_L=(1-\gamma_5)/2$ selects only one chirality state for a given massless lepton or anti-lepton. Note that the lepton thermal mass matrix will be diagonal in the basis where $y, M$ in \eqref{eq:lag} are diagonal,  provided that the corrections from neutrino Yukawa matrix $y'$ is smaller than $y$.  The quantity $f_{\alpha\beta}$ ($\bar f_{\alpha\beta}$) is \textit{occupation-number matrix} of leptons (anti-leptons). In the quasi-thermal limit, these quantities may be described by 
\begin{align}
 f_{\alpha\beta}(k_0,t)=\left[\frac{1}{e^{(k_0-\mu_{}(t))/T}+1}\right]_{\alpha\beta}\,,\quad  \bar f_{\alpha\beta}(-k_0,t)=\left[\frac{1}{e^{(-k_0+\mu_{}(t))/T}+1}\right]_{\alpha\beta}\,,
\end{align}
with $\mu_{\alpha\beta}/T\ll 1$. We expect that the momentum dependence of the diagonal entries $\mu_{\alpha\alpha}$ is weak for quasi-thermal particles, but this does not hold in general for the off-diagonal entries~$\mu_{\alpha\beta}$, with~${\alpha\neq \beta}$. For this reason, we will only assume that $\mu_{\alpha\beta}/T\ll 1$, when deriving the off-diagonal  correlations from~$f_{\alpha\beta}$, but we consider that they  still exhibit a strong dependence on the three-momentum norm 
$|{\bf k}|$.  

In the following, we will focus on the kinetic equation from \eqref{eq:S>-sim}. In Appendix~\ref{app:constraint-KB}, we will provide a consistent check of the constraint equation from \eqref{eq:S<-sim}. For thermally resummed leptons that receive corrections from gauge and charged-lepton Yukawa interactions in the thermal plasma, we realize that the commutator between $iS_{<,>}$ and $\bf k\cdot \boldsymbol\gamma$ vanishes. This also holds if we use $iS_{<,>}\propto \slashed{k}+ \slashed{u}$, where $u_\mu$ is the four-velocity of the background plasma; see Appendix~\ref{app:lepton-Wightman} for more details. Another easy way to see this is that we will finally take the Dirac trace to obtain the kinetic equation for distribution functions encoded in the Wightman propagators. The real part of the full retarded propagator, $\text{Re}S_{R}$, corresponds to off-shell propagation, which is independent of the occupation-number matrix $f_{\alpha\beta}$ and hence is diagonal in flavour space.  Then, by parametrizing $\Sigma_{<,>}, \text{Re}S_R\propto \slashed{k}+\slashed{u}$, it is easy to check that the commutator vanishes in the spinor space. After these treatments, the kinetic equation reduces to 
\begin{align}\label{eq:kinetic} i\partial_t[\gamma^0iS_{<(>)}]_{\alpha\beta}-\left[\text{Re}\Sigma_{R}\gamma^0,\gamma^0iS_{<(>)}\right]_{\alpha\beta}
=-\frac{i}{2}\left( \mathcal{C}_{\alpha\beta}+\mathcal{C}^\dagger_{\alpha\beta}\right),
\end{align}
where the commutation as the source of \textit{coherent lepton-flavour oscillations} will become clear later, and the collision rate is defined by
\begin{align}\label{eq:collision-rate}
   \mathcal{C}_{\alpha\beta}\equiv  [i\Sigma_>]_{\alpha\gamma}[iS_<]_{\gamma\beta}-[i\Sigma_<]_{\alpha\gamma}[iS_>]_{\gamma\beta}\,,
\end{align}
with the Hermitian conjugate of $\mathcal{C}$ acting on both the spinor and flavour space. Note that we can perform time-dependent unitary rotation $U$ to the basis where $S_<$ or $S_>$ is diagonal. In doing this, we can make all the other quantities diagonal in \eqref{eq:kinetic}~\cite{Prokopec:2003pj,Beneke:2010dz}, but at the cost of introducing a $\partial_t U$ term. Equivalently, without rotating  to the diagonal basis, we can take the trace over the flavour space in the end. This will correspond to evaluating the total lepton asymmetry, which is flavour invariant as $\text{Tr}[U^\dagger S_{<(>)}U]=\text{Tr}[S_{<(>)}]$.

\subsection{Kinetic equation for lepton-number asymmetry}\label{sec:lepton-asymmetry}

We can extract the lepton and anti-lepton number densities from Wightman propagators via
\begin{align}
    n_{\alpha\beta}(t)&=\int \frac{{\rm d}^3 {\bf k}}{(2\pi)^3}f_{\alpha\beta}({|{\bf k}|, t})=
        -\int  \frac{{\rm d}^3 {\bf k}}{(2\pi)^3}\int_0^\infty \frac{{\rm d}k_0}{2\pi} \text{Tr}[\gamma^0iS_<]_{\alpha\beta}\,,
        \\[0.2cm]
        \bar n_{\alpha\beta}(t) &=\int \frac{{\rm d}^3 {\bf k}}{(2\pi)^3}\bar f_{\alpha\beta}({|{\bf k}|, t})=\int \frac{{\rm d}^3 {\bf k}}{(2\pi)^3}\int_{-\infty}^0 \frac{{\rm d}k_0}{2\pi} \text{Tr}[\gamma^0 iS_>]_{\alpha\beta}\,.
\end{align}
Note that in the above equations, both $\gamma^0 [iS_<]_{\alpha\beta}=-\langle \gamma^0\bar \psi_\beta \psi_\alpha\rangle$ and $\gamma^0 [iS_>]_{\alpha\beta}=\langle \gamma^0 \psi_\alpha \bar \psi_\beta\rangle$ under the Dirac trace $\text{Tr}$ (not acting on the flavour space) have a clear correspondence to the generalized Noether charge $j^0_{\alpha\beta}=\bar \psi_\beta \gamma^0 \psi_\alpha$,  given that
\begin{align}
    -\text{Tr}[ \gamma^0\bar \psi_\beta \psi_\alpha]=\text{Tr}[ \gamma^0 \psi_\alpha\bar \psi_\beta ]=\text{Tr}j^0_{\alpha\beta}\,,
\end{align}
holds under the equal-time anti-commutation relation $\{\psi_\alpha(x), \bar \psi_\beta (y) \}=\gamma^0_{\alpha\beta} \delta^{(3)}({\bf x-y})$~\cite{Srednicki:2007qs}. In particular, the flavour-diagonal entries correspond to the approximately conserved lepton-number charge. 

The commutator in \eqref{eq:kinetic} under the four-momentum integration gives\footnote{We will not use the upper and lower indices in this paper. For clarity,  repeated indices on the one side of equations correspond to dummy indices only if they are not repeated on the other side; otherwise, they should be treated as free indices. For example, $\alpha, \beta$ are free indices in the equation: $X_{\alpha\beta}=Y_\alpha Z_{\alpha\beta}$.} 
\begin{align}
    & -\int  \frac{{\rm d}^3 {\bf k}}{(2\pi)^3}\int_0^\infty \frac{{\rm d}k_0}{2\pi} \text{Tr}\left[\text{Re}\Sigma_{R}\gamma^0,\gamma^0iS_{<}\right]_{\alpha\beta}
     \nonumber\\[0.2cm]
     &=-\int  \frac{{\rm d}^3 {\bf k}}{(2\pi)^3}\int_0^\infty \frac{{\rm d}k_0}{2\pi} (4\pi) \delta(k^2-2\tilde{m}_\alpha^2) f_{\alpha\beta}(k_0, t)I(a,b)_{\alpha\beta}\,,\label{eq:oscillation-12}
\end{align}
and 
\begin{align}
    & \int  \frac{{\rm d}^3 {\bf k}}{(2\pi)^3}\int_{-\infty}^0 \frac{{\rm d}k_0}{2\pi} \text{Tr}\left[\text{Re}\Sigma_{R}\gamma^0,\gamma^0iS_{>}\right]_{\alpha\beta}
     \nonumber\\[0.2cm]
     &=\int  \frac{{\rm d}^3 {\bf k}}{(2\pi)^3}\int_{-\infty}^0 \frac{{\rm d}k_0}{2\pi}  (4\pi) \delta(k^2-2\tilde{m}_\alpha^2) \bar f_{\alpha\beta}(k_0, t) I(a,b)_{\alpha\beta}\,,
\label{eq:oscillation-22}
\end{align}
where 
\begin{align}\label{eq:I_ab}
I(a,b)_{\alpha\beta}\equiv  \left[a_\alpha(k_0)-a_\beta(k_0)\right]k^2+\left[b_\alpha(k_0)-b_\beta(k_0)\right]k_0\,,
\end{align}
and we have defined 
\begin{align}\label{eq:ab-def}
[\text{Re}\Sigma_R(k)]_{\alpha\beta}\equiv -a_{\alpha\beta} P_R\slashed{k}P_L-b_{\alpha\beta} P_R\slashed{u}P_L\,,
\end{align}
with the minus sign for conventional purpose. The real coefficients $a_{\alpha\beta},b_{\alpha\beta}$  are diagonal in the lepton-flavour space in the approximation of small contributions from neutrino Yukawa couplings, and are well known in the Hard-Thermal-Loop (HTL) approximation~\cite{Weldon:1982bn,Li:2023ewv}, 
	\begin{align}\label{eq:aRi}
	a_{\alpha }(k_0, {\bf k})&=\frac{\tilde{m}^2_{\alpha}}{|{\bf{k}}|^2}\left[1+\frac{k_0}{2|{\bf {k}}|}\ln\left(\frac{k_0-|{\bf {k}}|}{k_0+|{\bf {k}}|}\right)\right],
		\\[0.2cm]
		b_{\alpha }(k_0, {\bf k})&=-\frac{\tilde{m}^2_{\alpha}}{|{\bf k}|}\left[\frac{k_0}{|{\bf k}|}-\frac{1}{2}\left(1-\frac{k_0^2}{|{\bf k}|^2}\right)\ln\left(\frac{k_0-|{\bf k}|}{k_0+|{\bf k}|}\right)\right],\label{eq:bRi}
	\end{align}
satisfying the  properties $a_{\alpha}(k_0)=a_{\alpha}(-k_0), b_{\alpha}(k_0)=-b_{\alpha}(-k_0)$. The thermal masses of leptons from SM interactions read
	\begin{align}\label{eq:thermalmass}
		\tilde{m}_{\alpha}^2=\left(\frac{3}{32}g_2^2+\frac{1}{32}g_1^2+\frac{1}{16}y_{\alpha}^2\right)T^2\,,
	\end{align}
with $g_2, g_1$ the $SU(2)_L$ and $U(1)_Y$ gauge couplings, respectively, and $T$ the plasma temperature. 

To further simplify \eqref{eq:oscillation-12} and \eqref{eq:oscillation-22}, we notice that in the HTL approximation, $k^2=2\tilde{m}_\alpha^2\sim \mathcal{O}(g^2) T^2$ and $|k_0|\sim T$, such that the $a_\alpha-a_\beta$ term from \eqref{eq:I_ab} is of higher order in gauge and Yukawa couplings. Using $b_\alpha(k_0=\sqrt{{|{\bf k}}|^2+2\tilde m_\alpha^2})\approx b_\alpha(k_0=|{\bf k}|)= -\tilde{m}_\alpha^2/|{\bf k}|$, we  define
\begin{align}\label{eq:tilde-b}
\quad \tilde{b}_\alpha (T)&\equiv -\frac{\tilde{m}_\alpha^2}{T}\,,
\\[0.2cm]
\langle n_{\alpha\beta} (T,t)\rangle &\equiv \int \frac{{\rm d}^3 {\bf k}}{(2\pi)^3}\frac{T}{|{\bf k}|}f_{\alpha\beta}({|{\bf k}|, t})\,,
\\[0.2cm]
\langle \bar n_{\alpha\beta} (T,t)\rangle &\equiv \int \frac{{\rm d}^3 {\bf k}}{(2\pi)^3}\frac{T}{|{\bf k}|}\bar f_{\alpha\beta}({|{\bf k}|, t})\,,
\end{align}
allowing us to simplify \eqref{eq:oscillation-12} and \eqref{eq:oscillation-22} as 
 \begin{align}
   -&\int  \frac{{\rm d}^3 {\bf k}}{(2\pi)^3}\left(b_\alpha (k_0=|{{\bf k}}|)-b_\beta (k_0=|{{\bf k}}|)\right) f_{\alpha\beta}(|{\bf k}|,t)
         \nonumber \\[0.2cm]
     &= -\left(\tilde b_\alpha (T)-\tilde b_\beta (T)\right)\langle n_{\alpha\beta}(T,t)\rangle 
       \nonumber     \\[0.2cm]
     &=-[\tilde b,\langle n\rangle]_{\alpha\beta}\,,\label{eq:oscillation-15}
\end{align}
and 
\begin{align}
&-\int  \frac{{\rm d}^3 {\bf k}}{(2\pi)^3} \left(b_\alpha(k_0=-|{{\bf k}}|)-b_\beta(k_0=-|{{\bf k}}|)\right)\bar f_{\alpha\beta}(|{\bf k}|,t)
      \nonumber    \\[0.2cm]
     &= \left(\tilde b_\alpha(T)-\tilde b_\beta(T)\right)\langle \bar n_{\alpha\beta}(T,t)\rangle
   \nonumber     \\[0.2cm]
     &=[\tilde b,\langle \bar n\rangle]_{\alpha\beta}\,, \label{eq:oscillation-25}
\end{align}
respectively. 

Consequently,  we obtain the kinetic equation for $n_{\alpha\beta}$  from \eqref{eq:kinetic} as
\begin{align}\label{eq:dn/dt}
    \frac{{\rm d}n_{\alpha \beta}}{{\rm d}t}-i [\tilde b,\langle n\rangle]_{\alpha\beta}=\frac{1}{2}\int  \frac{{\rm d}^3 {\bf k}}{(2\pi)^3}\int_0^\infty \frac{{\rm d}k_0}{2\pi}\text{Tr}\left(\mathcal{C}_{\alpha\beta}+\mathcal{C}^\dagger_{\alpha\beta}\right),
\end{align}
and analogously for $\bar n_{\alpha\beta}$,
\begin{align}\label{eq:dnb/dt}
    \frac{{\rm d}\bar n_{\alpha \beta}}{{\rm d}t}+i [\tilde b,\langle \bar n\rangle]_{\alpha\beta}=-\frac{1}{2}\int  \frac{{\rm d}^3 {\bf k}}{(2\pi)^3}\int_{-\infty}^0 \frac{{\rm d}k_0}{2\pi}\text{Tr}\left(\mathcal{C}_{\alpha\beta}+\mathcal{C}^\dagger_{\alpha\beta}\right).
\end{align}
Therefore, the lepton-number asymmetry $\Delta n\equiv n-\bar n$ yields
\begin{align}\label{eq:dDn/dt}
    \frac{{\rm d}\Delta n_{\alpha \beta}}{{\rm d}t}-i \left(\tilde b_\alpha-\tilde b_\beta\right)\langle \Sigma n_{\alpha\beta}\rangle=\frac{1}{2}\int  \frac{{\rm d}^3 {\bf k}}{(2\pi)^3}\int_{-\infty}^\infty \frac{{\rm d}k_0}{2\pi}\text{Tr}\left(\mathcal{C}_{\alpha\beta}+\mathcal{C}^\dagger_{\alpha\beta}\right),
\end{align}
and the equation for the sum $\Sigma n\equiv n+\bar n$ reads
\begin{align}\label{eq:dSn/dt}
    \frac{{\rm d}\Sigma n_{\alpha \beta}}{{\rm d}t}-i \left(\tilde b_\alpha-\tilde b_\beta\right)\langle\Delta n_{\alpha\beta}\rangle=\frac{1}{2}\int  \frac{{\rm d}^3 {\bf k}}{(2\pi)^3}\int_{-\infty}^\infty \frac{{\rm d}k_0}{2\pi}\text{sign}(k_0)\text{Tr}\left(\mathcal{C}_{\alpha\beta}+\mathcal{C}^\dagger_{\alpha\beta}\right).
\end{align}
Given that the integration of the distribution function over the spatial momentum ${\bf k}$ is  dominated by $|{\bf k}|\simeq T$, we expect that $\langle n(T,t)\rangle\approx n(t), \langle \bar n(T,t)\rangle\approx \bar n(t)$ and hence may regard the term, ${\tilde b_\alpha-\tilde b_\beta}$, as a source for enhancing the off‑diagonal correlations $\Delta n_{\alpha\beta}$, as will become clear soon. 

Equations${}$~\eqref{eq:dn/dt} and~\eqref{eq:dnb/dt} can be seen as an analogy of  neutrino oscillation in a plasma~\cite{Wolfenstein:1977ue,Mikheyev:1985zog,Sigl:1993ctk,Akhmedov:1998qx,BhupalDev:2014oar,BhupalDev:2014pfm}, but here the oscillation  is triggered by the flavour-dependent thermal mass rather than the vacuum mass. Such oscillations induced by lepton-flavour mixing in the thermal plasma was noticed earlier at high temperatures~\cite{Beneke:2010dz,Garbrecht:2012pq}, where the off-diagonal correlations $n_{\alpha\beta},\bar n_{\alpha\beta}$ are damped by gauge and charged-lepton Yukawa interactions. Such thermal lepton-flavour oscillations at high temperatures was also considered in \cite{Hamada:2016oft} at the post-inflationary epoch. The oscillation rate therein was determined by large lepton-flavour dependent Yukawa couplings, and as a result,  such off-diagonal correlation effects would not yield a large resonant enhancement from the $\tilde b_\alpha-\tilde b_\beta$ term, since the new-physics contributions to lepton thermal masses are sufficiently large. This underlies a crucial difference between the high- and low-scale leptogenesis, as will become more clear below.  

By calculating the one-loop collision rates in~\eqref{eq:dn/dt} and~\eqref{eq:dnb/dt}, we confirm in Appendix~\ref{app:collision-rate-1} the damping effects of $n_{\alpha\beta}$ and $\bar n_{\alpha\beta}$ correlations, which are consistent with former observations~\cite{Beneke:2010dz,Garbrecht:2012pq}.  It implies that oscillation effects among lepton flavours will quickly be erased,  if we focus on the one-loop collision rates. However, it is well known that in perturbative thermal QFT, thermal resummation effects can readily change the naive coupling counting such that higher-order loop corrections could be as important as the lower-loop corrections; see e.g.,~\cite{Elmfors:1997tt,Boyarsky:2020cyk,Li:2022dkc,Li:2022rde} for recent studies. This is especially the case for the evolution of $n_{\alpha\beta}$ and $\bar n_{\alpha\beta}$ correlations, where one-loop collision rates typically contribute to damping or washout effects while  two-loop collision rates contribute as driving force to increasing the correlations. A detailed discussion on the evolution of $\Delta n_{\alpha\beta}$ and $\Sigma n_{\alpha\beta}$, as well as collision rates at one and two loops, will also be presented in Section~\ref{sec:mix-osc-cor} and in Appendix~\ref{app:collision-rate-1}. 

Following~\eqref{eq:dDn/dt} and~\eqref{eq:dSn/dt}, we can outline the general evolution of $\Sigma n_{\alpha\beta}$ and $\Delta n_{\alpha\beta}$  as follows. $\Sigma n_{\alpha\beta}$ suffers from strong damping effects from gauge interactions while the dominant damping source in $\Delta n_{\alpha\beta}$ is from charged-lepton Yukawa interactions. At one-loop order, the source term that can create a non-zero $\Sigma n_{\alpha\beta}$ from an initial $\Sigma n_{\alpha\beta}=0$ is induced by the nonthermal Yukawa interactions. A non-zero $\Sigma n_{\alpha\beta}$ would  further source a non-zero $\Delta n_{\alpha\beta}$ via the coupled kinetic equations. However, at one-loop level, there is no other  source term for $\Delta n_{\alpha\beta}$ such that the damping force from charge-lepton Yukawa couplings will quickly dilute $\Delta n_{\alpha\beta}$. This observation is consistent with \cite{Garbrecht:2012pq}\footnote{In fact, by imposing kinetic equilibrium and local chemical equilibrium on leptons within the SM interactions at $T\ll 10^5$~GeV, one can reproduce the enhancement factor $1/(\tilde b_\alpha-\tilde b_\beta)$ in \eqref{eq:Dn_ab}; see e.g., (15)-(20) in \cite{Garbrecht:2012pq}.}. Nevertheless, in the asymptotic regime where $\Sigma n_{\alpha\beta}\to 0$, a new source term arising at two loops from non-thermal neutrino Yukawa interactions exists, which can still increase $\Delta n_{\alpha\beta}$ until $\text{d}\Delta n_{\alpha\beta}/\text{d}t\to  0$. This new two-loop source term, which has not been identified before in former studies~\cite{Garbrecht:2012pq}, is non-negligible, due to  thermal resummation effects of soft leptons~\cite{Kanemura:2024dqv}. In fact, it can drive $\Delta n_{\alpha\beta}$ to sizeable values leading to observable leptogenesis, as we will see in Seciton~\ref{sec:num}. 

In line with the discussion given above, we may well approximate $\Sigma n_{\alpha\beta}$ as follows:
\begin{equation}
    \label{eq:SigmaApprox}
\Sigma n_{\alpha\beta}\: =\: 2n{}^{\rm eq}_{\ell}\, \delta_{\alpha\beta}\: +\: \mathcal{O}(\mu^2/T^2)\,,
\end{equation}
where $n^{\rm eq}_\ell$ is the equilibrium number density of one lepton species.  The off-diagonal correlation $\Delta n_{\alpha\beta}$ is enhanced by $(\tilde b_{\alpha}-\tilde b_\beta)^{-1}$ at leading order of small chemical potentials. To see this, we derive from \eqref{eq:dSn/dt} with ${\rm d}\Sigma n_{\alpha\beta}/{\rm d}t=0$ for $\alpha\neq \beta$:
\begin{align}
   \langle \Delta n_{\alpha\beta}\rangle 
   &=\frac{-2i y'_{\beta j}y^{\prime \dagger}_{j \alpha}\Delta m_j^2}{\tilde b_\alpha-\tilde b_\beta}\label{eq:Dn_ab}
  \\
   &\times\int {\rm d}\Pi_{\ell}{\rm d}\Pi_{\phi}{\rm d}\Pi_{N}  (2\pi)^4 \tilde{\delta}^{(4)}(k_\ell+k_\phi-k_N)\delta f_{N_j}(E_N) \left(f_\phi^{\rm eq}(E_\phi)+f_\ell^{\rm eq}(E_\ell)\right), \nonumber 
\end{align}
where  the equilibrium distribution functions for the Higgs, leptons and neutrinos read,
\begin{align}
    f_\phi^{\rm eq}(E)=\frac{1}{e^{E/T}-1}\,,\quad f_\ell^{\rm eq}(E)=\frac{1}{e^{E/T}+1}\,,\quad f_N^{\rm eq}(E)=\frac{1}{e^{E/T}+1}\,.
\end{align}
As we mentioned before, for massless lepton doublets, we will neglect the thermal mass corrections in the distribution function such that $E_\ell\approx |{\bf k}_\ell|$. For later notational convenience, we define the following quantities
\begin{align}
    {\rm d}\Pi_i&\equiv \frac{{\rm d}^3 {\bf k}_i}{(2\pi)^3 2E_i}\,,  
               \\[0.2cm]
                 \delta f_{N_j}&\equiv f_{N_j}-f_{N_j}^{\rm eq}\,, \quad \delta f_{\alpha\beta}\equiv f_{\alpha\beta}-f_{\ell}^{\rm eq}\delta_{\alpha\beta}\,,
                        \\[0.2cm]
   \Delta m_j^2&\equiv m_\phi^2-\tilde{m}^2_\ell -M_j^2\,,\label{eq:Deltam-def}
       \\[0.2cm]
   \tilde{\delta}^{(4)}(k_\ell+k_\phi-k_N)&\equiv \delta^{(3)}({\bf k}_\ell+{\bf k}_\phi-{\bf k}_N)\delta(E_\ell- E_\phi+E_N) \,.\label{eq:tilde-delta}
\end{align}
In~\eqref{eq:Deltam-def}, $m_\phi^2$ denotes the squared mass of the Higgs doublet, including both the thermal and vacuum masses~\cite{Carrington:1991hz}. Instead,  $\tilde{m}_\ell$ corresponds to a common thermal lepton mass, which is obtained by neglecting the charged-lepton Yukawa couplings when evaluating the thermal mass difference between the SM Higgs and the lepton doublets.

We may now employ the relation, 
\begin{align}\label{eq:Delta-nab-def}
    \Delta n_{\alpha\beta}=2\int \frac{{\rm d}^3 {\bf k}_\ell}{(2\pi)^3}\,\delta f_{\alpha\beta}(E_\ell)\,,
\end{align}
in order to extract $\delta f_{\alpha\beta}$ (with $\alpha\neq \beta$) from \eqref{eq:Dn_ab} as 
\begin{align}\label{eq:delta-fab}
    \delta f_{\alpha\beta}(E_\ell)=\frac{-i y'_{\beta j}y^{\prime \dagger}_{j \alpha}\Delta m_j^2}{16\pi (\tilde b_\alpha-\tilde b_\beta)}\frac{1}{E_\ell T}\int_{E_{\rm low}}^{\infty}{\rm d}E'_N \,\delta f_{N_j}(E'_N)\left(f_\phi^{\rm eq}(E_\ell+E'_N)+f_\ell^{\rm eq}(E_\ell)\right).
\end{align}
In the above, the integration low limit results from the angular dependent part of the integration over the Dirac $\delta$-function, yielding the following:
\begin{align}\label{eq:Elow}
E_{\rm low}\approx \frac{\Delta m_j^2}{4|{\bf k_\ell}|}-M_j^2 \frac{|{\bf k_\ell}|}{\Delta m_j^2}\approx\frac{ m_\phi^2}{4|\bf {k_\ell}|}\,,
\end{align}
in the limit of $E_\ell\approx |\bf k_\ell|$. Note that \eqref{eq:delta-fab} is derived by neglecting the higher-order asymmetry between the neutrino and anti-neutrino distribution functions, i.e., $\delta f_{N_j}\approx \delta \bar f_{N_j}$.  This departure from thermal equilibrium can only be evaluated numerically for low-scale freeze-in leptogenesis,  since there is no simple kinetic‑equilibrium approximation analogous to the strong‑washout limit~\cite{Garbrecht:2012pq,BhupalDev:2014pfm,BhupalDev:2014oar}, and hence contributes as another difference between the high- and low-scale leptogenesis.  

The result given in \eqref{eq:delta-fab} for $\delta f_{\alpha\beta}$ shows clearly a resonant enhancement from the difference $\tilde b_\alpha-\tilde b_\beta\propto  y_\alpha^2-y_\beta^2$ due to the cancellation of flavour-universal gauge contributions and the smallness of charged-lepton Yukawa couplings.  We may further infer from the solution shown in \eqref{eq:delta-fab} that for nonthermal right-handed neutrinos, if their vacuum mass are not nearly degenerate, there would not be significant enhancement from the analogous term $b_i-b_j\propto M^2_i-M_{j}^2$ and hence sterile neutrino flavour oscillations become sub-dominant to modify the evolution of neutrino distribution functions $f_{{N_j}}$.

The evolution of the total lepton asymmetry from lepton doublets can be obtained from \eqref{eq:dDn/dt} by taking the trace in flavour space. Defining
\begin{align}
\Delta n_\ell\equiv \text{Tr}(\Delta n_{\alpha\beta})=\sum_{\alpha=1}^3\Delta n_{\alpha\alpha} \,,
\end{align}
we arrive at 
\begin{align}\label{eq:dDn/dt-2}
    \frac{{\rm d}\Delta n_{\ell}}{{\rm d}t}
   =&-2\sum_{\alpha,\gamma,j} y'_{\gamma j}y^{\prime \dagger}_{j \alpha}\Delta m_j^2 
  \nonumber  \\[0.2cm]
    &\times \int {\rm d}\Pi_{\ell} {\rm d}\Pi_{ \phi}{\rm d}\Pi_{N} (2\pi)^4  \tilde{\delta}^{(4)}(k_\ell+k_\phi-k_N)\left[\delta_{\gamma\alpha} (I^\phi_j-I^N_j)+I^{\ell}_{\gamma\alpha j}\right],
\end{align}
where 
\begin{align}
    I^\phi_j&\equiv \delta f_\phi (E_\phi) \left[1-f_\ell^{\rm eq}(E_\ell)-f_{N_j}^{\rm eq}(E_N)-\frac{1}{2}\left(\delta f_{N_j}(E_N)+\delta \bar f_{N_j}(E_N)\right)\right],
    \\[0.2cm]
   I^N_j &\equiv \frac{1}{2}\left(\delta f_{N_j}(E_N)-\delta \bar f_{N_j}(E_N)\right)\left(f_\ell^{\rm eq}(E_\ell)+f_\phi^{\rm eq}(E_\phi)\right),
   \\[0.2cm]
   I^{\ell}_{\gamma\alpha j}&\equiv \delta f_{\gamma\alpha}(E_\ell)\left[f_\phi^{\rm eq}(E_\phi)+f_{N_j}^{\rm eq}(E_N)+ \frac{1}{2}\left(\delta f_{N_j}(E_N)+\delta \bar f_{N_j}(E_N)\right)\right].
\end{align}
The presence of $\delta f_\phi$ in $I_j^\phi$ and $\delta f_{\gamma\alpha}$ in $I^\ell_{\gamma\alpha j}$ allows us to neglect the difference between $\delta f_{N_j}$ and $\delta \bar f_{N_j}$ by taking $\delta f_{N_j}+\delta\bar f_{N_j}\approx 2\delta f_{N_j}$. Nevertheless,  the difference $\delta f_{N_j} -\delta \bar f_{N_j}$ reflects the helicity-number asymmetry between neutrinos and anti-neutrinos, which could be at the similar order of $\delta f_\phi$ and $\delta f_{\alpha\alpha}$. 
It should be emphasized that when calculating the collision rates $\mathcal{C},\mathcal{C}^\dagger$ in~\eqref{eq:dDn/dt}, we should keep the full dependence of the off-diagonal correlation $f_{\gamma\beta}$ in the external lepton propagators $[iS_{<,>}]_{\gamma\beta}$, with the solution given by \eqref{eq:delta-fab}. In doing this, it turns out that  an effect of  $\Delta n_\ell\propto\mathcal{O}(y^{\prime 4})$ would arise from the CP-violating but $L_{\rm tot}$-conserving Higgs decay: a $y^{\prime 2}$ from the one-loop self-energy amplitudes and the other $y^{\prime 2}$ from the lepton off-diagonal correlations.  

For simplicity, let us now neglect the washout rate from the terms $I^\phi_j$ and $I^N_j$ in~\eqref{eq:dDn/dt-2}. In this approximation, we obtain  
\begin{align}
    \frac{{\rm d}\Delta n_{\ell}}{{\rm d}t}\:
&=\: \sum_{\substack{\alpha,\gamma \\ \alpha\neq \gamma}}\sum_{i,j}\frac{\Delta m_i^2\Delta m_j^2\,  \text{Im}(y'_{\gamma i}y^{\prime *}_{\alpha i}y'_{\alpha j}y^{\prime *}_{\gamma j})}{256\pi^4(\tilde{m}_\gamma^2-\tilde{m}_\alpha^2)}\int \frac{dE_\ell}{E_\ell}\int_{E_{\rm low}}^\infty dE_N \int_{E_{\rm low}}^\infty dE'_N  \mathcal{F}_{ij}\,,\label{eq:CPV-l}
\end{align}
where $E_{\rm low}$ is given in \eqref{eq:Elow}, and the statistics function $\mathcal{F}_{ij}$ reads
\begin{align}\label{eq:Fij-def}
  \mathcal{F}_{ij}\, \equiv\,  \Big(f_\phi^{\rm eq}(E_\ell+E_N)+f^{}_{N_i}(E_N)\Big)\,\Big(f_\ell^{\rm eq}(E_\ell)+f_\phi^{\rm eq}(E_\ell+E'_N)\Big)\,\delta f_{N_j}(E'_N)\,.
\end{align}
Before performing the numerical analysis, we provide a two-loop diagrammatic method in the next section   to evaluate the CP-violating source, building upon the KB kinetic equation for singlet neutrinos. In doing this, we confirm the source rate given in \eqref{eq:CPV-l}. In addition, we will also make a  comparison between the TRL with a CP-violating source given in~\eqref{eq:CPV-l} and the RL through quasi-degenerate singlet neutrinos, identifying therein the dominance regime of the former over the latter.

\subsection{Two-loop diagrammatic method for lepton-number asymmetry}\label{sec:2loo-diagram}
Due to approximate conservation of $L_{\rm tot}$ at $T\gg M$, we can calculate the lepton-number (helicity) asymmetry generated in the singlet neutrino sector. If the washout effects can be neglected, we will  obtain the SM lepton asymmetry simply via $\Delta L_{\rm SM}\approx -\Delta L_{N}$. In this section, we build the KB kinetic equations for singlet neutrinos and calculate the CP-violating source from two-loop self-energy amplitudes of neutrinos.  

At first sight, this method is different from that derived in the previous sections since the CP-violating source presented in \eqref{eq:CPV-l} results from the off-diagonal correlations $\delta f_{\alpha\beta}$, which is induced by the nonthermal neutrino Yukawa interactions at one-loop order and is significantly enhanced by the thermal-lepton oscillation factor $(\tilde b_\alpha-\tilde b_\beta)^{-1}$. However, both results will correspond to a CP-violating effect at $\mathcal{O}(y^{\prime 4})$. Indeed, we will confirm that both methods lead to a consistent CP-violating source obeying the basic equation,
\begin{align}
   \label{eq:L-conserved}
  \frac{{\rm d}\Delta n_\ell}{{\rm d}t}\:=\:-\,\frac{{\rm d}\Delta n_N}{{\rm d}t}\,.
\end{align}
More technical details have been relegated in Appendix~\ref{app:RFL-N}, where we further calculate the lepton-number asymmetry including scenarios with  quasi-degenerate neutrinos. These details can be used to allow for more direct comparisons between TRL and RL.  The two-loop diagrammatic method for calculating the CP-violating source may be seen as a cross check of the $L_{\rm tot}$ conservation, but we find that the consistency requires some non-trivial identifications. For instance, only when we take the asymptotic thermal mass of the leptons to be $\sqrt{2}\tilde m$ rather than $\tilde m$, can we achieve consistency.

We will first present the calculations of low-scale leptogenesis due to thermal lepton mixing in Section~\ref{sec:thermal-l-osc}, following the technical details presented in \cite{Kanemura:2024dqv,Kanemura:2024fbw}, and then provide the RL from CP- and $L_{\rm tot}$-violating Higgs decay, following the calculations and derivations documented in Appendix~\ref{app:RFL-N}. The followed simple comparison between the two channels aims to provide an order-of-magnitude estimate for the dominance regime where TRL dominates low-scale leptogenesis.

\subsubsection{Resonant thermal lepton mixing}\label{sec:thermal-l-osc}
Let us define the neutrino asymmetries as 
\begin{align}
    \Delta n_N\equiv \sum_{i} \Delta n_{N_i}\,,\quad \Delta n_{N_i}=n_{N_i}-\bar n_{N_i}\,,
\end{align}
summing over all singlet neutrinos contributing to leptogenesis. The evolution of $\Delta n_{N_i}$ is determined by the following KB equation~\cite{Kanemura:2024dqv}
\begin{align}\label{eq:dDeltanN/dt}
	\frac{{\rm d}\Delta n_{N_i}}{{\rm d}t}=\frac{1}{2}\int \frac{{\rm d}^4p}{(2\pi)^4}\text{Tr}\left(i {\Sigma}_{N_i}^>i{S}^<_{N_i}-i{\Sigma}_{N_i}^< i{S}^>_{N_i}\right),
\end{align}
where the CP-violating collision rates read
\begin{align}\label{eq:Sigma_N<}
i{\Sigma}_{N_i}^<(p)&=2\sum_{\substack{\alpha,\beta
    \\ \alpha\neq \beta}}\,\sum_{j}y'_{\alpha i} y_{\alpha j}^ {\prime*}y_{\beta i}^{\prime *}y'_{\beta j }\int \frac{{\rm d}^4 p_\ell}{(2\pi)^4}\frac{{\rm d}^4 p_\phi}{(2\pi)^4}(2\pi)^4 \delta^4(p+p_\phi-p_\ell)P_L [i{S}^<_{\ell}]_{\alpha\beta} iG_\phi^>\,,
	\\[0.2cm]
	i{\Sigma}_{N_i}^>(p)&=2\sum_{\substack{\alpha,\beta
    \\ \alpha\neq \beta}}\,\sum_{j}y_{\alpha i}^{\prime *} y'_{\alpha j}y'_{\beta i}y^{\prime *}_{\beta  j}\int \frac{{\rm d}^4 p_\ell}{(2\pi)^4}\frac{{\rm d}^4 p_\phi}{(2\pi)^4}(2\pi)^4 \delta^4(p+p_\phi-p_\ell)P_L [i{S}^>_{\ell}]_{\beta \alpha}  iG_\phi^<\,,\label{eq:Sigma_N>}
\end{align} 
with the factor of 2 taking into account the gauge-doublet degeneracy, and the relevant propagators are collected in Appendix~\ref{app:RFL-N}.  Here, we do not use the flavour-covariant formalism for the neutrino flavours. As mentioned below \eqref{eq:Elow}, for nonthermal neutrinos without  a quasi-degenerate mass spectrum, both the thermal corrections and the coherent flavour oscillation term are not the leading effect to induce the CP-violating source. For this reason, we consider only the flavour-diagonal part of the KB equations for singlet neutrinos.

\begin{figure}[t]
	\centering
\includegraphics[scale=0.6]{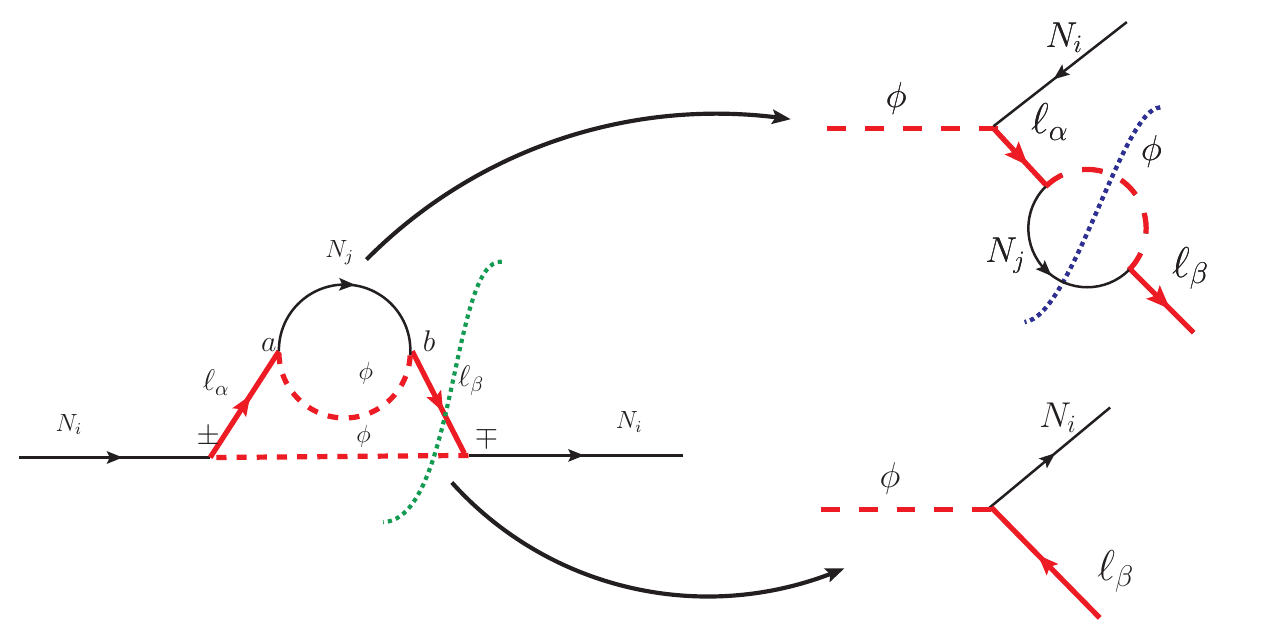} 
	\caption{\label{fig:2+1loop-Nself}The two-loop lepton-number conserving self-energy diagrams of relativistic singlet neutrinos contributing to CP asymmetry, with $\pm$ and $a,b=\pm$ being the CTP indices. A green-dashed thermal cut decomposes the two-loop diagram to a tree-level  decay diagram and the associated one-loop diagram. The blue-dashed thermal cut generates a purely thermal-induced absorptive part in the one-loop Higgs decay. The lepton-flavour correlation is caused by the out-of-equilibrium neutrino Yukawa interactions, significantly being enhanced by thermal resummation from the SM plasma. The red, thick lines for lepton and Higgs doublets reflect thermal corrections.}
\end{figure}

It is worthwhile to mention that if the lepton-doublet Wightman propagators $[iS^>_{\ell}]_{\beta\alpha},[iS^<_{\ell}]_{\alpha\beta}$ are independent of the lepton flavours, the Yukawa product after summing over all SM lepton flavours would  become $(y^{\prime\dagger} y')_{ji}(y^{\prime \dagger} y')_{ij}$, which is real and hence cannot provide a CP-violating phase. We can also infer that if all the charged-lepton flavours were degenerate, there would not be CP violation either.   On the other hand, if  $[iS^>_{\ell}]_{\beta\alpha},[iS^<_{\ell}]_{\alpha\beta}$  are independent of the neutrino flavours, the Yukawa product after summing over the neutrino flavours  would become $(y'y^{\prime \dagger} )_{\alpha\beta}(y'y^{\prime\dagger} )_{\beta \alpha}$, which is also real. Therefore,  the dependence of $i{S}_{\ell}^{<,>}$ on lepton and neutrino flavours is required for a non-vanishing CP-violating source, which is induced by thermal resummation of  lepton propagators $\ell_\alpha, \ell_\beta$ in Fig.~\ref{fig:2+1loop-Nself}  and by different neutrino Yukawa couplings among  $N_i$ flavours. 

Following the calculation steps we document in Appendix~\ref{app:RFL-N}, we can write down a compact result for the evolution of $\Delta n_{N}$ as
\begin{align}\label{eq:dDelta-nN/dt}
	\frac{{\rm d}\Delta n_{N}}{{\rm d}t}=\sum_{\substack{\alpha,\beta
    \\ \alpha\neq \beta}}\,\sum_{\substack{i,j
    \\ i\neq j}}\frac{\text{Im}\left(y'_{\alpha i} y_{\alpha j}^{\prime *}y_{\beta i}^{\prime *}y'_{\beta j}\right)}{256\pi^4 (\tilde{m}^2_\beta-\tilde{m}_\alpha^2)} \int_0^\infty {\rm d}E_\ell \int_{E_{\rm low}}^{\infty}{\rm d}E_N\int_{E_{\rm low}}^{\infty}{\rm d}E'_N\,\mathcal{Q}_\beta\, \mathcal{F}_{ij}\,,
\end{align}
where the momentum function reads
\begin{align}\label{eq:Q-N}
	\mathcal{Q}_\beta=\frac{m_\phi^2-2\tilde{m}_\beta^2}{E_\ell^3}\left[m_\phi^2 E_\ell^2+m_\phi^2\tilde{m}_\beta^2-2\tilde{m}_\beta^2E_\ell\left(E_N+E_\ell+E'_N\right)-2\tilde{m}_\beta^4\right],
\end{align} 
and the statistics function $\mathcal{F}_{ij}$ is given by~\eqref{eq:Fij-def}, with the energy threshold 
 \begin{align}
	E_{\rm low}=E_{\rm low}=\frac{m_\phi^2}{4E_\ell}\,.
\end{align}
We would like to emphasize that the same flavour case $i=j$ cannot provide a CP-violating phase from Yukawa couplings, and hence at least two singlet neutrino flavours are required to have a non-zero CP-violating source. 

In~\eqref{eq:L-conserved}, we derived a basic equation between the lepton-number asymmetries of right-handed neutrinos, $\Delta n_{N}$,  and SM leptons, $\Delta n_{\ell}$, from Higgs decay, in the relativistic limit of the singlet neutrinos. In fact, we have checked that the CP-violating source ${\rm d}\Delta n_{N}/{\rm d}t$ as given in \eqref{eq:dDelta-nN/dt} becomes equal and opposite in sign to the respective one ${\rm d}\Delta n_{\ell}/{\rm d}t$ in~\eqref{eq:CPV-l}, in the approximation of $m_\phi\gg \tilde{m}_\alpha, M_i$. In this way, we can confirm that the CP-violating source given in~\eqref{eq:CPV-l} is  equivalent to that derived  from the neutrino asymmetry via the two-loop diagram shown in Fig.~\ref{fig:2+1loop-Nself}.  Note that in \eqref{eq:L-conserved} the minus sign is guaranteed by the different imaginary part of the Yukawa product. This is nothing but the total lepton-number conservation in Higgs decay, which should work  as we do not include any effects from Majorana-mass insertion in Fig.~\ref{fig:2+1loop-Nself}.   A~condition of reaching such a consistency is that the modified dispersion relation for thermal leptons should be built upon the asymptotic mass $\sqrt{2\tilde{m}}$ rather than the thermal mass $\tilde{m}$, as also discussed  in \cite{Kiessig:2010pr,Kiessig:2011fw,Kiessig:2011ga,Li:2023ewv} that demonstrated some caveats when one uses the Boltzmann kinetic equations to calculate purely thermal-induced processes. 
The resonant enhancement appearing in the flavour-covariant  KB equations of leptons is induced through the one-loop retarded self-energy amplitude $\text{Re}\Sigma_R$ and hence through the commutators $[\tilde b,\langle n\rangle], [\tilde b,\langle\bar n\rangle]$  in \eqref{eq:oscillation-15} and \eqref{eq:oscillation-25}, while the same resonant enhancement appearing in the neutrino kinetic KB equation is induced by resumming the lepton propagators in the two-loop self-energy diagram shown in Fig.~\ref{fig:2+1loop-Nself}. 

Let us now consider the asymmetry yield 
\begin{align}
Y_N\equiv \sum_{i}\frac{\Delta n_{N_i}}{s_{}}\,,
\end{align}
where $s$ is the SM entropy density 
\begin{align}
s_{}= \frac{2\pi^2}{45}g_s(T)T^3\,,
\end{align}
with $g_s(T)$ the effective relativistic degrees of freedom and in particular $g_s(T)\approx 106.75$ at the electroweak scale. Note that it is a common practice to use the yield $Y=n/s$ such that the time derivative under cosmic expansion can be reduced to a total derivative. This treatment neglects the variation of  relativistic degrees of freedom at  electroweak temperatures, which appears in the kinetic equations through the Hubble expansion, through an overall scaling due to 
$\text{d}T/\text{d}t=-HT[1+\text{d}\ln g_s(T)/(3\text{d}\ln T)]^{-1}$ derived from entropy conservation $\text{d}(g_s(T) T^3 a^3)/\text{d}t=0$, as well as through a term proportional to $Y$: $\text{d}\ln g_s(T)/\text{d}\ln T\times Y$. In particular, including the  term proportional to $Y$ can lead to significant corrections for freeze-out leptogenesis and for $Y$ evolving from a large value to the observed baryon asymmetry $Y_B$ in the strong washout regime~\cite{daSilva:2022mrx,Karamitros:2023tqr}. Since we are considering freeze-in based leptogenesis with $Y$ evolving from a small value to $Y_B$, we will neglect this small correction.

We also define the dimensionless variables 
 \begin{align}\label{eq:x-def}
 	x^{(\prime)}_N\equiv \frac{E^{(\prime)}_N}{T}\,,\quad x^{(\prime)}_\ell\equiv \frac{E^{(\prime)}_\ell}{T}\,, \quad z\equiv\frac{m_h}{T}\,,
 \end{align}
 such that the Hubble parameter and entropy density read
 \begin{align}
 H(z)=\frac{m_h^2}{\bar M_{\rm Pl}z^2}\,,\quad s(z)\approx 46.8 \frac{m_h^3}{z^3}\,,
 \end{align}
with $\bar M_{\rm Pl}\equiv 17.5/M_{\rm Pl}\approx 6.97\times 10^{17}$~GeV.  Then from \eqref{eq:dDelta-nN/dt}, we can obtain a semi-analytic and ready-to-use result for the asymmetry yield $Y_N(z)$ from thermal lepton mixing as
   \begin{align}\label{eq:Yofz-N}
Y^{}_{N}(z)\approx 7.64\times 10^{-10}\sum_{\substack{\alpha,\beta
    \\ \alpha\neq \beta}}\,\sum_{\substack{i,j
    \\ i\neq j}}\text{Im}\left( \bar y'_{\alpha i} \bar y^{\prime*}_{\alpha j}\bar y_{\beta i }^{\prime *}\bar y'_{\beta j} \right)\left(\frac{10^{-4}}{y_{\beta}^2-y_{\alpha}^2}\right)I^{}_{ij}(z)\,.
  \end{align}
We will cut the asymptotic time at $z_{\rm sph}\equiv m_h/T_{\rm sph}\approx 0.95$ with $T_{\rm sph}\approx 132$~GeV the sphaleron decoupling temperature~\cite{DOnofrio:2014rug}, though a more precise treatment should also include the non-instantaneous sphaleron decoupling~\cite{Eijima:2017cxr}.  The normalized Yukawa couplings are defined through
\begin{align}
    \bar y'\equiv \frac{y'}{10^{-6}}\,, 
\end{align}
and the dimensionless 4D integral reads 
  \begin{align}\label{eq:4DI-ell}
  I^{}_{ij}(z)\equiv  \int_{0}^{z} {\rm d}z' &\int_0^\infty {\rm d}x_\ell \int_{x_{\rm low}}^{\infty}{\rm d}x_N  \int_{x_{\rm low}}^{\infty}{\rm d} x'_N 
  	\frac{\Theta_z-2c_{\beta}}{x_\ell^3} 
      \\[0.2cm]
  	&\times \left[\Theta_z(x_\ell^2+c_{\beta})-2c_{\beta}x_\ell (x_\ell+x_N+x'_N)-2c_{\beta}^2\right]
  		\nonumber \\[0.2cm]
  	&\times
  	\left(f^{\rm eq}_\phi(x_N+x_\ell)+f^{\rm eq}_{N_i}(x_N)\right)\left(f^{\rm eq}_\ell(x_\ell)+f^{\rm eq}_\phi(x_\ell+x'_N)\right)\delta f_{N_j}(x'_N)\,,\nonumber 
  \end{align}
where 
\begin{align}\label{eq:theta-z}
c_\beta \equiv  \frac{\tilde m_\beta^2}{T^2}\,, \quad x_{\rm low}=\frac{\Theta_z}{4x_\ell}\,,\quad \Theta_z\equiv z^2 \theta(z-z_c)+c_h\,,
\end{align} 
with $z_c\approx 0.78, c_h\approx 0.4$. Note here that $\Theta_z$ denotes the dimensionless Higgs squared mass $m_\phi^2$ in the approximation of instantaneous electroweak cross-over at $T_c\approx 160$~GeV, and we have applied $E_\ell\approx |{\bf p}_\ell|, E_N\approx |{\bf p}_N|$.

\subsubsection{Resonant heavy neutrino mixing}\label{sec:nu-mixing}

Following Appendix~\ref{app:RFL-N}, we may write a semi-analytic and more intuitive solution for $Y_N$ in low-scale leptogenesis with quasi-degenerate singlet neutrinos,
 \begin{align}\label{eq:Yofz-ell}
Y^{}_{N}(z)\approx -9.55\times 10^{-10}\text{Im}\left[(\bar y^{\prime \dagger} \bar y')_{12}^2\right]\left(\frac{10^{-5}}{\delta M/M}\right)I^{}_{N}(z)\,,
\end{align}
where $\delta M/M$ denotes the ratio of the neutrino mass difference to the mass scale. For definiteness, in \eqref{eq:Yofz-ell}, we have considered two singlet neutrinos $N_1, N_2$  and  assumed that $M_2>M_1$, with $M\equiv M_2=M_1+\delta M$ and $\delta M\ll M$. 

The result in~\eqref{eq:Yofz-ell} is deduced from \eqref{eq:dDelta-nell/dt} by taking lepton-number conservation $Y_\ell=-Y_N$ in each CP-violating Higgs decay. The 4D integral $I_N(z)$ denotes the integration over statistics functions,
\begin{align}\label{eq:4DI-N}
I^{}_N(z)\equiv\int_{0}^{z} {\rm d}z' &\int_0^\infty {\rm d}x_N \int_{x_{\rm low}}^{\infty}{\rm d}x_\ell  \int_{x_{\rm low}}^{\infty}{\rm d}x'_\ell  \frac{\Theta_z^2-2\Theta_z (x_\ell+x'_\ell)x_N}{x_N^3}
 \\[0.2cm]
&\times \left(f^{\rm eq}_\ell(x_\ell)+f^{\rm eq}_\phi(x_N+x_\ell)\right)\left(f^{\rm eq}_\ell(x'_\ell)+f^{\rm eq}_\phi(x_N+x'_\ell)\right)\delta f_{N}(x_N)\,,\nonumber
\end{align}
with $x_{\rm low}\equiv \Theta_z/(4x_N)$ and $\Theta_z$ given in \eqref{eq:theta-z}. In \eqref{eq:4DI-N}, we assumed for simplicity  that the Yukawa couplings for $N_1$ and $N_2$ are similar in magnitude such that  $f_{N_1}\approx f_{N_2}\equiv f_N$. Note that, however, this approximation cannot be applied to \eqref{eq:Yofz-N}, which would vanish under summation over $i,j=1,2$.  It is worthwhile to mention here that a minimal degeneracy degree of $ \delta M/M\sim 10^{-5}$ was found in \cite{Hambye:2016sby} to induce the required baryon asymmetry. The result shown in~\eqref{eq:Yofz-ell} provides a consistency check if we take the maximally allowed values from  $y'\sim 10^{-7}- 10^{-6}$ (the out-of-equilibrium condition) and a maximal CP violation ($\text{Im} (y^{\prime 2})\sim y^{\prime 2}$), together with the typical values from~\eqref{eq:4DI-N}: $I_N\sim 0.1-1$.

We  would like to emphasize again that the  $i=j$  channel in \eqref{eq:Yofz-N}  would give rise to a vanishing CP-violating source, but this channel still allows a non-zero asymmetry in \eqref{eq:Yofz-ell}. Furthermore, we should be cautious about the non-equilibrium conditions from right-handed neutrinos. While \eqref{eq:Yofz-N} suggests that an asymmetry should still be induced for flavour $i$ being in equilibrium due to the presence of $\delta f_{N_j}\neq 0$ with $i\neq j$, we should keep in mind that \eqref{eq:Yofz-N} is valid only when the washout rate can be neglected, which is not the case if the neutrino flavour $i$ is in thermal equilibrium. 

Finally, we present a simple comparison of the resonant CP-violating source from thermal lepton mixing and from heavy neutrino mixing. The purpose here is to identify,  in terms of the ratio $\delta M/M$, the regime where the latter is the dominant effect.  Given that both the 4D integration functions $I^{}_N(z)$ and $I^{}_{ij}(z)$ depend on the nonthermal neutrino distribution function, we can envisage that  $I^{}_N(z)$ and $I^{}_{ij}(z)$ are at the same order, which is numerically confirmed  by solving the kinetic equation of $f_N$ from  \eqref{eq:dfN/dz} with a  neutrino Yukawa coupling $y'$ at $[10^{-7},10^{-5}]$. Taking the largest resonance from \eqref{eq:Yofz-N}  at $\beta=\mu, 
\alpha=e$, with an enhancement factor $
10^{-4}/(y_{\mu}^2-y_{e}^2)\approx 275$, we can infer that RL through heavy neutrino mixing will dominate low-scale leptogenesis in Higgs decay, provided that  the degree of neutrino mass degeneracy   meets
\begin{align}
    \frac{\delta M}{M}\lesssim \mathcal{O}(10^{-6})\,.
\end{align}
Without the above degeneracy, we will show in Section~\ref{sec:num} that the  lepton-number asymmetry from TRL can still yield the required order for the BAU.

\subsection{Lepton flavour coherences: mixing versus oscillation effects}\label{sec:mix-osc-cor}

In this section, we wish to elucidate all key phenomena that may play an important role in the TRL mechanism, which  we have been studying here. Our aim is to clarify the potential impact of lepton-flavour coherences, lepton-flavour mixing, lepton-flavour oscillations, and thermal width effects on TRL.

Observe that we often used the terms \textit{flavour off-diagonal correlations} and  \textit{flavour coherences} interchangeably in discussing the TRL mechanis. For the use of the latter term, we simply follow the terminology from Quantum Mechanics. In particular, off-diagonal correlations  correspond to coherences in the density matrix, which vanish in the limit of thermal equilibrium as known in open quantum systems~\cite{OpenQuantumSystems}. This behaviour is also expected for lepton-flavour coherences in the occupation-number matrix $f_{\alpha\beta}$, where some out-of-equilibrium conditions must be present to maintain the off-diagonal correlations, for which $\alpha \ne \beta$.

The physical in-equivalence between flavour mixing and flavour oscillations that govern unstable\- heavy neutrinos in leptogenesis was debated more than a decade ago. In particular, in RL~\cite{BhupalDev:2014oar,BhupalDev:2014oar} 
resonant heavy-neutrino mixing and heavy-neutrino flavour oscillations were shown to be two physically distinct phenomena, which can be described by different collision terms in a unifying and flavour-covariant framework of transport equations. Instead, other studies assert that both phenomena are identical, which could be united in a single density-matrix formalism~\cite{Klaric:2020phc}. Even though there is still no consensus on the physical (in)-equivalence between the two pheno\-mena, a first-principle derivation is lacking as it significantly depends on certain aspects\-, including the specifics of the leptogenesis scenario considered~\cite{daSilva:2022mrx,Karamitros:2023tqr},  the resummation method under consideration~\cite{BhupalDev:2014oar}, and the approximations adopted to simplify the complicated system in non-equilibrium QFT.   

The TRL mechanism presented in this work should be seen as a combined effect from flavour mixing and flavour oscillations, which leads to resonant coherences between the lepton-doublet flavours in a thermal plasma. Although we do not address in this work the physical equivalence between flavour mixing and flavour oscillations, the two equivalent methods arrive at the same CP-violating source (see Sections~\ref{sec:lepton-asymmetry} and~\ref{sec:2loo-diagram}), which would seem to suggest that the presence of flavour-mixing and oscillation effects in TRL can be adequately captured by a single flavour-covariant KB equation.  

It is important to note that TRL is generated by off-diagonal lepton-flavour coherences in $\delta f_{\alpha\beta}$, as these are determined in~\eqref{eq:delta-fab}. To create such off-diagonal correlations or coherences, the following three conditions should be satisfied:
\begin{itemize}
    \item Flavour-changing neutrino Yukawa couplings.
    \item Coherent lepton-doublet oscillation sources.
    \item Out-of-equilibrium of some singlet neutrino species.
\end{itemize}
If any of the above does not hold, no lepton-flavour coherences can be produced, as we argue in more detail below.

The lepton-flavour mixing effect is described by \textit{non-vanishing,  lepton-flavour changing  self-energy amplitudes at one-loop level}, as shown in the right panel of Fig.~\ref{fig:Higgsdecay} (see also Fig.~\ref{fig:ellag-self}). The one-loop flavour-changing diagram, which is absent within the SM context, is also one of  the building blocks in generating the singlet neutrino asymmetry if we use the two-loop diagrammatic method to calculate the CP-violating source, as shown in Fig.~\ref{fig:2+1loop-Nself}.  In essence, the lepton-flavour mixing effect comes from both the non-diagonal neutrino Yukawa matrix and  the nonthermal neutrino Yukawa interactions, the latter of which should be present to maintain the off-diagonal correlations in the lepton occupation-number matrix $f_{\alpha\beta}$. Observe that if the neutrino Yukawa interactions establish thermal equilibrium, no lepton-flavour changing coherences can be created, for any
non-diagonal neutrino Yukawa matrix $y'$ in the weak basis where $y$ and $M$ are diagonal. This finding can be directly inferred from~\eqref{eq:delta-fab} by setting $\delta f_{N_j} = 0$.

The lepton-flavour oscillation phenomena are described by~\eqref{eq:dn/dt} and~\eqref{eq:dnb/dt} due to the commutation term, but not by~\eqref{eq:dDn/dt} and~\eqref{eq:dSn/dt}. Technically speaking, therefore, the evolution of $\Delta n_{\alpha\beta}$ or $\Sigma n_{\alpha\beta}$ does not exhibit the oscillation phenomena, and hence the TRL mechanism presented in this work is \textit{not a kind of leptogenesis through flavour oscillations}, such as the ARS mechanism where lepton asymmetry is first generated by CP-violating flavour oscillations. In particular, one cannot interpret the flavour transitions in the right panel of Fig.~\ref{fig:Higgsdecay}  as oscillations between two different lepton flavours. Nevertheless, once a non-vanishing off-diagonal correlation is induced by the flavour mixing effect, the coherent oscillation source term $\tilde b_\alpha-\tilde b_\beta$ that arises from thermal (background) corrections will enhance the off-diagonal correlation. In this sense, we contend that the TRL mechanism is aided by coherent oscillation effects. For our illustrative purposes outlined below, we denote the off-diagonal terms of $\Sigma n_{\alpha\beta}$ and $\Delta n_{\alpha\beta}$ (with $\alpha\neq \beta$) as $\Sigma' n_{\alpha\beta}$ and $\Delta' n_{\alpha\beta}$, so that they trivially satisfy the relation 
\begin{equation}
  \label{eq:Sigmaprime}  
  \Sigma' n_{\alpha\alpha}\: =\: \Delta' n_{\alpha\alpha}\: =\: 0\,.
\end{equation}

The CP-violating off-diagonal correlations $\Delta' n_{\alpha\beta}$ are instrumental for the generation of the diagonal lepton-number asymmetry after taking the summation over the entire flavour space. This can be understood by looking at the collision terms in the flavour-covariant KB kinetic equation given in~\eqref{eq:dDn/dt-2}, whose RHS contains a term of the
form,
\begin{align}
    \frac{\text{d}\Delta n_{\alpha\alpha}}{\text{d}t}\: \supset\:  \sum_{\gamma,j} y^{\prime *}_{\alpha j}y^{\prime }_{\gamma j} T\int \text{d}^3 {\bf k}\, \delta f_{\gamma\alpha}(|{\bf k}|,t)\, f^{\rm eq}_\phi (|{\bf k}|)\,,
\end{align}
where the factor $T$ was introduced for dimensional reasons. The neutrino Yukawa couplings arise from lepton self-energy amplitudes $\Sigma_{>,<}$, while the occupation-number matrix is from the lepton Wightman propagators $S_{<,>}$, both of which must have non-vanishing off-diagonal entries. The flavour mixing from self-energy amplitudes ($\alpha\neq \gamma$) ensures that the  off-diagonal flavour correlations from the Wightman propagators can contribute  to the generation of $\Delta n_{\alpha\alpha}$\footnote{If there is no lepton-flavour mixing, we have $[i\Sigma_{<,>}]_{\alpha\gamma}\propto \delta_{\alpha\gamma}$, which will select only the diagonal entries of $[iS_{>,<}]_{\gamma\alpha}$ in the collision rate~\eqref{eq:collision-rate}.}.

With the above clarification of lepton-flavour mixing and oscillations in mind,  two important questions now arise for the TRL mechanism. (i)~\textit{Will flavour coherences persist during lepton asymmetry generation?} and (ii)~\textit{Is the thermal resonance stable against finite-width effects?} To address these questions, one must understand the size of the damping effects on the CP-odd coherences $\Delta' n_{\alpha\beta}$. To this end, let us first specify the oscillation patterns of~\eqref{eq:dn/dt} and~\eqref{eq:dnb/dt}. For the off-diagonal entries ($\alpha\neq\beta$), we may schematically write the kinetic equations as follows:
\begin{align}\label{eq:dn/dt-2}
    \frac{{\rm d}n_{\alpha \beta}}{{\rm d}t}+ i\Delta \omega_{\alpha\beta} n_{\alpha\beta}&=y^{\prime *}_{\alpha j} y'_{\beta j}C_{N_j}  - (y_\alpha^2 +y_\beta^2) C_\ell n_{\alpha\beta}- g^4 C_V (n_{
    \alpha\beta}+\bar n_{\alpha\beta})\,,
    \\[0.2cm]
     \frac{{\rm d}\bar n_{\alpha \beta}}{{\rm d}t}- i\Delta \omega_{\alpha\beta} \bar n_{\alpha\beta}&=y^{\prime *}_{\alpha j} y'_{\beta j}C_{N_j}-(y_\alpha^2 +y_\beta^2) C_\ell \bar n_{\alpha\beta}- g^4 C_V (n_{
    \alpha\beta}+\bar n_{\alpha\beta})\label{eq:dbarn/dt-2}\,,
\end{align}
where we used the approximation $|{\bf k}|\approx T$ for illustration such that $\langle n_{\alpha\beta}\rangle= n_{\alpha\beta}$,  $\langle \bar n_{\alpha\beta}\rangle= \bar n_{\alpha\beta}$ and 
\begin{align}
\Delta\omega_{\alpha\beta}\equiv \frac{y_\alpha^2-y_\beta^2}{16}T\,.
\end{align}
We have explicitly factored out the coupling dependence from neutrino Yukawa, charged-lepton Yukawa, and gauge interactions.  The coefficient $C_{N_j}$ depends on the nonthermal neutrino distribution $f_{N_j}$ for flavour $j$,  while $C_\ell$ and $C_V$ denote the coefficients after phase-space integration in charged-lepton Yukawa and gauge collision rates, respectively. It is straightforward to calculate these coefficients, but the precise magnitudes are not relevant to our discussion here. Nevertheless, one can expect  $(y_\alpha^2+y_\beta^2) C_\ell< g^4 C_V$, especially for the muon and electron flavours that concern us.   Note that the sum $y_\alpha^2+y_\beta^2$ arises from the two collision terms $\mathcal{C_{\alpha\beta}}$, $\mathcal{C^\dagger_{\alpha\beta}}$ given in~\eqref{eq:kinetic} while the $g^4$ dependence arises from both $SU(2)_L$ and $U(1)_Y$ gauge interactions, with $g^2$ from amplitude vertex and the other $g^2$ from leading thermal mass corrections before the electroweak gauge symmetry breaking. 

It should be emphasized that the forms presented by~\eqref{eq:dn/dt-2} and~\eqref{eq:dbarn/dt-2} are not applicable to the diagonal entries. In particular, as we derived in Appendix~\ref{app:collision-rate-1}, we have neglected the chemical potential of  charged-lepton singlets, which is at the same order as that of lepton doublets in the diagonal entries $n_{\alpha\alpha}$. This is due to the fast charged-lepton Yukawa interactions at $T<10^{5}$~GeV, where the charged-lepton Yukawa collision rate vanishes up to leading order of small chemical potentials, owing to the generalized Kubo-Martin-Schwinger (KMS) relation; see e.g.,~\eqref{eq:gKMS-S1} and~\eqref{eq:KMS-Sigma}. Consequently, for off-diagonal entries $n_{\alpha\beta}$, contributions from the chemical potential of charged-lepton singlets are effects of higher order~\footnote{The reader may find useful to examine the derivation of~\eqref{eq:y-rate_Deltan}. Since our interest is in leptogenesis with freeze-in initial conditions, we always have: $\sum_\alpha\Delta n_{\alpha\alpha}\ll \sum_\alpha \Sigma n_{\alpha\alpha} \simeq 6 n^{\rm eq}_\ell$ [cf.~\eqref{eq:SigmaApprox}].}.

In~\eqref{eq:dn/dt-2} and~\eqref{eq:dbarn/dt-2}, the nonthermal neutrino Yukawa collision rate works as a driving force (source term) to generate the off-diagonal correlations, while the charged-lepton Yukawa collision rate is the friction force that damps the correlations.  Specifically speaking, the last term on the right-hand sides of~\eqref{eq:dn/dt-2} and~\eqref{eq:dbarn/dt-2} is not a direct damping source. If the last term is comparable to the second term, the gauge collision rate  will  effectively work as another damping source. However, if $\Sigma' n_{\alpha\beta}$ is never significant and vanishes with a rate much faster than $\Delta' n_{\alpha\beta}$, the leading damping effect for $n_{\alpha\beta}, \bar n_{\alpha\beta}$ can be dominated  by the charged-lepton Yukawa interactions. The former situation usually occurs in freeze-out like leptogenesis, where the lepton asymmetry evolves from a large value down to a smaller one that matches the observed BAU. For instance, in many freeze-out like leptogenesis, the initial lepton asymmetry can be larger than the BAU by several orders of magnitude. In this case, the chemical potential of leptons initially generated is large enough such that the last term cannot be neglected. In the latter case, which may be realized in freeze-in like leptogenesis,  the lepton asymmetry evolves from a negligibly small value up to the one that matches the observed BAU. In this case, we may anticipate that the $C_V$ terms in~\eqref{eq:dn/dt-2} and~\eqref{eq:dbarn/dt-2} are smaller than the $C_\ell$ terms. Then, given that the coherent oscillation source is comparable with the damping effect (decoherence), both the correlations $n_{\alpha\beta}$ and $\bar n_{\alpha\beta}$ can undergo certain flavour oscillations.

Nevertheless, the TRL mechanism does not depend directly on the oscillation patterns of $n_{\alpha\beta}$ and $\bar n_{\alpha\beta}$. Instead, it depends on the evolution of the off-diagonal CP-odd correlations
$\Delta n'_{\alpha\beta}$.  Based on~\eqref{eq:dn/dt-2} and~\eqref{eq:dbarn/dt-2}, we have
\begin{align}\label{eq:dDn/dt-3}
    \frac{{\rm d}\Delta' n_{\alpha \beta}}{{\rm d}t}+ i\Delta \omega_{\alpha\beta}\, \Sigma' n_{\alpha\beta}\: &=\: - (y_\alpha^2 +y_\beta^2) C_\ell\, \Delta' n_{\alpha\beta}\,,
    \\[0.2cm]
     \frac{{\rm d}\Sigma' n_{\alpha \beta}}{{\rm d}t}+ i\Delta \omega_{\alpha\beta}\, \Delta' n_{\alpha\beta}\: &=\: 2y^{\prime *}_{\alpha j} y'_{\beta j}\, C_{N_j}-(y_\alpha^2 +y_\beta^2) C_\ell\, \Sigma' n_{\alpha\beta}-2 g^4 C_V\, \Sigma' n_{\alpha\beta}\label{eq:dSn/dt-3}\,,
\end{align}
from which we clearly see that the damping of $\Sigma' n_{\alpha\beta}$ is stronger due to the  gauge collision rate $\propto -g^4 C_V$ that occurs in the last term of \eqref{eq:dSn/dt-3}. 

We note that from~\eqref{eq:dDn/dt-3} and~\eqref{eq:dSn/dt-3}, we can directly infer  the origin of CP violation by using the CP transformation properties of the number-density matrices, $n$ and $\bar n$. In detail, the CP-transformed number-density matrices, $n^{\rm CP}$ and $\bar n^{\rm CP}$, are given by~\cite{BhupalDev:2014pfm}:
\begin{align}
n^{\rm CP}=\bar n^{\sf T}\,,\quad \bar n^{\rm CP}=n^{\sf T}\,,
\end{align}
such that $\Delta n^{\rm CP}=-\Delta n^{\sf T}$ and $\Sigma n^{\rm CP}=\Sigma n^{\sf T}$. Here, the superscript ${\sf T}$ denotes the operation of matrix transposition. In the CP-symmetric limit of the theory, we have: $\Delta n=-\Delta n^{\sf T}$ and ${\Sigma n=\Sigma n^{\sf T}}$. 
Clearly, in the CP-invariant limit, 
any lepton asymmetry should vanish, meaning that $\Delta n_{\alpha\alpha}=0$, as they obviously do. Now, taking the lepton-flavour transpose on both sides of~\eqref{eq:dDn/dt-3} and~\eqref{eq:dSn/dt-3}, i.e., $\alpha\leftrightharpoons\beta$,  we find that the neutrino Yukawa collision rate will break the CP invariance if their respective couplings $y'$ are complex.

It is now interesting to notice that the non-thermal neutrino Yukawa collision rate does not appear as a source term in the evolution of $\Delta'n_{\alpha\beta}$, which may naively lead to the conclusion that $\Delta' n_{\alpha\beta}$ will rapidly damp away, as observed earlier in ~\cite{Beneke:2010dz,Garbrecht:2012pq,Garbrecht:2014iia}. However, such an analysis based on~\eqref{eq:dDn/dt-3} and~\eqref{eq:dSn/dt-3} is only valid when we consider the collision rates through one-loop order. At two-loop order, a new source term that violates CP is generated in a proper flavour-covariant consideration and, as such, it must be added to the RHS of~\eqref{eq:dDn/dt-3}. This can dramatically change the evolution of $\Delta' n_{\alpha\beta}$.
 As shown in Appendix~\ref{app:collision-rate-1}, the two-loop collision rate scales  as
 \begin{align}
   \label{eq:CPSource2loop}
 \frac{g^2\text{Im}(y'_{\beta j}y^{\prime *}_{\alpha j})}{y_\alpha^2-y_\beta^2}C'_{N_j}\,,
\end{align}
which modifies~\eqref{eq:dDn/dt-3} to
\begin{equation}
   \label{eq:dDn2loop/dt}
 \frac{{\rm d}\Delta' n_{\alpha \beta}}{{\rm d}t}+ i\Delta \omega_{\alpha\beta}\, \Sigma' n_{\alpha\beta}\: =\: - (y_\alpha^2 +y_\beta^2) C_\ell\, \Delta' n_{\alpha\beta}\: +\: 
 \frac{g^2\text{Im}(y'_{\beta j}y^{\prime *}_{\alpha j})}{y_\alpha^2-y_\beta^2}C'_{N_j}\,.   
\end{equation}
Here, $g^2$ denotes the gauge couplings from $SU(2)_L$ and $U(1)_Y$, and the coefficient $C'_{N_j}$ depends on the nonthermal neutrino distribution $f_{N_j}$ for flavour $j$. Hence, the new term is particularly non-negligible for the muon and electron flavours due to the enhancement of $(y_\mu^2-y_e^2)^{-1}$, which will  alter the evolution of the correlation $\Delta n_{\mu e}$ from simply being damped away through charged-lepton Yukawa interactions, maintaining thereby the existence of muon-electron coherence.

Next, let us consider the stability of the thermal resonance, following the analysis of  the  RL regulator from finite-width effects. Like the traditional RL, TRL also exhibits a resonant enhancement. Therefore, it is worth checking if the resonant enhancement may be suppressed by lepton thermal widths. Furthermore, when taking a width regulator,  one should also check if the regularization works properly. This double check should be warranted, given that  various finite-width regularization schemes are formulated and employed in different context; see e.g., Appendix~A in~\cite{BhupalDev:2014pfm} for discussions on these regularization schemes. 


To understand the appearance of resonance, it is instructive to consider the products of two Breit--Wigner (BW) propagator modes that typically occur in the calculation of loop amplitudes~\cite{Deppisch:2010fr},
\begin{align}
 \mathcal{D}_1&\equiv    \frac{1}{p^2-m_i^2-im_i \Gamma_i}    \frac{1}{p^2-m_j^2+im_j \Gamma_j}\approx \frac{i\pi\left[\delta(p^2-m_i^2)+\delta(p^2-m_j^2)\right]}{(m_i^2-m_j^2)+i(m_i\Gamma_i+ m_j\Gamma_j)}\,,
    \\[0.2cm]
       \mathcal{D}_2&\equiv   \frac{1}{p^2-m_i^2-im_i \Gamma_i}    \frac{1}{p^2-m_j^2-im_j \Gamma_j}\approx \frac{i\pi\left[\delta(p^2-m_i^2)-\delta(p^2-m_j^2)\right]}{(m_i^2-m_j^2)+i(m_i\Gamma_i- m_j\Gamma_j)}\,,
\end{align}
 where we used the decomposition
 \begin{align}
     \frac{1}{AB}=\frac{1}{B-A}\left(\frac{1}{A}-\frac{1}{B}\right),
 \end{align}
 and then applied the narrow-width approximation such that $1/A-1/B$ approximates two Dirac $\delta$-functions. Clearly, the difference between $\mathcal{D}_1$ and $\mathcal{D}_2$ is the sign between the two widths. The above decomposition allows us to see whether the sum or the difference of two widths should appear in the denominator. For instance, the sum of $\Gamma_i+\Gamma_j$ would appear in the limit of $m_i\to m_j$ if the two BW propagator modes carry the opposite sign in the imaginary part. An important difference between $\mathcal{D}_1$ and $\mathcal{D}_2$ is that in the limit of $m_i\to m_j$, the flavour-universal contribution to the finite widths is cancelled in $\mathcal{D}_2$ but not in $\mathcal{D}_1$. In particular, it implies that large flavour-universal contributions from gauge interactions to the finite widths will suppress the resonant enhancement if the CP-violating source carries only the structure of $\mathcal{D}_1$. 

In the context of TRL, we found that both of the regularization structures  similar to $\mathcal{D}_1$ and $\mathcal{D}_2$ will appear if we take into account the lepton thermal width. To see this, let us consider the two-loop diagram shown in Fig.~\ref{fig:2+1loop-Nself}, since it predicts the same CP-violating source as that induced by the flavour-covariant KB equation. Then we will make a simple, qualitative identification of the thermal resonance and thermal width effects. The lepton doublets in Fig.~\ref{fig:2+1loop-Nself} are thermally resummed, where  off-shell propagation appearing in $[iS_\ell^{<,>}]_{\alpha\beta}$ of~\eqref{eq:Sigma_N<} and~\eqref{eq:Sigma_N>}  can be constructed by the resummed retarded and advanced  propagators (see also Appendix~\ref{app:RFL-N} for more detailed calculations)
\begin{align}
[S_R(p)]_{\alpha}=&\sum_{s=\pm}\frac{1}{[\text{Re}\Sigma_s(p)]_\alpha+[i\text{Im}\Sigma_s(p)]_\alpha}P_s\,,
    \\[0.2cm]
[S_A(p)]_{\alpha}=&\sum_{s=\pm}\frac{1}{[\text{Re}\Sigma_s(p)]_\alpha-[i\text{Im}\Sigma_s(p)]_\alpha}P_s\,,
\end{align}
where $P_\pm$ denotes the decomposition of lepton helicity eigenstates~\cite{Braaten:1990wp},
\begin{align}
    P_\pm (p) =P_L\frac{\gamma^0\pm {\bf p}\cdot {\boldsymbol{ \gamma}}/|{\bf p}|}{2}P_R\,,
\end{align}
and
\begin{align}
    \text{Re}\Sigma_\pm&\equiv(1+\text{Re}a)(p_0\pm |{\bf p}|)+\text{Re}b\,,
    \\[0.2cm]
    \text{Im}\Sigma_\pm&\equiv\text{Im}a(p_0\pm |{\bf p}|)+\text{Im}b\,,
\end{align}
with the coefficients $a,b$ defined via~\eqref{eq:ab-def}. From Appendix~\ref{app:lepton-Wightman}, 
the resummed Wightman functions satisfy 
\begin{align}
    S_{<,>}\propto S_R-S_A\,.
\end{align} 
Since the thermal resonance appears when one of the leptons in Fig.~\ref{fig:2+1loop-Nself} goes on-shell while the other one is off-shell, we envisage that the CP-violating source can be constructed by the product
\begin{align}\label{eq:SR-S<}
    [S_R(p)]_\alpha\times [S_{<(>)}(p)]_\beta\,.
\end{align}
A structure similar to $\mathcal{D}_1$ would arise from the product of $ [S_R(p)]_\alpha$ and $ [S_A(p)]_\beta$, since  $ [S_R(p)]_\alpha$ and $ [S_A(p)]_\beta$ carry the opposite sign of $i\text{Im}\Sigma_\pm(p)$, whereas a structure similar to  $\mathcal{D}_2$  would arise from the product of $ [S_R(p)]_\alpha$ and $ [S_R(p)]_\beta$.  Given that the leading contribution to the lepton thermal width is at order of $g^4 T$, as can also be inferred from the gauge damping rate given in~\eqref{eq:dn/dt-2} and~\eqref{eq:dbarn/dt-2}, the product $ [S_R(p)]_\alpha\times [S_R(p)]_\beta$ will become the origin of the thermal resonance, where the flavour-universal contributions to lepton thermal width cancel. 

More explicitly, we follow the analysis of~\cite{Kanemura:2024dqv} and  found that the thermal resonance evaluated from~\eqref{eq:SR-S<} contains a denominator of the form 
\begin{align}
   \label{eq:D3}
\mathcal{D}_3=\frac{1}{\tilde m_\alpha^2-\tilde m_\beta^2
+\mathcal{O}(g^6)+ i\mathcal{O}[g^2 (\tilde m_\alpha^2-\tilde m_\beta^2)]}\,,
\end{align}
which does not diverge in the limit of $\tilde m_\alpha\to \tilde m_\beta$ (or $y_\alpha\to y_\beta$). 
Recall that the non-degeneracy of charged-lepton masses is a \textit{necessary} condition for leptonic CP violation~\cite{Branco:2011zb}. It implies that in the limit of $y_\alpha \to y_\beta$, there should not be CP violation. It is easy to verify this expectation. In this limit, we arrive at $\mathcal{D}_3\propto g^{-6}$, which is independent of  lepton flavours. Given this flavour independence, we have $\text{Im}[(y^{\prime\dagger} y')_{ji}(y^{\prime \dagger} y')_{ij}]=0$ after summing over the lepton flavours  in~\eqref{eq:dDelta-nN/dt}, and hence there is indeed no  CP-violating source. Therefore, after including thermal width effects, we see from~\eqref{eq:D3} that the thermal resonance is stable and 
exhibits a regular analytic behaviour in the limit of $y_\alpha\to y_\beta$.

\section{Numerical estimates in a simple flavour model}\label{sec:num}

In this section, we provide numerical estimates of TRL predictions within a simple flavour model. This numerical analysis proves useful in delineating the parameter space for which TRL can account for the observable BAU. To this end, it is instructive to discuss lepton and singlet-neutrino flavour effects caused by hierarchies of neutrino Yukawa couplings, which, on the one hand, can provide significant differences among the evolution of non-thermal neutrino distribution functions, the washout rates of lepton flavours and the lepton-flavour dependent CP-violating sources, and, on the other hand, may give rise to effects observable in laboratory experiments.  A~full fledged numerical analysis requires performing a multi-dimensional parameter scan and should take into account the SM neutrino masses and mixing. We have presented such an analysis in~\cite{Li:2026rqg} that also includes possible phenomenological signatures in 
collider experiments.

The lepton-number asymmetry from lepton doublets will be converted into baryon asymmetry via thermal sphaleron processes following the chemical relation~\cite{Harvey:1990qw}
\begin{align}
    \mu_B=-\frac{2}{3}\sum_{\alpha=1}^3\mu_{\alpha}\,.
\end{align}
It gives rise to 
\begin{align}\label{eq:YB-Ya}
    Y_B\equiv \frac{n_B-\bar n_B}{s}=-\frac{2}{3}\frac{n_\ell-\bar n_\ell}{s}\equiv-\frac{2}{3}Y_\ell=-\frac{2}{3}\sum_\alpha Y_\alpha\,.
\end{align}
Note here that we have defined the chemical potential $\mu_\alpha \equiv \mu_{\alpha\alpha}$ for the whole lepton doublet $\ell_\alpha$, and we have accounted for the gauge degeneracy from lepton doublets by multiplying the collision rates by a factor of 2 whenever necessary. The final $Y_B$ should match the observed value~\cite{Planck:2018vyg}
\begin{align}
    Y_B\approx 8.75\times 10^{-11}\,.
\end{align}
The identity confirmed in \eqref{eq:L-conserved} does not guarantee the same washout effect between lepton doublets and right-handed neutrinos. In fact, once lepton asymmetries are created in the lepton doublets from thermally induced Higgs decays, these asymmetries will quickly be redistributed, via spectator processes, among all the lepton doublets and right-handed charged leptons. As a result, the total lepton-number asymmetry in the SM sector is given by~\footnote{In this section, we will use $e,\mu,\tau$ and numbers 1, 2, 3 interchangeably for charge-lepton flavours, provided there is no confusion. }
\begin{align}
    \mu_L\,\equiv\, \sum_{e,\ell}(\mu_{e}+\mu_{\ell})\: =\: -\frac{51}{28}\,\mu_B\,.
\end{align}
Nevertheless, under kinetic evolution, $\mu_L$ would not correspond to the total lepton-number asymmetry stored in right-handed neutrinos. Instead, due to $B+L$ violating effects and different washout rates between $Y_\ell$ and $Y_N$, the final baryon asymmetry, when expressed in terms of $Y_N$, can be given via the conservation relation $\Delta (B-L_{\rm SM})=\Delta L_N$ such that
\begin{align}
    Y_B=\frac{28}{79}Y_N\,.
\end{align}
The above discussion implies two equivalent methods to calculate the final $Y_B$. The first one uses the KB equation of lepton doublets, and the second one focuses on the total lepton-number asymmetry stored in the $N$-sector. The first method  can be regarded as a direct method following the techniques presented in Section~\ref{sec:0thKB} and Section~\ref{sec:lepton-asymmetry}, while the second method is a more indirect one that follows Section~\ref{sec:2loo-diagram}. Here, we use the first method to estimate the final~$Y_B$.

In Section~\ref{sec:lepton-asymmetry}, we did not include the washout effects in the evolution of $\Delta n_\ell$. The washout effects contain both lepton-number violating and conserving  processes which are distinguished by singlet neutrino mass insertion in Fig.~\ref{fig:2+1loop-Nself} (see also Fig.~\ref{fig:N-self} for a detailed diagrammatic depiction). In~addition\- to Higgs decay/inverse decay, there are also scattering processes associated with quarks and gauge bosons, which, in the KB formalism, can be estimated from the absorptive part of the resummed Higgs propagator.  Unlike the CP-violating source, the washout effects are not induced by plasma effects and hence can also be determined by tree-level scattering amplitudes, as have been evaluated explicitly in~\cite{Anisimov:2010gy,Besak:2012qm}. Here, we will only take Higgs decay and inverse decay for simple estimate. On the other hand, since we are considering relativistic Majorana neutrinos at temperatures above 100~GeV, we will neglect the washout rate induced by lepton-number breaking diagrams, i.e., those with Majorana mass insertion.  Note that in doing this, we will also neglect the CP-violating source induced by flipping the neutrino arrow in the inner loop of the two-loop diagram shown in  Fig.~\ref{fig:2+1loop-Nself}.

From \eqref{eq:dDn/dt-2}, we may read off the terms that contribute to the washout rate,  $\mathcal{W}\equiv\sum_{\alpha}\mathcal{W}_{\alpha}$:
\begin{align}
\mathcal{W}_\alpha\, =\, -2\sum_j |y'_{\alpha j}|^2 \Delta m_j^2 \int {\rm d}\Pi_{\ell,\phi,N}    (2\pi)^4 \tilde\delta^{(4)}({ k}_\ell+{k}_\phi-{ k}_N) 
\left(K^\phi_j-K^N_j+K^\ell_{\alpha j}\right),\label{eq:dDn/dt-washout}
\end{align}
where ${\rm d}\Pi_{\ell,\phi,N}={\rm d}\Pi_{\ell}{\rm d}\Pi_{\phi}  {\rm d}\Pi_{N}$ and the $K$-functions are defined as 
\begin{align}
    K^\phi_j&\equiv \delta f_\phi (E_\phi) \left(1-f_\ell^{\rm eq}(E_\ell)-f_{N_j}(E_N)\right),
    \\[0.2cm]
    K^N_j&\equiv \frac{1}{2} \left(\delta f_{N_j}(E_N)-\delta \bar f_{N_j}(E_N)\right)\left(f_\ell^{\rm eq}(E_\ell)+f_\phi^{\rm eq}(E_\phi)\right),
        \\[0.2cm]
        K^\ell_{\alpha j}&\equiv\delta f_{\alpha\alpha}(E_\ell)\left(f_\phi^{\rm eq}(E_\phi)+f_{N_j}(E_N)\right).
\end{align}
The above washout rate is equivalent to that obtained from the Boltzmann equation in the limit $m_\phi\gg \tilde{m}, M$. Such equivalence between the KB (one-loop collision rates) and Boltzmann (tree-level collision rates) equations suggests that one can add the purely plasma-induced CP-violating source into their Boltzmann equations already built upon traditional leptogenesis. In doing this, one may estimate the contribution from SM lepton coherences to the scenarios under consideration in an easier way.

The term $K^\phi_j$ may be interpreted as the spectator effect from the redistribution of SM lepton-number asymmetry among lepton doublets and right-handed charged-lepton singlets.  The term $K^\ell_{\alpha j}$ can be recast as $\Delta n_{\alpha\alpha}$ due to quasi-thermal leptons. For the  term $K^N_j$ that depends on $\delta f_N-\delta \bar f_N$, if right-handed neutrinos have not established kinetic equilibrium near the end of leptogenesis, one should solve the kinetic equations ${\rm d}(f_{N_j}-\bar f_{N_j})/{\rm d} t$ from the unintegrated KB equation up to two-loop level. In doing this,   we should expect that both the Boltzmann equation with tree-level amplitudes and the KB equation with one-loop self-energies will give the same result: $\text{d}f_{N_j}/\text{d}t=\text{d}\bar f_{N_j}/\text{d}t$, provided that the  asymmetry between $f_{N_j}$ and $\bar f_{N_j}$ can be neglected. However, since the difference $f_{N_j}-\bar f_{N_j}$ is only generated by a thermal effect, it is not straight\-forward to determine ${\rm d}(f_{N_j}-\bar f_{N_j})/{\rm d} t$ from the Boltzmann equation at one-loop order. In this case, the unintegrated version of \eqref{eq:dDeltanN/dt} is required. 

Comparing to  those leptogenesis scenarios where kinetic equilibrium approximately holds for singlet neutrinos, the calculation of $\delta f_{N_j}-\delta\bar f_{N_j}$ in $K^N_j$ would constitute a leading technical difficulty in the precise calculation of washout effects if all $f_{N_j}$ functions are far away from thermal equilibrium.  Nevertheless, one can  expect that the contribution from the $K_j^N$ term should  at most approach the order of the $K^\phi_j$ spectator effect when some (but not all) neutrino flavours reach quasi-thermal equilibrium.
Given the relation of the chemical potentials $\mu_\phi=4\mu_L/21=4\sum_\alpha \mu_\alpha/21$~\cite{Harvey:1990qw}, we will neglect both the $K^\phi_j$ spectator effect  and the $K_j^N$ term. The numerical results shown below imply that before sphaleron decoupling, the $Y_N$ asymmetry will eventually be stored in a single neutrino flavour, whereas the other flavours will enter (quasi) thermal equilibrium. Consequently, the thermalized neutrino flavours start to play the spectator role in redistributing the lepton-number asymmetry in the thermal plasma. 

Taking the $K^\ell_{\alpha j}$ term to estimate the washout rate with $f_N\approx f_N^{\rm eq}$ for inclusion of any potentially thermalized neutrino flavour(s), we use the following quasi-thermal relation\footnote{Here, we used the approximation of momentum independence for the diagonal chemical potentials $\mu_{\alpha\alpha}$, which, however, does not hold for the off-diagonal $\mu_{\alpha\beta}$.},
\begin{align}
\delta f_{\alpha\alpha}(E)&=\frac{\mu_\alpha}{T}\frac{e^{E/T}}{(e^{E/T}+1)^2} = \frac{ \Delta n_{\alpha\alpha}}{T^3}\frac{6e^{E/T}}{(e^{E/T}+1)^2} \,,
\end{align}
to obtain
\begin{align}\label{eq:dDn/dt-washout-approx}
\mathcal{W}_{\alpha}&\approx-\frac{3(y'y^{\prime \dagger})_{\alpha\alpha} \Delta m^2}{8\pi^3} \frac{\Delta n_{\alpha\alpha}}{T^3}\int \text{d}E_\ell \int_{E_{\rm low}}^\infty \text{d} {E_N} \frac{e^{E_\ell/T}\left(f_{\phi}^{\rm eq}(E_\ell+E_N)+f_N^{\rm eq}(E_N)\right)}{(e^{E_\ell/T}+1)^2}\,,
\end{align}
 where we used $E_\ell\approx |{\bf k}_\ell|,E_N\approx |{\bf k}_N|$, and
 \begin{align}
     E_{\rm low}=\frac{m_\phi^2}{4E_\ell}\,,\quad   \Delta m^2\approx m_h^2\theta(T_c-T)+0.3T^2\,,
 \end{align}
with  $T_c\approx 160$~GeV being the cross-over temperature~\cite{DOnofrio:2014rug}  and $m_h=125$~GeV the vacuum Higgs mass. Using $Y_\alpha\equiv \Delta n_{\alpha\alpha}/s$ and applying   the Boltzmann distribution approximation, we arrive at the semi-analytic result for the washout rate,
 \begin{align}\label{eq:Wa-approx}
\mathcal{W}_{\alpha}\approx -0.2 |y'_\alpha|^2 T^4 Y_{\alpha}\,,
\end{align}
with the shorthand
\begin{align}\label{eq:yalpha-def}
    |y'_\alpha|^2&\equiv (y'y^{\prime \dagger})_{\alpha\alpha} \,.
\end{align}

From \eqref{eq:CPV-l}, we notice that the maximal CP-violating rate occurs in
\begin{align}
    \mu~\text{channel}&:\ \alpha=2\,(\text{muon~flavour})\,,\qquad  \gamma=1\,(\text{electron~flavour})\,,
    \\[0.1cm]
       e~\text{channel}&:\  \alpha=1\,(\text{electron~flavour})\,,\quad  \gamma=2\,(\text{muon~flavour})\,.
\end{align}
Correspondingly, the expressions for $\mathcal{S}_{\alpha=2}\equiv \mathcal{S}_{\mu}$ and  $\mathcal{S}_{\alpha=1}\equiv \mathcal{S}_{e}$ take on a simpler form
\begin{align}\label{eq:CPmax}
 \mathcal{S}_{\mu}\: =\: \mathcal{S}_{e}\:\approx\: 158.5 T^4  \sum_{\substack{i,j=1\\ i \neq j}} \text{Im}(y'_{\mu i}y^{\prime *}_{ei}y'_{ej}y^{\prime *}_{\mu j})\, I_{ij}\,.
\end{align}
To simplify our numerical analysis, we will rely on the working hypothesis that only two singlet neutrinos contribute significantly to TRL, even though the generalisation of this hypothesis to more right-handed neutrinos is straightforward. In~\eqref{eq:CPmax}, the 3D integrals $I_{ij}$ may be conveniently expressed as 
\begin{align}
\label{eq:Iijz}
    I_{ij}(z)\: \equiv\: \int \frac{dx_\ell}{x_\ell}&\int_{x_{\rm low}}^\infty dx_N \int_{x_{\rm low}}^\infty dx'_N 
    \\[0.2cm]
    &\times \Big(f_\phi^{\rm eq}(x_\ell+x_N)+f^{\rm eq}_{N_i}(x_N)\Big)\Big(f_\ell^{\rm eq}(x_\ell)+f_\phi^{\rm eq}(x_\ell+x'_N)\Big)\,\delta f_{N_j}(z,x'_N)\,,\nonumber
\end{align}
where the dimensionless variables $x^{(\prime)}_i$ are defined in~\eqref{eq:x-def}, and  $x_{\rm low}$ is the dimensionless version of \eqref{eq:Elow}, i.e., $x_{\rm low}\equiv \Theta_z/(4x_\ell)$, with $\Theta_z$ stated in~\eqref{eq:theta-z}. The evolution $\delta f_{N_j}$ is determined by Higgs decay and inverse decay, while including additional soft gauge interactions associated with  $\ell$ and $\phi$  amounts to an enhancement by a factor of $\mathcal{O}(1)$~\cite{Anisimov:2010gy,Besak:2012qm,Garbrecht:2013bia,Laine:2013lka}. 
 For instance, the thermal production rate of  the right-handed neutrino flavour $j$: $\Gamma_j\approx 4.5\times 10^{-3}(y^{\prime \dagger} y')_{jj} T$ was considered in \cite{Akhmedov:1998qx}, but after including soft gauge-boson scattering, it yields $\Gamma_j\approx 0.01 (y^{\prime \dagger} y')_{jj} T$~\cite{Anisimov:2010gy,Besak:2012qm,Garbrecht:2013bia,Laine:2013lka}, which is enhanced by nearly a factor of 3.  For simplicity, we use the Boltzmann equation with tree-level decay/inverse decay and  absorb  the corrections from soft gauge-boson interactions  by effectively scaling  the tree-level result by a factor of 3. 
 
For relativistic Majorana neutrinos, the evolution of $f_{N_j}$ in terms of the dimensionless variable $z\equiv m_h/T$ gives 
\begin{align}\label{eq:dfN/dz}
	\frac{{\rm d}f_{N_j}(z,x_N)}{{\rm d}z}=\frac{\bar M_{\rm Pl}|y'_{j}|^2}{32\pi m_h}\frac{\Theta_z}{x_N^2}\int_{x'_{\rm low}}^\infty  \mathcal{F}_j(x_N,x') {\rm d}x'\,,
\end{align}
where 
\begin{align}\label{eq:yi-def}
|y'_j|^2\equiv (y^{\prime \dagger} y')_{jj}\,,\quad x'_{\rm low}=\frac{\Theta_z}{4x_N}\,,
\end{align}
with $\Theta_z$ given in~\eqref{eq:theta-z}, and 
\begin{align}
  \label{eq:calFj}
	\mathcal{F}_j(x_N,x')=f^{\rm eq}_\phi(x'+x_N)\left(1-f^{\rm eq}_\ell(x')\right)- f_{N_j}(x_N)\left(f^{\rm eq}_\phi(x'+x_N)+f^{\rm eq}_\ell(x')\right).
\end{align}
Notice that  in the thermal equilibrium limit  in which $\delta f_{N_j}=0$, the function $\mathcal{F}_j(x_N,x')$ given in~\eqref{eq:calFj} can be shown to vanish after first rearranging it to become: ${\cal F}^{\rm eq}_j = f_\phi^{\rm eq}(1-f_\ell^{\rm eq})(1-f^{\rm eq}_{N_j})-f_\ell^{\rm eq}f^{\rm eq}_{N_j}(1+f_\phi^{\rm eq})$ and then employing the detailed balance relations among the thermal distribution functions of the particles involved.  

Including washout effects, the Boltzmann equation for the yield $Y_\alpha$ may be approximately written as 
\begin{align}
  \label{eq:dYdz-sim}
  sHz \frac{\text{d} Y_{\alpha}}{\text{d}z}\: \approx\: \mathcal{S}_{\alpha}\, +\, \mathcal{W}_{\alpha}\,.
    \end{align}
The washout rate $\mathcal{W}_{\alpha}$ depends on 
the Yukawa combination
\begin{align}
\mathcal{W}_\alpha\propto \sum_i|y'_{\alpha i}|^2 = |y'_\alpha|^2\,. 
\end{align}
Instead, $\mathcal{S}_{\alpha}$ exhibits a different dependence on the neutrino Yukawa couplings. Let us concentrate on the maximal CP-violating rates given in \eqref{eq:CPmax}. 
Since the dependence of $I_{ij}$ is weak with respect to the singlet neutrino flavour $i$ if all heavy neutrinos are relativistic, we expect that  
\begin{align}\label{eq:CP-max-2}
   \mathcal{S}_{\mu} = \mathcal{S}_{e}\: &\propto\:   \sum_{\substack{i,j=1\\ i \neq j}} \text{Im}(y'_{\mu i}y^{\prime *}_{e i}y'_{e j}y^{\prime *}_{\mu j})\, \mathcal{P}_{j}\,,
\end{align}
where $\mathcal{P}_j$ is a function of $|y'_j|^2$ defined in~\eqref{eq:yi-def}, i.e.
\begin{align}\label{eq:P-def}  \mathcal{P}_{j}=\mathcal{P}\left(|y'_j|^2\right),
\end{align}
arising from the equation of $f_{N_j}$ in \eqref{eq:dfN/dz}. From the above discussion, it is obvious that the source function $I_{ij} \propto {\cal P}_j$ has a different Yukawa dependence than the washout rate $\mathcal{W}_\alpha$. Thus, it is possible to enhance the CP-violating sources ${\cal S}_{e,\mu}$ through a larger $\mathcal{P}_j$, while reducing $\mathcal{W}_\alpha$ at the same time, since a smaller $|y'_j|^2$ predicts a larger $\mathcal{S}_{e,\mu}$ because of a larger $|\delta f_{N_j}|$. This phenomenon was also noticed  in \cite{Canetti:2014dka}, but it is challenging when one introduces only two singlet neutrino flavours in the type-I seesaw framework. Moreover, for $\alpha,\beta=1,2$, if  $\mathcal{W}_\beta\gg \mathcal{W}_\alpha$, the asymmetry of flavour $\beta$ will be smaller than that of flavour $\alpha$ since both $Y_\mu$ and $Y_e$ depend on the same CP-violating rate given in \eqref{eq:CP-max-2}.  We can then deduce an important flavour effect in the $\mu e$ channel. If $\mathcal{W}_\beta\gg \mathcal{W}_\alpha$, the $\beta$-flavour asymmetry is suppressed by a large washout rate, whilst the $\alpha$-flavour asymmetry is enhanced  not only by the function $\mathcal{P}_{j}$ but also by the hierarchy $|y'_{\beta i}|\gg |y'_{\alpha j}|$ encoded in the CP-odd Yukawa-coupling expression: $\text{Im}(y'_{\beta i}y^{\prime *}_{\alpha i}y'_{\alpha j}y^{\prime *}_{\beta j})$.
 
There is a caveat when we consider flavour effects caused by hierarchies of neutrino Yukawa matrix elements.  In particular, the special case
\begin{align}\label{eq:W-hierarchy}
\mathcal{W}_{\tau},\mathcal{W}_{\mu}\gg \mathcal{W}_{e}\,,
\end{align}
was often considered in high-scale flavoured leptogenesis. The above condition implies that 
\begin{align}\label{eq:y-con}
|y'_{e}|^2\ll |y'_{\mu}|^2, |y'_{\tau}|^2\,,
\end{align}
such that sufficient electron asymmetry can be protected from large washout before sphaleron decoupling. However,  the above hierarchy can hardly be tuned in the parameter space if one adopts a $3\times 2$ matrix for the neutrino Yukawa matrix $y'$. This fact has led to the observation that two singlet neutrinos are difficult to realize lepton-flavoured leptogenesis unless the two neutrino flavours are strongly quasi-degenerate~\cite{Pilaftsis:2004xx,Pilaftsis:2005rv} or one uses a $3\times n$ matrix for $y'$ with the number of neutrino flavours $n$ greater than two~\cite{Canetti:2012kh,Shuve:2014zua,Canetti:2014dka}. 
Nevertheless, we should emphasize that while the condition~\eqref{eq:W-hierarchy} is strictly selected in lepton-flavoured leptogenesis for a weaker washout effect in the electron flavour,  it is not a necessary condition in low-scale leptogenesis based on \eqref{eq:dYdz-sim}, since we are concerned with a total SM lepton-number asymmetry rather than a specific flavour asymmetry.  Furthermore,  since the CP-violating source already exhibits a large hierarchy in lepton-flavour space, and predicts  $\mathcal{S}_{\mu}= \mathcal{S}_{e}$,  one may alternatively consider $|y'_{e}|^2\gg |y'_{\mu}|^2$ such that the total lepton-number asymmetry will be dominantly stored in the muon flavour.

For definiteness, let us now consider a simple flavour model with $|y'_{e}|^2\ll |y'_{\mu}|^2$, and thus focus on the $e$-number abundance~$Y_e$, which is numerically dominant. The corresponding Boltzmann equation for $Y_e$ reads
\begin{align}\label{eq:dYedz-sim}
  sHz \frac{\text{d} Y_{e}}{\text{d}z}\: \approx\: \mathcal{S}_{e}\, +\, \mathcal{W}_{e}\,.
    \end{align}
Here and in the following, we parametrize the CP-violating source and washout rates as follows:
\begin{align}\label{eq:We-Se}
\mathcal{S}_{e}(z)\ &\approx\:  79.3 \, |y^{\prime}_e|^2\,|y^{\prime}_\mu|^2\,\frac{m^4_h}{z^4} \sum_{\substack{j>i=1}} \xi^{(i,j)}_{\rm CP}\,    \Big( I_{ij}(z) - I_{ji}(z) \Big)\,,\\
\mathcal{W}_{e}(z)\ &\approx\:  -\,0.2\, |y^{\prime}_e|^2\,\frac{m^4_h}{z^4}\,Y_{e}(z)\,,
\end{align}
where $I_{ij}(z)$  is given by \eqref{eq:Iijz}, and the CP-odd parameter $\xi^{(i,j)}_{\rm CP}$ is defined as
\begin{align}\label{eq:xi-def}
    \xi^{(i,j)}_{\rm CP}\, \equiv\, \frac{2\,\text{Im}(y'_{\mu i}y^{\prime *}_{e i}y'_{e j}y^{\prime *}_{\mu j})}{|y^{\prime}_e|^2\, |y^{\prime}_\mu|^2}\, =\, -\,\xi^{(j,i)}_{\rm CP}\,,
\end{align}
which encodes the net effect of CP violation, the relative magnitude of $|y'_{\alpha i}|$ and $|y'_{\alpha j}|$, and the hierarchy $|y'_{\mu i}|\gg |y'_{e j}|$, for all $i,j$ singlet neutrino flavours. Notice that for two singlet neutrino models, $\xi_{\rm CP} \equiv \xi^{(1,2)}_{\rm CP}$ as defined in~\eqref{eq:xi-def} is an arbitrary free quantity, constrained to take values
in the interval: $-1 \le \xi_{\rm CP} \le 1$.

Let us set the stage for our numerical evaluation of a simple flavour model. In the three neutrino-flavour case, 
as one may anticipate, a wider range of the seesaw parameter space can open after one adopts flavour symmetries or the Casas-Ibarra parameter\-isation~\cite{Casas:2001sr} to numerically scan the $3\times 3$ matrix~$y'$, which conveniently describes all neutrino masses and mixing~\cite{Hernandez:2015wna,Abada:2015rta,Drewes:2016gmt,Abada:2018oly,Drewes:2022kap,Fuyuto:2025feh}.
To better exemplify the significance of TRL, we consider only two non-degenerate GeV-scale singlet neutrinos contributing to leptogenesis, while the third singlet neutrino has a negligible contribution by either making it sufficiently heavy to decouple, or by suppressing its neutrino Yukawa interactions\footnote{As can be inferred from the seesaw relation, this condition can reflect a light sterile neutrino below the GeV scale, if the lightest SM neutrino does not have a vanishingly small mass.}. 
To maximise the value of the CP-violating source $\mathcal{S}_e$ in~\eqref{eq:We-Se}, we would need $|\xi_{\rm CP}| \sim 1$, while maintaining a sufficient hierarchy between $I_{12}$ and $I_{21}$, e.g.,~demanding $|y'_2|^2 \gg |y'_1|^2$ or vice versa.

There may be an additional enhancement  factor from neutrino Yukawa couplings. As mentioned below \eqref{eq:4DI-N}, $f_{N_1}$ and $f_{N_2}$ must be different to produce  non-vanishing $\mathcal{S}_{\mu}$ and $\mathcal{S}_{e}$ in \eqref{eq:CPmax}, which can be easily realized by different neutrino Yukawa matrix elements. It is also possible that one of the singlet neutrinos, e.g.~$N_2$, enters thermal equilibrium before sphaleron decoupling, whereas the singlet neutrino $N_1$ does not. In this case,  certain enhancement of  $\mathcal{S}_{e}$ can be realized by larger $N_2$-related Yukawa couplings. Nevertheless, the washout rate $\mathcal{W}_e$ given in \eqref{eq:We-Se} would be enhanced as it also depends on  the $N_2$-related Yukawa couplings. It is non-trivial to analyse the competition between $\mathcal{S}_{e}$ and  $\mathcal{W}_e$, and in particular the net effect on $Y_{e}$ by increasing the $N_2$-related Yukawa couplings, unless the washout effect is always smaller than the CP-violating source~\cite{Kanemura:2025rsy}.  For this reason, we will not detail this kind of enhancement in the current work. 

\begin{figure}[t]
	\centering
\includegraphics[scale=0.45]{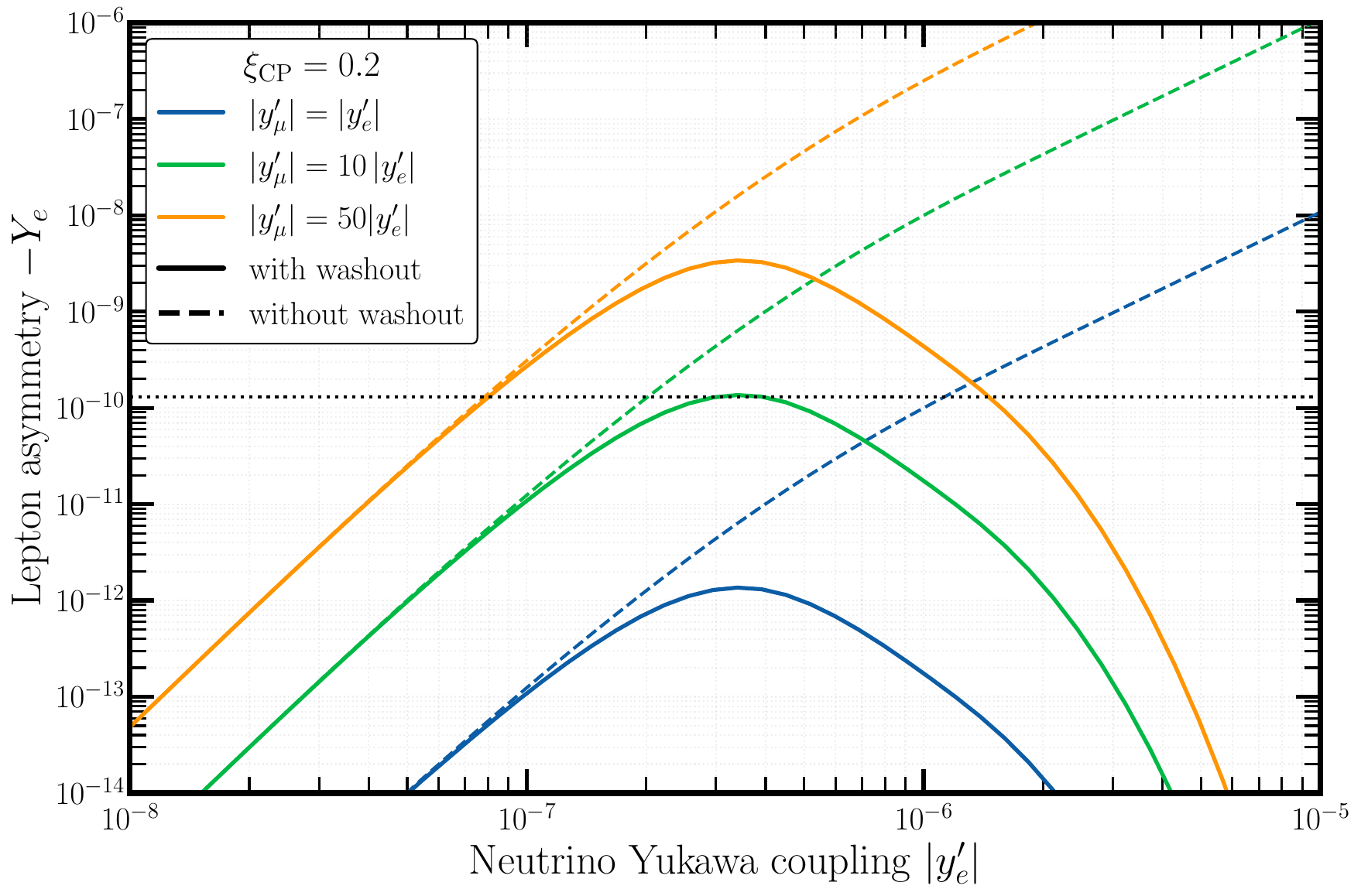}
	\caption{\label{fig:Ye-plot} The lepton asymmetry yield $|Y_e|=-Y_{e}$ from a maximised CP-violating source $\mathcal{S}_{e}$ as a function of the neutrino Yukawa coupling expression $|y'_e|$, for discrete choices of $|y'_\mu |$, where $|y'_\alpha| \equiv \sqrt{(y'y'^\dagger )_{\alpha\alpha}}$, with $\alpha =e,\mu$. The solid (dashed) lines denote the evolution with (without) washout effects~${\cal W}_{e,\mu}$. The parameter $\xi_{\rm CP} \equiv \xi^{(1,2)}_{\rm CP}$ [cf.~\eqref{eq:xi-def}] characterizes the size of CP violation from neutrino Yukawa couplings, and it is taken to be $\xi_{\rm CP} = 0.2$ for this numerical analysis. 
    In addition, we limit the masses of the two singlet neutrinos $N_{1,2}$ to lie within the interval: $1~\text{GeV} < M_1 < M_2  < 50~\text{GeV}$, so as to obey the relativistic  approximation considered here.  
    Finally, we assume $|y'_1|=|y'_e|$ and a large  $|y'_2|>10^{-6}$ such that $I_{12}$ is negligible, where $|y'_j| \equiv \sqrt{(y'^\dagger y')_{jj}}$, with $j =1,2$.
    Note that the horizontal dotted line denotes the value 
    of~${-Y_{e}}$ needed to explain the BAU via \eqref{eq:YB-Ya}. }
\end{figure}

To further simplify our numerical analysis, we set $\delta f_{N_2}=0$, such that $I_{12}=0$. We also neglect the different Yukawa dependences of $\mathcal{P}_1$ and $\mathcal{W}_e$ that may further enhance the final $Y_e$, as explained below \eqref{eq:P-def}. Simply assuming a similar dependence of $\mathcal{P}_1$ and $\mathcal{W}_e$  on Yukawa couplings, i.e., $|y'_1|\approx|y'_e|$, we can solve $Y_{e}$ in terms of $|y'_e|,|y'_\mu|$ and $\xi_{\rm CP}$ defined in \eqref{eq:xi-def}. A positive $\mathcal{S}_{e}$ would then require $\xi_{\rm CP}=(0,1]$ as $I_{21}<0$. For a sufficient $\xi_{\rm CP}=\mathcal{O}(0.1)$, the variation of $\mathcal{S}_{e}$ mainly comes from the hierarchy between $|y'_e|$ and $|y'_\mu|$. Taking this into account, we moderately set $\xi_{\rm CP}=0.2$ and consider three simple benchmark scenarios for $|y'_\mu|$: $|y'_\mu|=|y'_e|$, $|y'_\mu|=10|y'_e|$, and $|y'_\mu|=50|y'_e|$, when showing the predicted values of $Y_e=-|Y_e|$ in terms of $|y'_e|$ in Fig.~\ref{fig:Ye-plot}. 

At this point, we should stress that the precise values of the two singlet neutrino masses, $M_{1,2}$, are not as relevant for the numerical analysis presented here, as long as they obey the relativistic approximation we used throughout this work. For this to happen, we restrict $M_1,M_2\lesssim 50$~GeV, since the Higgs mass at $T\sim 200$~GeV yields $m_\phi\sim 80$~GeV. For heavier neutrinos, the thermally induced Higgs decay becomes kinematically forbidden at earlier times, and the lepton-number violating processes become important. For lighter neutrinos, on the other hand, the seesaw relation suggests smaller neutrino Yukawa couplings and consequently the CP-violating source becomes suppressed. Never\-theless, in~\cite{Li:2026rqg} it will be shown that a viable TRL with $M_1,M_2\gtrsim 1$~GeV can still be feasible. 

We then show the results in Fig.~\ref{fig:Ye-plot} by varying $|y'_e|$ in the interval: $[10^{-8},10^{-5}]$, where $|y'_e|$ is the square root of $|y'_e|^2$ defined in~\eqref{eq:yalpha-def}. We see from Fig.~\ref{fig:Ye-plot} that a certain hierarchy of the neutrino Yukawa matrix elements is preferred to yield the right amount of the BAU. We should mention that the required hierarchy $|y'_\mu|=10|y'_e|$ shown in Fig.~\ref{fig:Ye-plot} applies for the presumed  $\mathcal{W}_e$ and $\mathcal{P}_1$ having the same dependence on~$|y'_e|$ and $|y'_1|$, where $|y'_e|=|y'_1|$ was used.
Nevertheless, for a given washout rate $\mathcal{W}_e$ and hence a given {\it fixed} Yukawa combination $|y'_e|^2 \equiv \sum_i|y'_{e i}|^2$, the final $|Y_e|=-Y_e$ may be further enhanced~if
\begin{equation}
   \label{eq:enhance-2}
|y'_1|^2\equiv \sum_{\beta}|y'_{\beta 1}|^2\ <\ |y'_e|^2\,.
\end{equation}
This situation corresponds to a larger $\mathcal{S}_e$ through the $\mathcal{P}_1$ function defined in~\eqref{eq:P-def}.

It is evident from Fig.~\ref{fig:Ye-plot} that $|Y_e|$ increases for higher values of the ratio $|y'_\mu|/|y'_e|$, where we found  $|Y_e|\propto |y'_\mu|/|y'_e|$. Since the SM neutrino mass scale in the type-I seesaw is proportional to the neutrino Yukawa couplings, we should caution that considering an even larger ratio $|y'_\mu|/|y'_e|$ could necessitate a stronger arrangement of cancellation  among the neutrino Yukawa couplings, either by invoking approximate flavour symmetries or numerical fine-tuning.
The scaling $|Y_e|\propto |y'_\mu|/|y'_e|$ is consistent with the result presented in \cite{Kanemura:2025rsy}, where the simple scaling was observed in the weak washout regime ($\text{d}|Y_e|/\text{d}|y'_e|>0$). The result shown in  Fig.~\ref{fig:Ye-plot} further implies that the simple scaling is also valid at $\text{d}|Y_e|/\text{d}|y'_e|<0$ when the washout rate becomes important.  Finally, we  see that when $|y'_e|\gtrsim  10^{-7}$, the washout rate cannot be neglected, which  is expected  as the  decay rate of  $\phi \to \bar\ell+N$ will become larger than the Hubble expansion before sphaleron decoupling.

Thus far, we have analysed the lepton asymmetry from two neutrino-flavour contributions without specifying the potential textures of the neutrino Yukawa matrix. Applying the above numerical results, let us now consider a simple flavour scenario for the $3\times 3$ Yukawa matrix $y'$, and see how the general numerical results can give hints at the neutrino Yukawa textures. For definiteness, we take the Yukawa matrix $y'$ to have the following structure~\cite{Pilaftsis:2004xx}: 
\begin{align}\label{eq:y-structure}
y'=\begin{pmatrix}
\varepsilon_{e}   & a e^{-i\delta} & a e^{i\delta} \\
\varepsilon_{\mu} & b e^{-i\delta} & b e^{i\delta} \\
\varepsilon_{\tau}& c e^{-i\delta} & c e^{i\delta}
\end{pmatrix},
\end{align}
with $\varepsilon_{e,\mu,\tau}\ll a,b,c$ and $\delta\simeq \pi/4$. Such a flavour structure for $y'$ can accommodate the light neutrino masses and mixing observed in low-energy oscillation experiments. As studied in~\cite{Pilaftsis:2004xx}, it allows for successful electroweak-scale RL with large Yukawa couplings testable in laboratory experiments, such as collider searches and lepton-flavour or lepton-number violating processes. Subsequently, similar neutrino-Yukawa textures have also been considered in low-scale RL and ARS leptogenesis; e.g.,~see \cite{Pilaftsis:2005rv,Canetti:2010aw,Abada:2018oly,Drewes:2021nqr}.

With the Yukawa structure stated in \eqref{eq:y-structure}, let us first consider the case where only two singlet neutrinos participate in the formulae of  $\mathcal{W}_e$ and $\mathcal{S}_e$, fixing  $\delta=\pi/4$. In this case, we drop the $\varepsilon_i$ entries such that there are only two non-vanishing columns of $y'$, i.e., the second and third columns.  The washout rates would scale as
\begin{align}
    \mathcal{W}_e\propto 2 |a|^2\,,\quad \mathcal{W}_\mu\propto 2|b|^2\,.
    \end{align}
It indicates that a hierarchical washout rate can be realized by $a\ll b$ or $b\ll a$. Considering the limit $\mathcal{W}_\mu\gg \mathcal{W}_e$, we have $|y'_\mu|^2=2|b|^2\gg |y'_e|^2=2|a|^2$, which is also preferred for a correct order of $|Y_e|$, as shown in Fig.~\ref{fig:Ye-plot}.  However, the out-of-equilibrium condition provided by at least one singlet neutrino flavour requires that $|y'_i|^2$ should also be small.
This turns out to be not the case since the structure of~\eqref{eq:y-structure} predicts
\begin{align}
|y'_1|^2=|y'_2|^2=|a|^2+|b|^2+|c|^2\,.
\end{align}
Note that $|y'_1|^2$ and  $|y'_2|^2$ in our notation correspond  to the second and third columns of \eqref{eq:y-structure}, respectively. It indicates that a hierarchical washout rate between $ \mathcal{W}_e$ and $ \mathcal{W}_\mu$ cannot be achieved, since there is no hierarchy between $|y'_1|$ and $|y'_2|$. More precisely, the equality between $|y'_1|$ and $|y'_2|$ implies that $I_{12}=I_{21}$, so that the CP-violating source $\mathcal{S}_e$ given in~\eqref{eq:We-Se} vanishes. Therefore, we expect that it would be challenging to yield the correct order of $Y_B$.  

When we go to the three-neutrino case, where the summation over the  neutrino flavour index in the washout and CP-violating source rates runs from 1 to 3, we expect that there are more viable parameters for successful leptogenesis if one performs a numerical parameter scan. Without going to the details of the full parameter scan, one can also consider the generic Yukawa structure given in \eqref{eq:y-structure}.  In this case, we can make $a\ll b$ or $b\ll a$, and at the same time have small Yukawa couplings for one neutrino flavour keeping out of equilibrium, since the structure predicts 
\begin{align}
    |y'_1|^2=\sum_{i=1}^3|\varepsilon_i|^2\,, |y'_2|^2=|y'_3|^2=|a|^2+|b|^2+|c|^2\,.
\end{align}
For example, successful leptogenesis may be realized in the doubly hierarchical scenario:
\begin{align}
    \varepsilon_i\ll a\ll b,c\,,
\end{align}
provided that the SM neutrino masses and mixing are well described by current data~\cite{ParticleDataGroup:2024cfk}. In this doubly hierarchical scenario, the out-of-equilibrium neutrino $N_1$ can enhance $\mathcal{S}_e$  without increasing the washout rate $\mathcal{W}_e$ simultaneously, as indicated by~\eqref{eq:enhance-2}.

We conclude this section by commenting on the minimal number of  two singlet-neutrino flavours that we have considered in our numerical analysis. Longstanding discussions are ongoing about whether two or three singlet  neutrino flavours are needed  in low-scale leptogenesis, compatible with other cosmological or experimental requirements. From the theory perspective,  if only two heavier flavours are assumed, the lightest one could be a candidate for DM, and the scenario is rather predictive after accounting for the active neutrino masses and mixing~\cite{Hernandez:2016kel,Drewes:2016jae,Klaric:2020phc,Klaric:2021cpi}. If~all three flavours contribute to leptogenesis, this would generally require a demanding parameter scan~\cite{Canetti:2014dka,Drewes:2016gmt,Abada:2018oly,Drewes:2021nqr}. On the experimental side, if the lightest sterile neutrino is the DM  (or any very light, long-lived particle), then the required stability of the DM implies that the lightest active neutrino would have a mass much smaller than the other two heavier ones; see e.g.,~\cite{Canetti:2012vf,Drewes:2016upu}, which can be tested in upcoming neutrino experiments by measuring the lightest neutrino mass scale.  However, for all flavours participating in leptogenesis without imposing a DM restriction or a strong mass hierarchy among the three flavours, there is still no concrete result about the lightest SM neutrino mass. It would therefore be interesting to test whether TRL can be realized in the type-I seesaw, where two singlet neutrinos contribute to low-scale leptogenesis while the lightest one is the dark matter candidate.

\section{Conclusions}
\label{sec:con}

We have studied a new dominant thermal resonant mechanism for low-scale leptogenesis, termed Thermal Resonant Leptogenesis, within the minimal
Type-I seesaw framework. TRL goes well beyond the frequently discussed scenarios of mixing and oscillation of heavy neutrinos, as happens in RL and ARS models. The mechanism under study is generated by thermally induced Higgs decays to lepton doublets and singlet neutrinos, which can yield a dominant contribution to the BAU in large regions of the seesaw parameter space. In particular, a resonant phenomenon can occur in the thermal plasma through quantum coherences between lepton-doublet flavours, which can become dominant at the two-loop level. A Feynman-diagrammatic illustration of this pheno\-menon is shown in~Fig.~\ref{fig:2+1loop-Nself}. As a consequence, these resonant thermal lepton-flavour coherences can enhance the lepton asymmetries to the observable level. In addition, the mechanism we have been studying is very predictive, as it is mainly specified by SM-mediated interactions. Unlike RL and ARS scenarios, TRL does not require~us to assume a quasi-degenerate spectrum for the heavy neutrino masses, where the lightest singlet neutrino mass could naturally be as low as 1~GeV. Finally\-, the TRL mechanism does not depend on the nature of the singlet neutrinos, whether these neutrinos are Dirac or Majorana fermions.    

To consistently compute the  coherences between thermal lepton-doublet flavours, we build the appropriate flavour-covariant KB formalism which is employed to derive the relevant transport equations. To verify the consistency of our findings, we have presented two different methods to calculate the proper CP-violating source driving the TRL. We have confirmed the validity of the basic equation~\eqref{eq:L-conserved} in the relativistic approximation for the singlet neutrinos, thereby proving the equivalence of the two methods and their results obtained through~\eqref{eq:CPV-l} and~\eqref{eq:dDelta-nN/dt}. 

We should reiterate that there are two key ingredients in obtaining an observable TRL. The first is the enhancement factor $(\tilde b_\alpha-\tilde b_\beta)^{-1}$ [cf.~\eqref{eq:dSn/dt}] that comes from thermal lepton-doublet flavour coherences. This factor is fully determined, to leading order, by known SM charged-lepton Yukawa couplings, thereby rendering TRL a predictive scenario. The second key ingredient is the existence of a new CP-violating source that emerges at two loops [cf.~\eqref{eq:CPSource2loop}] which drastically modifies the flavour off-diagonal transport equations for $\Delta n_{\alpha\beta}$, as illustrated in~\eqref{eq:dDn2loop/dt}. The~presence\- of this CP-violating source, which was not taken into account in previous studies${}$~\cite{Beneke:2010dz,Garbrecht:2012pq,Garbrecht:2014iia}, is of utmost importance to avoid the suppression of lepton-flavour coherences by the interplay between charged-lepton Yukawa and gauge interactions, and therefore to obtain successful scenarios of TRL. Such constructions of viable TRL can be guided by the simple parametric formula of $Y_N \sim Y_\ell$ in~\eqref{eq:Yofz-N}. This can be contrasted with the corresponding formula in~\eqref{eq:Yofz-ell}, which results from heavy-neutrino mixing within the usual RL framework. Unless it is~${\delta M/M\lesssim 10^{-6}}$ in the singlet neutrino mass spectrum, TRL can naturally become the dominant mechanism for low-scale leptogenesis~\cite{Li:2026rqg}.

To explicitly demonstrate the significance of TRL with feeze-in initial conditions, in  Section~\ref{sec:num}  we consider a simple flavour model for the neutrino Yukawa couplings~$y'_{\alpha i}$, which is motivated by a flavour hierarchy that maximises the CP-violating source, ${\cal S}_e = {\cal S}_\mu$, while reducing the washout rates ${\cal W}_{e,\mu}$. In this simple model, we find in Fig.~\ref{fig:Ye-plot} that a wide range of values for~$y'_{\alpha i}$ can give rise to the correct order of magnitude for the leptonic asymmetry $Y_e$ leading to the observable~BAU. A~more detailed numerical analysis, taking into account the SM neutrino masses and mixing, as well as potential phenomenological implications for low-energy experiments and high-energy colliders, is  presented in a companion paper~\cite{Li:2026rqg}.

The present study opens up a number of new research directions in the construction of more holistic cosmological models that aspire to simultaneously address the DM problem and the CMB data by virtue of cosmic inflation. It would be interesting to investigate whether minimal and viable TRL models exist in which one of the singlet neutrinos could be the (warm) DM, within a suitable framework of Higgs inflation.

\section*{Acknowledgments}
We thank Marco Drewes and  Bj\"orn Garbrecht for discussions.  S.-P.~Li was supported by the JSPS Grant-in-Aid for JSPS Research Fellows No. 24KF0060, whilst AP's work is supported in part by
the STFC research grant: ST/X00077X/1.

\vfill\eject

\appendix

\section{From Schwinger-Dyson to Kadanoff-Baym equations}\label{app:full-KB}

To keep the discussion at a more intuitive and transparent level, our starting point is the Schwinger-Dyson equation which will be used to derive the Kadanoff-Baym (KB) equation relevant\- to flavour-covariant leptogenesis. Further technical details may be found in~\cite{Calzetta:1986cq,Prokopec:2003pj,Prokopec:2004ic,Beneke:2010dz,BhupalDev:2014pfm}, concerning  the application of  non-equilibrium QFT to models of baryogenesis and leptogenesis.  

The Schwinger-Dyson equation in spacetime coordinate $(x,y)$ reads\footnote{The minus sign is based on the convention of  the self-energy amplitude $-i\Sigma$, which will be consistently taken into account when we write down the self-energy amplitudes. }
\begin{align}\label{eq:SD-xy}
    [S_{ab}^{-1}(x,y)]_{\alpha\beta}=    [S_{0,ab}^{-1}(x,y)]_{\alpha\beta}-[\Sigma_{ab}(x,y)]_{\alpha\beta}\,,
\end{align}
where $S (S_0)$ is the full (free) propagator and $\Sigma$ is the self-energy correction. In this appendix, we use $\alpha,\beta,\gamma,...$ for flavour indices and $a,b,c,...$ for the Closed-Time-Path (CTP) indices.  After convolution with the full propagator from the right on both sides of \eqref{eq:SD-xy}, the first term on the right-hand side (RHS) yields
\begin{align}\label{eq:S0-1S}
        \int d^4 z[S_{0,ac}^{-1}(x,z)]_{\alpha\gamma}[S_{cb}(z,y)]_{\gamma\beta}&=g_{ad}g_{dc}\int d^4 z (\delta_{\alpha\gamma}i\slashed{\partial}_x-m_{\alpha\gamma})\delta^{(4)}(x-z)[S_{cb}(z,y)]_{\gamma\beta}
        \\[0.2cm]
        &=g_{ad}g_{dc} (\delta_{\alpha\gamma}i\slashed{\partial}_x-m_{\alpha\gamma})[S_{cb}(x,y)]_{\gamma\beta}\,,
\end{align}
where $m_{\alpha\gamma}$ denotes the vacuum mass matrix and for definiteness we will always work in the diagonal basis such that $m_{\alpha\gamma}=m_\alpha \delta_{\alpha\gamma}$.  $g_{ab}=(1,-1)$ is the `metric' in the CTP space featuring $X_{ac}Y_{cb}=g_{cd}X_{ac} Y_{db}$ for propagator or self-energy amplitude $X,Y$.  From here on, the repeated indices on both sides of equations represent the dummy indices summed in the corresponding space. 
With the same convolution, the left-hand side (lhs) of \eqref{eq:SD-xy} becomes\footnote{This can be seen as a matrix generalization of the propagator convolution both in flavour and CTP space; see Michael E. Peskin and Daniel V. Schroeder, An Introduction to Quantum Field Theory, Eq (2.56) and (3.118).}
\begin{align}\label{eq:SD-lhs}
    \int d^4 z[S_{ac}^{-1}(x,z)]_{\alpha\gamma}[S_{cb}(z,y)]_{\gamma\beta}=\delta_{ab}\delta_{\alpha\beta}\delta^{(4)}(x-y)\,.
\end{align}

We can rewrite  \eqref{eq:S0-1S} in the Wigner space as 
\begin{align}\label{eq:SD-1strhs}
    \int d^4 r e^{ik r } \int d^4 z[S_{0,ac}^{-1}(x,z)]_{\alpha\gamma}[S_{cb}(z,y)]_{\gamma\beta}&=g_{dc}e^{-i\Diamond}[S^{-1}_{0,ad}(k,u)]_{\alpha\gamma}\circ [S_{cb}(k,u)]_{\gamma\beta}\,,
\end{align}
by performing the Wigner transform,
\begin{align}
    F(k,u)&=\int d^4 r e^{ik r}F\left(u+\frac{r}{2},u-\frac{r}{2}\right),
    \\[0.2cm]
 e^{-i\Diamond}  A(k, u)\circ B(k,u)  &= \int d^4 r e^{ik r }\int d^4 z A(x,z)B(z,y)\,,
\end{align}
where $u\equiv (x+y)/2, r=x-y$,  and the diamond operator $\Diamond$ is defined as
\begin{align}
    \Diamond A \circ B\equiv \frac{1}{2}\left(\partial_u^{A}\partial_k^{B}-\partial_k^{A}\partial_u^{B}\right) A  B\,,
\end{align}
with $\partial_j^i$ acting on object $i=A,B$ with respect to momentum or spacetime $j=k, u$. We can further simplify the rhs of \eqref{eq:SD-1strhs} by using the Wigner-transformed inverse free propagator
\begin{align}
    [S^{-1}_{0,ad}(k,u)]_{\alpha\gamma}&=\int d^4 r e^{ik r}g_{ad}(\delta_{\alpha\gamma}i\slashed{\partial}_x-m_{\alpha\gamma})\delta^{(4)}(r)
    \\
    &=g_{ad} (\delta_{\alpha\gamma}\slashed{k}-m_{\alpha\gamma})\,,
\end{align}
where we have used $\partial_x \delta^{(4)}(r)=\partial_r\delta^{(4)}(r)$ and the $\delta$-function identity
\begin{align}\label{eq:delta-1}
    \int dx \delta'(x-x_0)\mathcal{F}(x)=-\mathcal{F}'(x_0)\,.
\end{align}
Then, expanding the diamond operator in \eqref{eq:SD-1strhs} with a spacetime-independent real mass, we arrive at
\begin{align}
 \left(\delta_{\alpha\gamma}\slashed{k}-m_{\alpha\gamma}+\frac{i}{2}\delta_{\alpha\gamma}\slashed{\partial}_u\right) g_{ad}g_{dc} [S_{cb}(k,u)]_{\gamma\beta}\,,
\end{align}
after identifying the fact that the second and higher order terms vanish because of 
\begin{align}
    \partial_u S_0^{-1}(k,u)=0\,, \quad \partial^2_k S_0^{-1}(k,u)=0\,. 
\end{align}
After  the same Wigner transform for the self-energy term, the Wigner-transformed Schwinger-Dyson equation reduces to 
\begin{align}
 \left(\delta_{\alpha\gamma}\slashed{k}-m_{\alpha\gamma}+\frac{i}{2}\delta_{\alpha\gamma}\slashed{\partial}_u\right) &g_{ad}g_{dc} [S_{cb}(k,u)]_{\gamma\beta}
\nonumber \\[0.2cm]
&=\delta_{ab}\delta_{\alpha\beta}+e^{-i\Diamond} [\Sigma_{ac}(k,u)]_{\alpha\gamma}\circ [S_{cb}(k,u)]_{\gamma\beta}\,.
\end{align}

Let us consider the full Wightman propagator 
\begin{align}
S_{<}(k,u)\equiv S_{12}(k,u)\,, \quad  S_{>}(k,u)\equiv S_{21}(k,u)\,,
\end{align}
where the kinetic equation for the distribution function can be derived.  Taking $a=1, b=2$, and  $a=2, b=1$, respectively,  we have 
\begin{align}\label{eq:S<}
\left(\delta_{\alpha\gamma}\slashed{k}-m_{\alpha\gamma}+\frac{i}{2}\delta_{\alpha\gamma}\slashed{\partial}_u\right)[S_<(k,u)]_{\gamma\beta}
 &=e^{-i\Diamond}[\Sigma_{T}(k,u)]_{\alpha\gamma}\circ [S_{<}(k,u)]_{\gamma\beta}
 \nonumber \\[0.2cm]
 &-e^{-i\Diamond} [\Sigma_{<}(k,u)]_{\alpha\gamma}\circ [S_{\bar T}(k,u)]_{\gamma\beta}\,,
\\[0.2cm]
\left(\delta_{\alpha\gamma}\slashed{k}-m_{\alpha\gamma}+\frac{i}{2}\delta_{\alpha\gamma}\slashed{\partial}_u\right)[S_>(k,u)]_{\gamma\beta}
&=e^{-i\Diamond}[\Sigma_{>}(k,u)]_{\alpha\gamma}\circ [S_{T}(k,u)]_{\gamma\beta}
 \nonumber \\[0.2cm]
 &-e^{-i\Diamond} [\Sigma_{\bar T}(k,u)]_{\alpha\gamma}\circ [S_{>}(k,u)]_{\gamma\beta}\,,\label{eq:S>}
\end{align}
where the minus sign on the rhs appears due to the contraction $\Sigma_{ac}S_{cb}=g_{cd} \Sigma_{ac} S_{db}$. Note that we can replace the time and anti-time ordered $X_{T}\equiv X_{11}, X_{\bar T}\equiv X_{22}$ as
\begin{align}
    X_{T}(k,u)&=\frac{1}{2}\left(X_>(k,u)+X_<(k,u)+X_{R}(k,u)+X_{A}(k,u)\right),
    \\
     X_{\bar T}(k,u)&=\frac{1}{2}\left(X_>(k,u)+X_<(k,u)-X_{R}(k,u)-X_{ A}(k,u)\right),
\end{align}
for $X=S,\Sigma$, where $X_{R(A)}\equiv X_T-X_{<(>)}=X_{>(<)}-X_{\bar T}$ denotes the retarded (advanced) propagator  or self-energy amplitude, with the second relation derived by unitarity relations~\cite{Millington:2012pf},
\begin{align}
X_T+X_{\bar T}=X_>+X_<\,.
\end{align}  
Further given that $X_A(k,u)=X^*_R(k,u)$\footnote{Precisely speaking, this condition holds only in the spatially homogeneous regime; see e.g., \cite{Millington:2012pf}.}, we can rewrite \eqref{eq:S<}-\eqref{eq:S>} as
\begin{align}
&\left[\slashed{k}-m+\frac{i}{2}\slashed{\partial}_u\right]_{\alpha\gamma}[iS_{<(>)}(k,u)]_{\gamma\beta}
\\[0.2cm]
&-e^{-i\Diamond} [\text{Re}\Sigma_{R}(k,u)]_{\alpha\gamma}\circ [iS_{<(>)}(k,u)]_{\gamma\beta}
- e^{-i\Diamond}[i\Sigma_{<(>)}(k,u)]_{\alpha\gamma}\circ [\text{Re}S_{R}(k,u)]_{\gamma\beta}
    \nonumber \\[0.2cm]
    &=-\frac{i}{2}e^{-i\Diamond} \left([i\Sigma_{>}(k,u)]_{\alpha\gamma}\circ [iS_{<}(k,u)]_{\gamma\beta} -[i\Sigma_{<}(k,u)]_{\alpha\gamma}\circ [iS_{>}(k,u)]_{\gamma\beta}\right),\nonumber
\end{align}
and the Hermitian-conjugate equation (acting both on the spinor and flavour space)
\begin{align}
  [i S_{<(>)}\gamma^0]_{\alpha\gamma}&\left[\gamma^0\slashed{k}\gamma^0-m -\frac{i}{2}\overleftarrow{\slashed{\partial}_u}\right]_{\gamma\beta}
 \\[0.2cm]
 &-e^{i\Diamond}[iS_{<(>)}\gamma^0]_{\alpha\gamma}\circ [\gamma^0\text{Re}\Sigma_{R}\gamma^0]_{\gamma\beta}
-e^{i\Diamond}[\text{Re}S_{R}\gamma^0]_{\alpha\gamma}\circ [\gamma^0 i\Sigma_{<(>)}\gamma^0]_{\gamma\beta}
  \nonumber  \\[0.2cm]
    &=\frac{i}{2}e^{i\Diamond} \left([iS_{<}\gamma^0]_{\alpha\gamma}\circ [\gamma^0i\Sigma_{>}\gamma^0]_{\gamma\beta}-[iS_{>}\gamma^0]_{\alpha\gamma}\circ [\gamma^0i\Sigma_{<}\gamma^0]_{\gamma\beta}\right),  \nonumber
\end{align}
where we have used the spinor-space Hermitian conditions\footnote{This can be checked explicitly by using the Hermitian distribution matrix defined in appendix~\ref{app:lepton-Wightman}: $f^\dagger=f$, and the Dirac matrix property: $\gamma^0\gamma^\mu\gamma^0=(\gamma^\mu)^\dagger$.}
\begin{align}
(iS_{<(>)})^\dagger=\gamma^0(iS_{<(>)})\gamma^0\,, \quad (i\Sigma_{<(>)})^\dagger=\gamma^0(i\Sigma_{<(>)})\gamma^0\,,\quad(\text{Re}\Sigma_{R})^\dagger=\gamma^0(\text{Re}\Sigma_{R})\gamma^0\,.
\end{align}

In spatially homogeneous regime, we can write the above two equations by rearranging the $\gamma^0$ matrix as\footnote{Here, we take the upper and lower index notation as: $\slashed{k}=k_0\gamma^0-{\bf k}\cdot\boldsymbol{ \gamma}=k_0\gamma_0-k_i \gamma^i=k_0\gamma^0+k_i \gamma_i$.}
\begin{align}\label{eq:KB-full}
&\left[k_0 -{\bf k}\cdot{\boldsymbol{ \gamma}}\gamma^0-m_{}\gamma^0+\frac{i}{2}\partial_t\right]_{\alpha\gamma}[\gamma^0iS_{<(>)}]_{\gamma\beta}
\\[0.2cm]
&-e^{-i\Diamond}[\text{Re}\Sigma_{R}\gamma^0]_{\alpha\gamma}\circ [\gamma^0 iS_{<(>)}]_{\gamma\beta}-e^{-i\Diamond}[i\Sigma_{<(>)}\gamma^0]_{\alpha\gamma}\circ [\gamma^0\text{Re}S_{R}]_{\gamma\beta}
    \nonumber \\[0.2cm]
    &=\frac{-i}{2}e^{-i\Diamond} \left([i\Sigma_{>}\gamma^0]_{\alpha\gamma}\circ [\gamma^0 iS_{<}]_{\gamma\beta}-[i\Sigma_{<}\gamma^0]_{\alpha\gamma}\circ[\gamma^0iS_{>}]_{\gamma\beta}\right),\nonumber 
\end{align}
and its Hermitian conjugate (multiplying from the left by the Dirac matrix $\gamma^0$)
\begin{align}\label{eq:KB-full-hc}
[\gamma^0iS_{<(>)}]_{\alpha\gamma}&\left[k_0 -\frac{i}{2}\overleftarrow{\partial_t}+{\bf k}\cdot\boldsymbol{ \gamma}\gamma^0-m_{}\gamma^0\right]_{\gamma\beta}
\\[0.2cm]
&-e^{i\Diamond}[\gamma^0iS_{<(>)}]_{\alpha\gamma}\circ[\text{Re}\Sigma_{R}\gamma^0]_{\gamma\beta}-e^{i\Diamond}[\gamma^0\text{Re}S_{R}]_{\alpha\gamma}\circ [i\Sigma_{<(>)}\gamma^0]_{\gamma\beta}
    \nonumber \\[0.2cm]
    &=\frac{i}{2}e^{i\Diamond} \left([\gamma^0iS_{<}]_{\alpha\gamma}\circ [i\Sigma_{>}\gamma^0]_{\gamma\beta}-[\gamma^0iS_{>}]_{\alpha\gamma}\circ [i\Sigma_{<}\gamma^0]_{\gamma\beta}\right).  \nonumber
\end{align}

The sum of \eqref{eq:KB-full} and \eqref{eq:KB-full-hc} yields the constraint equation
\begin{align}\label{eq:KB-H}
   2k_0[\gamma^0iS_{<(>)}]_{\alpha\beta}&-\left\{{\bf k}\cdot\boldsymbol{ \gamma}\gamma^0+m\gamma^0,\gamma^0iS_{<(>)}\right\}_{\alpha\beta}
    \\[0.2cm]
    &-\cos\Diamond\left\{\text{Re}\Sigma_{R}\gamma^0,\gamma^0iS_{<(>)}\right\}_{\alpha\beta}+i\sin\Diamond\left[\text{Re}\Sigma_{R}\gamma^0,\gamma^0iS_{<(>)}\right]_{\alpha\beta}
       \nonumber  \\[0.2cm]
    & -\cos\Diamond\left\{i\Sigma_{<(>)}\gamma^0,\gamma^0\text{Re}S_{R}\right\}_{\alpha\beta}+i\sin\Diamond \left[i\Sigma_{<(>)}\gamma^0,\gamma^0\text{Re}S_{R}\right]_{\alpha\beta}
  \nonumber   \\[0.2cm]
    &=\frac{i}{2}\cos\Diamond \left(\left[\gamma^0 iS_<,i\Sigma_>\gamma^0\right]_{\alpha\beta}-\left[\gamma^0iS_>,i\Sigma_<\gamma^0\right]_{\alpha\beta}\right)
 \nonumber    \\[0.2cm]
    &-\frac{1}{2}\sin\Diamond\left(\left\{\gamma^0 iS_<,i\Sigma_>\gamma^0\right\}_{\alpha\beta}-\left\{\gamma^0iS_>,i\Sigma_<\gamma^0\right\}_{\alpha\beta}\right),\nonumber 
\end{align}
and the difference gives the kinetic equation
\begin{align}\label{eq:KB-antiH}
 i\partial_t[\gamma^0iS_{<(>)}]_{\alpha\beta}&-\left[{\bf k}\cdot\boldsymbol{ \gamma}\gamma^0+m\gamma^0,\gamma^0iS_{<(>)}\right]_{\alpha\beta}
 \\[0.2cm]
  &-\cos\Diamond\left[\text{Re}\Sigma_{R}\gamma^0,\gamma^0iS_{<(>)}\right]_{\alpha\beta}+i\sin\Diamond\left\{\text{Re}\Sigma_{R}\gamma^0,\gamma^0iS_{<(>)}\right\}_{\alpha\beta}
\nonumber   \\[0.2cm]
     &-\cos\Diamond\left[i\Sigma_{<(>)}\gamma^0,\gamma^0\text{Re}S_{R}\right]_{\alpha\beta}
    +i\sin\Diamond \left\{ i\Sigma_{<(>)}\gamma^0,\gamma^0\text{Re}S_{R}\right\}_{\alpha\beta}
    \nonumber     \\[0.2cm]
  &  =-\frac{i}{2}\cos\Diamond \left(\left\{\gamma^0 iS_<,i\Sigma_>\gamma^0\right\}_{\alpha\beta}-\left\{\gamma^0iS_>,i\Sigma_<\gamma^0\right\}_{\alpha\beta}\right)
  \nonumber   \\[0.2cm]
    &+\frac{1}{2}\sin\Diamond\left(\left[\gamma^0 iS_<,i\Sigma_>\gamma^0\right]_{\alpha\beta}-\left[\gamma^0iS_>,i\Sigma_<\gamma^0\right]_{\alpha\beta}\right),\nonumber 
\end{align}
where $[A,B],\{A,B\}$ denote the commutation and anti-commutation, respectively.

\section{Wightman propagators for quasi-thermal leptons}
\label{app:lepton-Wightman}

Following \cite{Millington:2012pf,BhupalDev:2014pfm}, in this appendix we derive the CTP propagators for SM leptons assuming spatial homogeneity, which works well for models of low-scale leptogenesis. Furthermore, we derive the free and resummed Wightman propagators that are consistent with the results based on the KB and on-shell approximations adopted in \cite{Prokopec:2003pj,Prokopec:2004ic,Beneke:2010dz,Garbrecht:2012pq}. The derivation presented below makes clear the approximations that were made to reach consistency. 

Let us start by writing down the Wightman propagator
\begin{align}
 \left[iS_{>}(x,y)\right]_{\alpha \beta}\equiv \langle\psi_\alpha(x)\bar\psi_\beta(y)\rangle\,,
\end{align}
where  $\alpha,\beta$ denote the flavour indices. The fermion field $\psi_\alpha$ in an arbitrary flavour basis can be obtained by performing rotation of the mass eigenstate basis. However, it is always convenient to express it in a special basis that removes most of the matrix manipulation. For SM leptons with the Yukawa interactions given in \eqref{eq:lag},  one is free to work in the basis where both the charged-lepton Yukawa matrix $y$ and the Majorana neutrino mass matrix $M$ are diagonal. Then, in this mass eigenstate basis, the free lepton field can be expanded as
\begin{align}
    \psi_\alpha(x)=\sum_{s}\int\frac{{\rm d}^3 \bf{k}}{(2\pi)^3\sqrt{2E_\alpha(\bf k)}}\left(e^{i {\bf k\cdot x}}u_\alpha({\bf{k}},s)b_\alpha({\bf{k}},s,x_0)+e^{-i {\bf k\cdot x}} v_\alpha({\bf{k}},s)d^\dagger_\alpha({\bf{k}},s,x_0)\right),
\end{align}
where the energy $E_\alpha({\bf k})=\sqrt{|{\bf k}|^2+m_\alpha^2}$, spinor $u_\alpha,v_\alpha$, and operators $b_\alpha, d_\alpha$ are all given in the diagonal-matrix form\footnote{In doing this, there is no summation over the repeated flavour indices on the rhs.}. The above interaction-picture particle annihilation and antiparticle creation operators are related to the Schr\"odinger-picture operators via 
\begin{align}
b_\alpha({\bf{k}},s,x_0)=b^S_\alpha({\bf{k}},s)e^{-iE_\alpha({\bf k})x_0}\,, \quad d_\alpha^\dagger({\bf k},s, x_0)=d_\alpha^{S\dagger }({\bf k},s)e^{iE_\alpha({\bf k})x_0}\,,
\end{align}
with $b^{S(\dagger)}_\alpha({\bf{k}},s)=b^{(\dagger)}_\alpha({\bf{k}},s,0)$, and the anti-commutation relation
\begin{align}\label{eq:commute-tf}
    \left\{a_\alpha({\bf{k}},s_1,x_0),a^\dagger_\beta({\bf{k'}},s_2,y_0)\right\}=(2\pi)^2\delta^{(3)}({\bf k-k'})\delta_{s_1s_2}\delta_{\alpha\beta}e^{-i E_\alpha x_0+iE_\beta y_0}\,,
\end{align}
for $a=b,d$, derived from the equal-time anti-commutation relation of Schr\"odinger-picture operators
\begin{align}\label{eq:commute-S}
    \left\{a^S_\alpha({\bf{k}},s_1,t),a^{S\dagger}_\beta({\bf{k'}},s_2,t)\right\}=(2\pi)^2\delta^{(3)}({\bf k-k'})\delta_{s_1s_2}\delta_{\alpha\beta}\,.
\end{align}
It is straightforward to evaluate the product $\psi_\alpha \bar\psi_\beta$ using the field expansion, which yields four combinations of creation and annihilation operator products, $b_\alpha b_\beta^\dagger$, $b_\alpha d_\beta$, $d_\alpha^\dagger b_\beta^\dagger$, and $d_\alpha^\dagger d_\beta$. In spatially homogeneous and isotropic background, we expect $b_\alpha d_\beta$, and  $d_\alpha^\dagger b_\beta^\dagger$ to vanish within the ensemble average, which may be verified explicitly through $\langle\mathcal{O}\rangle=\text{Tr}(\rho \mathcal{O})/\text{Tr}(\rho)$ with $\rho$ the density matrix of the system in a spatially homogeneous environment~\cite{Calzetta:1986cq,Millington:2012pf}.

Define the \textit{occupation-number matrix}\footnote{We may also call it  \textit{statistical distribution matrix}. This is a matrix generalization of occupation numbers or distribution functions in the flavour space. However, we should mention that it is the trace rather than a diagonal entry  that is flavour invariant.},
\begin{align}
    [f_{s_1s_2}({\bf{k}},{\bf{k'}},x_0,y_0)]_{\alpha\beta}&\equiv\langle b_\beta^\dagger({\bf{k'}}, s_2,y_0)b_\alpha({\bf{k}}, s_1,x_0)\rangle\,,
    \\[0.2cm]
        [\bar f_{s_1s_2}({\bf{k}},{\bf{k'}},x_0,y_0)]_{\alpha\beta}&\equiv \langle d_\alpha^\dagger({\bf{k}}, s_1,x_0)d_\beta({\bf{k
    '}}, s_2,y_0)\rangle\,,
\end{align}
where the inverse order of the flavour indices in $f_{s_1s_2}$ is due to the fact that $b_\alpha^\dagger$ transfers differently from $b_\alpha$ in flavour space.  Note that the occupation-number matrix  satisfies the following Hermitian condition in flavour space. 
\begin{align}
    [f_{s_1s_2}({\bf k},{\bf k'},x_0,y_0)]_{\alpha\beta}^*&=[f_{s_2s_1}({\bf k'}, {\bf k},x_0,y_0)]_{\beta\alpha}\,,  
    \\[0.2cm]
    [\bar f_{s_2s_1}({\bf k'},{\bf k},x_0,y_0)]_{\beta\alpha}&=[f_{s_1s_2}({\bf k},{\bf k'},x_0,y_0)]_{\alpha\beta}\,.
\end{align}
Using the anti-commutation relation, we have 
\begin{align}
    \langle b_\alpha({\bf k}, s_1,x_0,y_0)& b_\beta^\dagger({\bf k'}, s_2,x_0,y_0)\rangle
    \\[0.2cm]
    &=(2\pi)^3 \delta^{(3)}({\bf k-k'}) \delta_{s_1s_2}\delta_{\alpha\beta}-[f_{s_1s_2}({\bf k},{\bf k'},x_0,y_0)]_{\alpha\beta}\,,\nonumber 
\end{align}
yielding 
\begin{align}\label{eq:2pf-expand}
    \langle\psi_\alpha(x) \bar \psi_\beta(y)\rangle &=\sum_{s_1,s_2}\int\frac{{\rm d}^3 {\bf k}}{(2\pi)^3\sqrt{2E_\alpha(\bf k)}}\frac{{\rm d}^3 {\bf k'}}{(2\pi)^3\sqrt{2E_\beta(\bf k')}}
  \\[0.2cm]
    &\times \Big(e^{i{\bf k\cdot x}-i{\bf k' \cdot y}}u_\alpha({\bf k},s_1)\bar u_\beta({\bf k'},s_2)(2\pi)^3 \delta^{(3)}({\bf k-k'})\delta_{s_1s_2}\delta_{\alpha\beta}e^{-i E_\alpha x_0+i E_\beta y_0}
   \nonumber\\[0.2cm]
   &-e^{i{\bf k\cdot x}-i{\bf k' \cdot y}}u_\alpha({\bf k},s_1)\bar u_\beta({\bf k'},s_2)[f_{s_1s_2}({\bf{k}},{\bf{k'}},x_0,y_0)]_{\alpha\beta}
\nonumber    \\[0.2cm]
    &+
    e^{-i{\bf k\cdot x}+i{\bf k'\cdot y}}v_\alpha({\bf k},s_1)\bar v_\beta({\bf k'},s_2)[\bar f_{s_1s_2}({\bf{k}},{\bf{k'}},x_0,y_0)]_{\alpha\beta}\Big).  \nonumber
\end{align}
In a spatially homogeneous and isotropic background, we make the following correspondence:
\begin{align}\label{eq:corres}
[f(\bar f)_{s_1s_2}({\bf{k}},{\bf{k'}},x_0,y_0)]_{\alpha\beta}\to \delta_{s_1s_2} (2\pi)^3 \delta^{(3)}({\bf k-k'})f_{\alpha\beta}({\bf k}, u_0)e^{-(+)iE_\alpha({\bf k})(x_0-y_0)}\,,
\end{align}
where $f_{\alpha\beta}({\bf k},u_0)$ with $u_0\equiv(x_0+y_0)/2$ reflects the fact that the spatial evolution of the occupation-number matrix  depends only on ${\bf x-y}$ after Fourier transform, and the time dependence arises from the spacetime average in a  temporally inhomogeneous regime.  This spacetime correspondence is also expected if one performs the Wigner transform.  Further using the spin sum~\cite{BhupalDev:2014pfm}
\begin{align}
    \sum_s u_\alpha({\bf k},s)\bar u_\beta({\bf k},s)&=[\slashed{p}+m]_{\alpha\beta}\,,
    \\[0.2cm]
    \sum_s v_\alpha({\bf k},s)\bar v_\beta({\bf k},s)&=[\slashed{p}+m]_{\alpha\beta}\,,
\end{align}
and integrating over ${\bf k'}$ in \eqref{eq:2pf-expand}, we arrive at 
\begin{align}\label{eq:S>-1}
    \left[iS_{>}(x,y)\right]_{\alpha \beta}=\int\frac{{\rm d}^3 {\bf k}}{(2\pi)^32E_\alpha (\bf k)}&\Big[e^{-ik (x-y)}(\slashed{k}+m_\alpha)\left(\delta_{\alpha\beta}-f_{\alpha\beta}({\bf k},u_0)\right)
    \nonumber\\[0.2cm]
    &+e^{ik (x-y)}(\slashed{k}-m_\alpha) \bar f_{\alpha\beta}({\bf k}, u_0)\Big].
\end{align}
Using
\begin{align}
    \int\frac{{\rm d}^3 {\bf k}}{(2\pi)^3 2E_\alpha({\bf k})}=\int \frac{{\rm d}^4 k}{(2\pi)^4}(2\pi)\delta(k^2-m_\alpha^2)\theta(k_0)\,,
\end{align}
we can rewrite the Wightman propagator as
\begin{align}
        \left[iS_{>}(x,y)\right]_{\alpha \beta}=\int\frac{{\rm d}^4 k}{(2\pi)^4}&e^{-i k (x-y)}(2\pi)\delta(k^2-m_\alpha^2)(\slashed{p}+m_\alpha)
        \\[0.2cm]
        &\times \left[\theta(k_0)\left(\delta_{\alpha\beta}-f_{\alpha\beta}({\bf k}, u_0)\right)-\theta(-k_0)\bar f_{\alpha\beta}(-{\bf k},u_0)\right]\,,\nonumber
\end{align}
after taking $k\to -k$ in the second term of \eqref{eq:S>-1}. Comparing to the inverse Wigner transform 
\begin{align}
    F(x,y)=\int \frac{{\rm d}^4k}{(2\pi)^4}  e^{-ik r}F\left(k,u\right),
\end{align}
with $r=x-y, u=(x+y)/2$,   we can extract the single-momentum representation of the free Wightman propagator as
\begin{align}\label{eq:S>-free}
       \left[iS_{>}({ k},t)\right]_{\alpha \beta}=(2\pi)&\delta(k^2-m_\alpha^2)(\slashed{k}+m_\alpha)
       \\[0.2cm]
       &\times \left[\theta(k_0)\left(\delta_{\alpha\beta}-f_{\alpha\beta}(k_0,{\bf k}, t)\right)-\theta(-k_0)\bar f_{\alpha\beta}(-k_0,-{\bf k},t)\right]\,,\nonumber
\end{align}
with $t\equiv u_0$. Note that we have explicitly spelled out the dependence of $\bar f_{\alpha\beta}$ on $k_0$ since the on-shell condition $k^2-m_\alpha^2=0$ has two frequency modes and $\theta(-k_0)$ selects the negative mode. 
For massless leptons in quasi-thermal equilibrium, we take $m_\alpha=0$, and may write
\begin{align}
 f_{\alpha\beta}(k_0,{\bf k},t)=\left[\frac{1}{e^{(E({\bf k})-\mu_{}(t))/T}+1}\right]_{\alpha\beta}\,,\quad  \bar f_{\alpha\beta}(-k_0,-{\bf k},t)=\left[\frac{1}{e^{(E({\bf k})+\mu_{}(t))/T}+1}\right]_{\alpha\beta}\,,
\end{align}
with $E({\bf k})=|{\bf k}|$. It is trivial to see that the small chemical matrix  $\mu_{}(t)$ transforms in the same way as $f_{\alpha\beta}$ in flavour space. Recalling that the field rotation $\psi'\to V \psi$ or more explicitly  $b'_\alpha\to V_{\alpha\gamma}b_\gamma, d'_\alpha\to d_\beta V^\dagger_{\beta \alpha}$ dictates the transformation rules of $f_{\alpha\beta}$ and $\bar f_{\alpha\beta}$  as
\begin{align}
   f'_{\alpha\beta}\to V_{\gamma\beta}^\dagger V_{\alpha \sigma} f_{\sigma \gamma} \,,\quad \bar f'_{\alpha\beta}\to V_{\gamma\beta}^\dagger V_{\alpha \sigma}\bar f_{\sigma \gamma}\,,
\end{align}
from which we can infer the same transformation rule for  the chemical potential matrix after expanding $f_{\alpha\beta}$ in terms of a small $\mu_{}$,
\begin{align}
    f_{\alpha\beta}(|{\bf k}|,t)=\frac{1}{e^{|{\bf k}|/T}+1}\delta_{\alpha\beta}+\frac{\mu_{\alpha\beta}(t)}{T}\frac{e^{|{\bf k}|/T}}{(e^{|{\bf k}|/T}+1)^2}+\mathcal{O}(\mu^2/T^2)\,.
\end{align}
There is  caution when we use the above expansion. While we generally expect a sufficiently weak dependence of diagonal $\mu_{\alpha\alpha}$ on momentum, as we have seen from~\eqref{eq:delta-fab}, the off-diagonal entries  $\mu_{\alpha\beta}$ are momentum dependent if we match them to $\delta f_{\alpha\beta}$. The derivation of  the off-diagonal correlations $\delta f_{\alpha\beta}$ in this work does not assume a momentum-independent $\mu_{\alpha\beta}$.  As we can anticipate, using a momentum-independent $\mu_{\alpha\beta}$ in~\eqref{eq:Delta-nab-def} would lead to inconsistent results, if one simply factors  $\mu_{\alpha\beta}$ out of the momentum integral. 

Analogous derivation can be done for  $[iS_<(x,y)]_{\alpha\beta}\equiv -\langle \bar \psi_\beta(y)\psi_\alpha(x)\rangle$\footnote{The minus sign comes from anti-commutation relation of fermions.}, which yields
\begin{align}\label{eq:S<-free}
    [iS_{<}(k,t)]_{\alpha\beta}=-(2\pi)&\delta(k^2-m_\alpha^2)(\slashed{k}+m_\alpha)
 \\[0.2cm]
       &\times \left[\theta(k_0)f_{\alpha\beta}(k_0,{\bf k}, t)-\theta(-k_0) \left(\delta_{\alpha\beta}-\bar f_{\alpha\beta}(-k_0,-{\bf k}, t)\right)\right].   \nonumber  
\end{align}
From \eqref{eq:S>-free} and \eqref{eq:S<-free}, we reproduce the free Wightman propagators that are derived from the KB and on-shell approximations, after we make the correspondence~\eqref{eq:corres}. Note that the two Wightman propagators satisfy the generalized  KMS relations only in the diagonal entries,
\begin{align}\label{eq:gKMS-S1}
   [iS_>(k)]_{\alpha\alpha}+e^{(k_0-\mu_{\alpha\alpha})/T} [iS_<(k)]_{\alpha\alpha}&=\mathcal{O}(\mu^2_{\alpha\alpha})\,, \quad \text{for}~\alpha=\beta\,,
   \\[0.2cm]
      [iS_>(k)]_{\alpha\beta}&= [iS_<(k)]_{\alpha\beta}\,, \quad \text{for}~\alpha\neq \beta\,.\label{eq:gKMS-S2}
\end{align}

Next, let us consider the resummed Wightman propagators for SM leptons. Including the loop corrections, we write the resummed propagator as
\begin{align}\label{eq:Sab-matrix}
    S_{ab}^{-1}=S_{0,ab}^{-1}-\Sigma_{ab}
=P_R\begin{pmatrix} \slashed{k}-\Sigma_T & -\Sigma_< \\ -\Sigma_> & -\slashed{k}+\Sigma_T^*\\
    \end{pmatrix}
    P_L\,.
\end{align}
where $\Sigma_T\equiv \Sigma_{11}=-\Sigma_{22}^*$, and the chirality operators $P_R, P_L$ dictate the left-handed property of massless lepton doublets, with $m=0$ applied. In general, the self-energy amplitudes can be parametrized as $\Sigma=-a \slashed{k}-b\slashed{u}$\footnote{Note that the chirality operators have been factored out in \eqref{eq:Sab-matrix} such that we will write all the self-energy amplitudes without the chirality projection. The minus sign in $\Sigma$ is a conventional choice.}  with $u_\mu$ the four-velocity of the plasma~\cite{Weldon:1982bn}, such that the matrix elements in $S_{ab}^{-1}$ do not commute in the spinor space.  
Using the unitarity and causality relations
\begin{align}
 \Sigma_>+\Sigma_<&=\Sigma_T-\Sigma_{T}^*=2i\text{Im}\Sigma_{T}\,,
 \\[0.2cm]
    \Sigma_>-\Sigma_<&=\Sigma_{R}-\Sigma_{\rm A}=2i\text{Im}\Sigma_{R}\,,
\end{align}
and the generalized KMS relation
\begin{align}\label{eq:KMS-Sigma}
    \Sigma_>(k)=-e^{(k_0-\mu)/T}\Sigma_<(k)\,,
\end{align}
with $\mu$ denoting the diagonal chemical potentials\footnote{As shown in~\eqref{eq:gKMS-S1}-\eqref{eq:gKMS-S2}, the KMS relation only holds for the diagonal entries. For thermally resummed leptons, we will only take into account the corrections from charged-lepton Yukawa and gauge interactions, but neglect that from nonthermal neutrino Yukawa interactions.}, we have
\begin{align}
    \Sigma_<(k)&= -2if(k_0,\mu)\text{Im}\Sigma_{R}(k)\,,
    \\[0.2cm]
        \Sigma_>(k)&=2i\left(1-f(k_0,\mu)\right)\text{Im}\Sigma_{R}(k)\,,
        \\[0.2cm]
        \text{Im}\Sigma_T(k)&=\left(1-2f(k_0,\mu)\right)\text{Im}\Sigma_ R(k)\,,
\end{align}
where $f(k_0,\mu)=(e^{(k_0-\mu)/T}+1)^{-1}$. Further given the identity $\text{Re}\Sigma_T=\text{Re}\Sigma_{R}$, we can rewrite all the self-energy amplitudes in terms of the retarded one. Moreover, observing the fact that $\text{Im}\Sigma_R$  will be a higher-order effect with respect to $\text{Re}\Sigma_R$ near the pole~\cite{Kanemura:2024dqv}, the commutation among the $S_{ab}^{-1}$ elements can then be reduced to scalar product up to $\mathcal{O}(\text{Im}a), \mathcal{O}(\text{Im}b)$ corrections. Neglecting these corrections, the inverse of the resummed propagator reduces to simple matrix manipulation in spinor space. We arrive at 
\begin{align}
    S_{ab}&=P_L\frac{1}{(\slashed{k}-\Sigma_T)(\slashed{k}-\Sigma_T^*)+\Sigma_{<}\Sigma_{>}}
\begin{pmatrix} \slashed{k}-\Sigma_T^* & -\Sigma_< \\ -\Sigma_> & -\slashed{k}+\Sigma_T \\
    \end{pmatrix}P_R
  \\[0.2cm]
  &  =P_L\frac{1}{(\slashed{k}-\text{Re}\Sigma_{R})^2+\text{Im}\Sigma_{ R}^2} \begin{pmatrix} \slashed{k}-\Sigma_T^* & -\Sigma_< \\ -\Sigma_> & -\slashed{k}+\Sigma_T \\
    \end{pmatrix}P_R\,,\label{eq:S-full}
\end{align}
where we used $P_R^{-1}=P_R, P_L^{-1}=P_L$. 

In the following, we will specify the form of the resummed Wightman propagator $S_{<}$, and then make a comparison with the free case.  The 12-component of \eqref{eq:S-full} yields 
\begin{align}\label{eq:S<-inm}
    S_<(k)=P_L\frac{2if(k_0,\mu)\text{Im}\Sigma_{ R}(k)}{(\slashed{k}-\text{Re}\Sigma_{R})^2+\text{Im}\Sigma_{R}^2}P_R\,.
\end{align}
Since $\text{Im}\Sigma_{R}(k)$ carries higher-order couplings with respect to the real part near the pole in perturbation theory, we treat $\text{Im}a, \text{Im}b$ as infinitesimal variables via $-\text{Im}a, -\text{Im}b\to \epsilon$. Now, if we keep only the $\slashed{k}$ structure in $\text{Im}\Sigma_{R}(k)$ and apply  the narrow-width approximation, 
\begin{align}
    \frac{\text{sign}(k_0)\epsilon}{x^2+[\text{sign}(k_0)\epsilon]^2}\approx \pi \delta (x)\,,
\end{align}
we would arrive at
\begin{align}
    S_<(k)&=2\pi i \text{sign}(k_0) f(k_0,\mu) P_L\slashed{k}P_R\delta[(\slashed{k}-\text{Re}\Sigma_{ R})^2]\label{eq:S<-full-approx-0}
    \\[0.2cm]
    &=-2\pi \left[\theta(k_0)f(k_0,\mu)-\theta(-k_0) \left(1-\bar f(-k_0,-\mu)\right)\right]P_L\slashed{k}P_R \delta(k^2-2\tilde{m}^2)\,,\label{eq:S<-full-approx}
\end{align}
which has the same form as the free propagator, except for the modified dispersion relation within the $\delta$ function. In deriving \eqref{eq:S<-full-approx}, we have spelled out the Dirac $\delta$-function as
\begin{align}
\delta[(\slashed{k}-\text{Re}\Sigma_{R})^2]&=\delta\left[(1+\text{Re}a)^2k^2+2(1+\text{Re}a)\text{Re}b \,k\cdot u+\text{Re}b^2\right]
\\[0.2cm]
&\approx \delta(k^2-2\tilde{m}^2)\,,
\end{align}
where we expressed the modified dispersion with the approximate asymptotic pole equation $k^2-2\tilde{m}^2$ with $\sqrt{2}\tilde{m}$  the asymptotic thermal mass of thermal leptons. This asymptotic thermal mass holds to a very good approximation at $k_0, |{\bf k}|\gtrsim T$~\cite{Weldon:1982bn,Kiessig:2010pr,Drewes:2013iaa,Li:2023ewv}. 

In the above derivations, we have implicitly assumed that the lepton thermal mass matrix $\tilde{m}$ is diagonal. This is based on the consideration that the contributions to $\tilde{m}$ from new physics are smaller than from the gauge and charged-lepton Yukawa interactions. In general, the self-energy amplitude of leptons can be formally written as~\cite{Weldon:1982bn}
\begin{align}
    \Sigma_{\alpha\beta}\propto  \mathcal{O}(g^2)\delta_{\alpha\beta}+ \mathcal{O}([y y^\dagger]_{\alpha\beta})+ \mathcal{O}([y' y^{\prime\dagger}]_{\alpha\beta})\,.
\end{align}
The first term denotes the flavour-diagonal gauge interactions,  the second term comes from charged-lepton Yukawa interactions, and the last term is from neutrino Yukawa couplings.  In the basis where $y_{}$ is diagonal, the second term would also be diagonal but the third term is not. However, for low-scale leptogenesis, the neutrino Yukawa couplings are dictated to be smaller than $\mathcal{O}(10^{-6})$ by the out-of-equilibrium condition, which is indeed smaller than the charged-lepton Yukawa couplings. More precisely, the maximal resonant enhancement from thermal lepton-flavour correlations occurs in the difference $y_\mu^2-y_e^2\approx y_\mu^2\sim 10^{-7}$ (see e.g., Section~\ref{sec:num}), and  hence the third term can be safely neglected. However, this will not be the case if new-physics interactions are comparable with the charged-lepton Yukawa interactions, which generally renders $\Sigma_{\alpha\beta}$ non-diagonal in lepton flavour space. The impact of non-diagonal $\Sigma_{\alpha\beta}$ on the resonant enhancement was discussed in \cite{Kanemura:2024dqv,Kanemura:2024fbw}, and considered in \cite{Hamada:2016oft} for high-scale leptogenesis, where the third term may  induce thermal lepton-flavour oscillations but without significant resonant enhancement. 

In  previous studies, one practically assumes that the resummed Wightman propagators also take the same form as the free propagators by using the KB ansatz and the on-shell approximation. From the above derivations, we see that the resummed Wightman propagator of quasi-thermal leptons can indeed exhibit the same form as the free case if one neglects the thermal corrections to the spin structure, which should be interpreted as an approximation given that $\text{Re}a,\text{Re}b=\mathcal{O}(g^2),\text{Im}a,\text{Im}b=\mathcal{O}(g^4)$ near the pole resonance. Such an approximation, i.e., dropping the $\slashed{u}$ term in $\text{Im}\Sigma_R$, may neglect the potential spin-correlation contribution caused by the plasma-velocity dependent contributions. In fact, when deriving the free Wightman propagator, we  applied the correspondence given in \eqref{eq:corres}, where we have implicitly assumed a helicity-independent $f_{\alpha\beta}({\bf k},u_0)$. Following the derivation presented in Section~\ref{sec:lepton-asymmetry}, the spin-correlation contribution may also give rise to resonant enhancement when commuting with the retarded self-energy amplitude. Including these terms goes beyond the scope of the current work, and may be taken into account via chiral kinetic theory~\cite{Stephanov:2012ki,Huang:2018wdl},  if one aims at more precise calculation of the BAU. 

\section{Constraint Kadanoff-Baym equation and on-shell approximation} \label{app:constraint-KB}

The kinetic KB equation used in Section~\ref{sec:FC-leptons} and the  calculation of the CP-violating source from the two-loop diagrammatic method use the resummed lepton Wightman propagators from~\eqref{eq:S<-approx}-\eqref{eq:S>-approx}, where we have not discussed the consistency with the constraint equation given in \eqref{eq:S<-sim}. In this appendix, we check if the resummed lepton Wightman propagators  are consistent with the following leading-order constraint KB equation:
\begin{align}\label{eq:constraint}
  2k_0[\gamma^0iS_{<(>)}]_{\alpha\beta}&-\left\{{\bf k}\cdot{ \boldsymbol{ \gamma}}\gamma^0+\text{Re}\Sigma_{ R}\gamma^0,\gamma^0iS_{<(>)}\right\}_{\alpha\beta}-\{i\Sigma_{<(>)}\gamma^0,\gamma^0 \text{Re}S_R\}_{\alpha\beta}
    \nonumber \\[0.2cm]
   & =\frac{i}{2}\left(\left[\gamma^0 iS_<,i\Sigma_>\gamma^0\right]_{\alpha\beta}-\left[\gamma^0iS_>,i\Sigma_<\gamma^0\right]_{\alpha\beta}\right).
\end{align}

For nonthermal neutrino Yukawa interactions, we can neglect the $y'$ contribution to $\text{Re}\Sigma_R$, so that  $\text{Re}\Sigma_R$ becomes diagonal in lepton flavour space, as we did for the kinetic equation.  Furthermore, the real part of retarded propagator $\text{Re}S_R$ is an off-shell propagation mode (see also \eqref{eq:SNR}), which is independent of the occupation-number matrix and hence is time independent. Recall that in Appendix~\ref{app:lepton-Wightman} the on-shell approximation of resummed $i S_{<,>}$ is obtained by using the hierarchy $\text{Re}\Sigma_{R}\gg \text{Im}\Sigma_R$ near the pole. With this on-shell approximation, the anti-commutator $\{i\Sigma_{<,>}\gamma^0,\gamma^0 \text{Re}S_R\}_{\alpha\beta}$ would contribute to  off-shell effects of the resummed Wightman propagators. Nevertheless, the off-shell effects should be regarded as additional sub-dominant contributions to the CP-violating source, which can be specified by rewriting \eqref{eq:S<-inm} as
\begin{align}\label{eq:S<-decom}
    S_<(k)=S_<(k)\Big|_{\rm onshell}+S_<(k)\Big|_{\rm offshell}\,,
\end{align}
with the on-shell mode arising from $\text{Re}\Sigma_{R}\gg \text{Im}\Sigma_R$ near the pole $\slashed{k}\approx \text{Re}\Sigma_R$ and the off-shell one away from the pole. The inclusion of off-shell effects goes beyond the scope of the current work, which requires a more sophisticated numerical analysis as analytic approximations are generally not available therein.
 
With the aforementioned approximations,  \eqref{eq:constraint} will reduce to 
\begin{align}
     \left(\slashed{k}-[\text{Re}\Sigma_R]_\alpha-\frac{[\text{Re}\Sigma_R]_\beta-[\text{Re}\Sigma_R]_\alpha}{2}\right)& [iS_{<(>)}]_{\alpha\beta}
    \nonumber \\[0.2cm]
    &=-\frac{i}{4}\left(\mathcal{C}_{\alpha\beta}(y^{\prime 2},\delta f_N)-\mathcal{C}^\dagger_{\alpha\beta}(y^{\prime 2},\delta f_N)\right).
\end{align}
The leading collision rate is determined by nonthermal neutrino Yukawa interactions that depend on quadratic $y'$ and the nonthermal neutrino distribution function $\delta f_N$. Using \eqref{eq:S<-full-approx-0}, we see that $(\slashed{k}-[\text{Re}\Sigma_R]_\alpha)[iS_{<(>)}]_{\alpha\beta}=0$. The difference $[\text{Re}\Sigma_R]_\beta-[\text{Re}\Sigma_R]_\alpha$ cancels the flavour-universal gauge contributions, and leaves the small lepton-flavour dependent couplings. After summing over $\alpha,\beta$ and taking the Dirac trace on both sides, we can formally extract the following scaling:
\begin{align}
\sum_{\alpha,\beta}   (y_\beta^2-y_\alpha^2) f_{\alpha\beta}\sim \sum_{\alpha,\beta}\text{Tr}\left(\mathcal{\tilde {C}}_{\alpha\beta}(y^{\prime 2},\delta f_N)\right)-\text{Tr}\left(\mathcal{\tilde {C}}^\dagger_{\alpha\beta}(y^{\prime 2},\delta f_N)\right)\,,
\end{align}
where $\mathcal{\tilde {C}}$ denotes the corresponding dimensionless collision rate. The above result implies that the off-diagonal correlation of the lepton occupation-number matrix $f_{\alpha\beta}$ arises from the non-equilibrium neutrino Yukawa interactions, and is significantly enhanced by the small charged-lepton Yukawa coupling $y_\beta^2-y_\alpha^2$. This picture is consistent with what we observed in \eqref{eq:Dn_ab}.  For the diagonal channel $\alpha=\beta$, we have $\text{Tr}(\mathcal{\tilde C}_{\alpha\alpha})=\text{Tr}(\mathcal{\tilde C}^\dagger_{\alpha\alpha})$ as can be inferred from  \eqref{eq:TrC} and \eqref{eq:TrCdag}, and hence the above result vanishes  on both sides. Therefore, the resummed Wightman propagators in the on-shell approximation given in \eqref{eq:S<-approx}-\eqref{eq:S>-approx} yield consistent kinetic and constraint KB equations at leading order.

\section{Collision rates in lepton kinetic equations}\label{app:collision-rates}

In this appendix, we provide further technical details about the evolution of the off-diagonal correlations $\Delta n_{\alpha\beta}$ and the diagonal entries $\Delta n_{\alpha\alpha}$, as used in Section~\ref{sec:lepton-asymmetry}. We also present the general structures of  the collision rates for $\Delta n_{\alpha\beta}$  and $\Sigma n_{\alpha\beta}$ given in \eqref{eq:dDn/dt} and \eqref{eq:dSn/dt}, highlighting the importance of going beyond the one-loop order.  A more precise calculation of the full collision rates may be given elsewhere. Nonetheless, their basic structure suffices to help us qualitatively understand the general features and, in particular, identify the quasi-steady state of~$\Delta n_{\alpha\beta}$. We~begin in Appendix~\ref{app:collision-rate-1} with the identification of damping and source terms  for $\Delta n_{\alpha\beta}$  and~$\Sigma n_{\alpha\beta}$. Then, in  Appendix~\ref{app:collision-rate-2}, we present the calculation of source and washout rates for $\Delta n_{\alpha\alpha}$ using the results obtained in  Appendix~\ref{app:collision-rate-1} .

\begin{figure}[t]
	\centering
\includegraphics[scale=0.7]{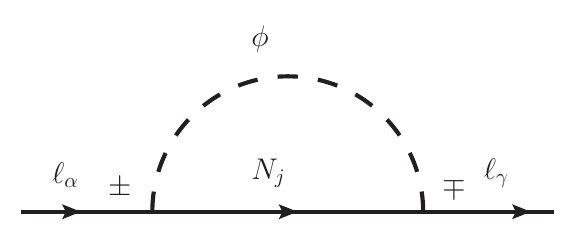} 
	\caption{\label{fig:ellag-self} The flavour-changing but $L_{\rm tot}$-conserving one-loop self-energy diagrams of leptons contributing to the collision rate $\mathcal{C}_{\alpha\beta}$ defined in \eqref{eq:collision-rate}, with $a,b=\pm$ being the CTP indices and $\Sigma_<\equiv \Sigma_{+-}, \Sigma_>\equiv \Sigma_{-+}$. }
\end{figure}

\subsection{Source and damping effects in off-diagonal correlations}\label{app:collision-rate-1}
From the collision rate $\mathcal{C}_{\alpha\beta}$ defined in \eqref{eq:collision-rate}, we first calculate the one-loop self-energy amplitudes $[i\Sigma_{<,>}]_{\alpha\gamma}$ from neutrino Yukawa interactions, as shown in Fig.~\ref{fig:ellag-self}. We then discuss the one-loop collision rates from charged-lepton Yukawa and gauge interactions that determine the evolution of $\Delta n_{\alpha\beta}$ and $\Sigma n_{\alpha\beta}$. After that, we show that going beyond the one-loop level can induce a significant source term for $\Delta n_{\alpha\beta}$.  For simplicity, we take the approximation $\langle \Delta n_{\alpha\beta}\rangle\approx 
\Delta n_{\alpha\beta}, \langle \Sigma n_{\alpha\beta}\rangle\approx 
\Sigma n_{\alpha\beta}$, but the conclusions drawn in this appendix, as well as the results presented in Section~\ref{sec:lepton-asymmetry}, do not depend on this approximation.

Given our convention for the self-energy amplitude $-i\Sigma$, we can write down the amplitudes of Fig.~\ref{fig:ellag-self} as
\begin{align}
    [i\Sigma_>(k_\ell)]_{\alpha\gamma}&=y^{\prime *}_{\alpha j}y'_{\gamma j}\int \frac{\text{d}^4 k_\phi}{(2\pi)^4}\frac{\text{d}^4 k_N}{(2\pi)^4}(2\pi)^4 \delta^{(4)}(k_\ell+k_\phi-k_N)P_R[iS_>(k_N)]_j P_L iG_<(k_\phi)\,,
    \\[0.2cm] 
    [i\Sigma_<(k_\ell)]_{\alpha\gamma}&=y^{\prime *}_{\alpha j}y'_{\gamma j}\int \frac{\text{d}^4 k_\phi}{(2\pi)^4}\frac{\text{d}^4 k_N}{(2\pi)^4}(2\pi)^4 \delta^{(4)}(k_\ell+k_\phi-k_N)P_R[iS_<(k_N)]_j P_L iG_>(k_\phi)\,,
\end{align}
where we neglected flavour correlations for the nonthermal and non-degenerate singlet neutrinos, and simply used the free Wightman propagators $iS_{<,>}$ for neutrinos. However, we included the thermal mass corrections to the Higgs propagators $iG_{<,>}$. Using the  neutrino and Higgs propagators from~\eqref{eq:SN<}-\eqref{eq:SNTbar} and \eqref{eq:G<}-\eqref{eq:GTbar}, we can simplify the amplitudes as 
\begin{align}
    [i\Sigma_>(k_\ell)]_{\alpha\gamma} &=-y^{\prime *}_{\alpha j}y'_{\gamma j}\int \frac{\text{d}^3 {\bf k}_\phi}{(2\pi)^3} \frac{\text{d}^3 {\bf k}_N}{(2\pi)^3}(2\pi)^4 \delta^{(4)}(k_\ell+k_\phi-k_N)
 \\[0.2cm]
    &\phantom{-y^{\prime *}_{\alpha j}y'_{\gamma j}\int \frac{\text{d}^3 {\bf k}_\phi}{(2\pi)^3} \frac{\text{d}^3 {\bf k}_N}{(2\pi)^3}}\times \int \text{d}k_{\phi0}\text{d}k_{N0}\delta(k_N^2-M_j^2)\delta(k_\phi^2-m_\phi^2)P_R\slashed{k}_N P_L\mathcal{F}_j^>\,,    \nonumber
    \\[0.2cm] 
    [i\Sigma_<(k_\ell)]_{\alpha\gamma} &=-y^{\prime *}_{\alpha j}y'_{\gamma j}\int \frac{\text{d}^3 {\bf k}_\phi}{(2\pi)^3} \frac{\text{d}^3 {\bf k}_N}{(2\pi)^3}(2\pi)^4 \delta^{(4)}(k_\ell+k_\phi-k_N)
\\[0.2cm]
    &\phantom{-y^{\prime *}_{\alpha j}y'_{\gamma j}\int \frac{\text{d}^3 {\bf k}_\phi}{(2\pi)^3} \frac{\text{d}^3 {\bf k}_N}{(2\pi)^3}}\times \int \text{d}k_{\phi0}\text{d}k_{N0}\delta(k_N^2-M_j^2)\delta(k_\phi^2-m_\phi^2)P_R\slashed{k}_N P_L\mathcal{F}_j^<\,,    \nonumber 
\end{align}
where $\mathcal{F}^{>,<}_j$ are defined as
\begin{align}
    \mathcal{F}_j^>\equiv  &-\theta(k_{N0})\theta(k_{\phi 0})\left(1-f_{N_j}(k_{N0})\right)f_\phi(k_{\phi0})
 \nonumber  \\[0.2cm]
 &+\theta(-k_{N0})\theta(-k_{\phi 0})\bar f_{N_j}(-k_{N0})\left(1+\bar f_\phi(-k_{\phi0})\right)\,,
    \\[0.2cm]
        \mathcal{F}_j^<\equiv  & \,~\theta(k_{N0})\theta(k_{\phi 0})f_{N_j}(k_{N0})\left(1+f_\phi(k_{\phi0})\right)
         \nonumber  \\[0.2cm]
 &-\theta(-k_{N0})\theta(-k_{\phi 0})\left(1-\bar f_{N_j}(-k_{N0})\right)\bar f_\phi(-k_{\phi0})\,.
\end{align}
 The presence of the two Dirac $\delta$-functions, $\delta(k_\phi^2-m_\phi^2)$, $\delta(k_N^2-M_j^2)$, and the on-shell condition $\delta(k_\ell^2-2\tilde{m}^2)$ from the resummed lepton propagators $[iS_{<,>}]_{\alpha\beta}$  in $\mathcal{C}_{\alpha\beta}$ dictate that the frequencies should satisfy 
\begin{align}\label{eq:kinetics}
    k_{N0}<0\,, k_{\phi0}<0\,,k_{\ell0}>0\,, \quad \text{or}\quad k_{N0}>0\,, k_{\phi0}>0\,,k_{\ell0}<0\,,
\end{align}
which produces $\mathcal{F}_j^{<,>}$. We can then evaluate the Dirac trace 
\begin{align}\label{eq:Tr1}
    \text{Tr}\left([i\Sigma_>(k_\ell)]_{\alpha\gamma}[iS_<(k_\ell)]_{\gamma\beta}\right)&=-2\pi y^{\prime *}_{\alpha j}y'_{\gamma j}\Delta m_j^2\delta(k_\ell^2-2\tilde{m}_\gamma^2)
      \\[0.2cm]
    &\times \int \text{d}\Pi_{\phi} \text{d}\Pi_{ N}(2\pi)^4\tilde\delta^{(4)}(k_\ell+k_\phi-k_N)\left([\Theta_{-k_\ell}]_{\gamma\beta}+[\Theta_{k_\ell}]_{\gamma\beta}\right)\,,  \nonumber
    \\[0.2cm]
        \text{Tr}\left([i\Sigma_<(k_\ell)]_{\alpha\gamma}[iS_>(k_\ell)]_{\gamma\beta}\right)&=-2\pi y^{\prime *}_{\alpha j}y'_{\gamma j}\Delta m_j^2\delta(k_\ell^2-2\tilde{m}_\gamma^2)\label{eq:Tr2} 
    \\[0.2cm]
    &\times \int \text{d}\Pi_{\phi} \text{d}\Pi_{ N}(2\pi)^4\tilde \delta^{(4)}(k_\ell+k_\phi-k_N)\left([\tilde\Theta_{-k_\ell}]_{\gamma\beta}+[\tilde\Theta_{k_\ell}]_{\gamma\beta}\right)\,,   \nonumber
\end{align}
where  $\Delta m_j^2$ is from the Dirac trace $\text{Tr}(P_R\slashed{k}_N P_L \slashed{k}_\ell)=2k_\ell \cdot k_N$, and we have defined the following quantities,
\begin{align}
\text{d}\Pi_i&\equiv \frac{\text{d}^3 {\bf k}_i}{(2\pi)^3 2E_i}\,,i=\phi, N\,,
\\[0.2cm]
\Delta m_j^2&\equiv m_\phi^2-M_j^2-2\tilde{m}^2\,,
\\[0.2cm]
\tilde\delta^{(4)}(k_\ell+k_\phi-k_N)&\equiv\delta^{(3)}({\bf k}_\ell+{\bf k}_\phi-{\bf k}_N)\delta(E_\ell+E_N-E_\phi)\,.
\end{align}
The $\Theta$ functions are given by
\begin{align}
[\Theta_{-k_\ell}]_{\gamma\beta}&\equiv \theta(-k_{\ell0})f_\phi(E_\phi)\left(1-f_{N_j}(E_N)\right)\left(\delta_{\gamma\beta}-\bar f_{\gamma\beta}(E_\ell)\right)\,,\label{eq:theta-k}
\\[0.2cm]
[\Theta_{k_\ell}]_{\gamma\beta}&\equiv \theta(k_{\ell0})\bar f_{N_j}(E_N)f_{\gamma\beta}(E_\ell)\left(1+\bar f_{\phi}(E_\phi)\right)\,,\label{eq:thetak}
\\[0.2cm]
[\tilde\Theta_{-k_\ell}]_{\gamma\beta}&\equiv \theta(-k_{\ell0})f_{N_j}(E_N)\bar f_{\gamma\beta}(E_\ell)\left(1+f_\phi(E_\phi)\right)\,,\label{eq:ttheta-k}
\\[0.2cm]
[\tilde\Theta_{k_\ell}]_{\gamma\beta}&\equiv \theta(k_{\ell0})\bar f_{\phi}(E_\phi)\left(1-\bar f_{N_j}(E_N)\right)\left(\delta_{\gamma\beta}-f_{\gamma\beta}(E_\ell)\right)\,.\label{eq:tthetak}
\end{align}

Taking the difference of  \eqref{eq:Tr1} and \eqref{eq:Tr2}, we arrive at  $\text{Tr}(\mathcal{C}_{\alpha\beta})$.  For the Hermitian conjugate of  $\mathcal{C}_{\alpha\beta}$, since
\begin{align}
    \text{Tr}\left(\mathcal{C}^\dagger_{\alpha\beta}\right)&=\text{Tr}\left([iS_<]^\dagger[i\Sigma_>]^\dagger-[iS_>]^\dagger[i\Sigma_<]^\dagger\right)
  \nonumber  \\[0.2cm]
    &=\text{Tr}\left([iS_<]_{\alpha\gamma}[i\Sigma_>]_{\gamma\beta}-[iS_>]_{\alpha\gamma}[i\Sigma_<]_{\gamma\beta}\right),
\end{align}
we can readily obtain $\text{Tr}(\mathcal{C}^\dagger_{\alpha\beta})$ from $\text{Tr}(\mathcal{C}_{\alpha\beta})$ by changing the indices $\alpha\gamma\to \gamma\beta$ in $i\Sigma_{<,>}$ and $\gamma\beta\to \alpha \gamma$ in $iS_{<,>}$.  

To get the final collision rates in \eqref{eq:dDn/dt} and \eqref{eq:dSn/dt}, we can flip all the momenta $k\to -k$ such that $\theta(-k_{\ell0})$ in $\Theta_{-k_\ell}$ and $\tilde\Theta_{-k_\ell}$ is converted to $\theta(k_{\ell 0})$. In doing this, we notice that the sign function $\text{sign}(k_{\ell 0})$ in \eqref{eq:dSn/dt} leads to different damping sources for the off-diagonal correlations  $\Delta n_{\alpha\beta}$ and $\Sigma n_{\alpha\beta}$. In particular, the dominant damping source for $\Delta n_{\alpha \beta}$ comes from charged-lepton Yukawa couplings while that for $\Sigma_{\alpha\beta}$ results from gauge interactions. Consequently, $\Sigma n_{\alpha\beta}$ will damp with a much faster rate and hence becomes smaller than $\Delta n_{\alpha\beta}$ after some short period of time. 

At one-loop order, the evolution of $\Delta n_{\alpha\beta}$ and $\Sigma n_{\alpha\beta}$ can be traced qualitatively via the dependence on the distribution functions (matrices). To this end, we define
\begin{align}
    f_{\gamma\alpha}&=f_\ell^{\rm eq}\delta_{\gamma\alpha}+\delta f_{\gamma\alpha}\,,\quad 
   \bar f_{\gamma\alpha}=f_\ell^{\rm eq}\delta_{\gamma\alpha}+\delta \bar f_{\gamma\alpha}\,, \label{eq:fga}
    \\[0.1cm]
 f_{e}&=f_e^{\rm eq} +\delta  f_{e}\,, \quad \qquad \bar f_{e}=f_e^{\rm eq} +\delta \bar f_{e}\,,\label{eq:fe}
        \\[0.1cm]
        f_{\phi}&=f_\phi^{\rm eq}+\delta f_{\phi}\,,\quad\qquad
        \bar f_{\phi}=f_\phi^{\rm eq}+\delta \bar f_{\phi}\,,\label{eq:fphi}
    \\[0.1cm]
f_{N_j}&=f_N^{\rm eq}+\delta f_{N_j}\,,\quad\quad~
\bar f_{N_j}=f_N^{\rm eq}+\delta  \bar f_{N_j}\,.\label{eq:fN}
\end{align}
In general, $\delta f_\phi, \delta \bar f_\phi$ arise from the nonthermal neutrino Yukawa interactions. However, due to fast gauge-scalar interactions,  the relation $\delta f_\phi+\delta \bar f_\phi\approx 0$ will quickly establish. Analogously,  $\delta f_e+\delta \bar f_e\approx 0$ from  right-handed charged-lepton singlets will quickly establish via charged-lepton Yukawa interactions, though in a speed slower than the gauge-scalar interaction rate. These interaction timescales can be seen as instantaneous events when compared with  the nonthermal neutrino Yukawa interactions, where $\delta f_N+\delta \bar f_N\approx 0$ does not hold. Instead, we should expect $n_N\approx \bar n_{N}$ and hence $\delta f_N\approx \delta \bar f_N$ from $\mathcal{O}(y^{\prime 2})$ interaction rates. In addition,  $\delta f_N$ and $\delta \bar f_N$ are in general larger than $\delta f_{\gamma\beta}, \delta f_\phi, \delta f_e$ since right-handed  neutrinos are not present initially at which $\delta f_N\simeq -f_N^{\rm eq}$.  Finally, the relation between the off-diagonal correlation $\delta f_{\gamma\alpha}$ and $\delta \bar f_{\gamma\alpha}$ is determined by  \eqref{eq:dDn/dt} and \eqref{eq:dSn/dt}. 
  
For gauge  interactions, the dependence of $\mathcal{C}_{\alpha\beta}, \mathcal{C}^\dagger_{\alpha\beta}$ on the distribution functions (matrices) can be simply obtained by replacing $f_N$ in \eqref{eq:theta-k}-\eqref{eq:tthetak} with $f_{\alpha\gamma}$ and replacing the scalar distribution function with gauge-boson ones $f_V$ ($V=W_\mu, B_\mu$). With \eqref{eq:fga}-\eqref{eq:fN}, we can infer from the structures shown in \eqref{eq:theta-k}-\eqref{eq:tthetak} that the collision rates from gauge  interactions for $\Delta n_{\alpha\beta}$ and $\Sigma n_{\alpha\beta}$, up to linear order of small chemical potentials, scale as
\begin{align}
    \text{gauge~rate~for}~\Delta n_{\alpha\beta} &:~\delta_{\alpha\gamma}(f_\ell^{\rm eq}+f_V^{\rm eq})(\delta f_{\gamma\beta}-\delta \bar f_{\gamma\beta})+\delta_{\gamma\beta}(f_\ell^{\rm eq}+f_V^{\rm eq})(\delta \bar f_{\alpha\gamma}-\delta f_{\alpha\gamma })\,,\label{eq:gauge-Delta_nab}
    \\[0.2cm]
        \text{gauge~rate~for}~\Sigma n_{\alpha\beta} &:~\delta_{\alpha\gamma}(f_\ell^{\rm eq}+f_V^{\rm eq})(\delta f_{\gamma\beta}+\delta \bar f_{\gamma\beta})+\delta_{\gamma\beta}(f_\ell^{\rm eq}+f_V^{\rm eq})(\delta \bar f_{\alpha\gamma}+\delta f_{\alpha\gamma })\,,\label{eq:gauge-Sigma_nab}
\end{align}
where summation over $\gamma=1,2,3$ is taken. The above sign difference between $\delta f_{}$ and $\delta \bar f_{}$ for $\Delta n_{\alpha\beta}$ and $\Sigma n_{\alpha\beta}$ arises from the momentum flip in $\Theta_{-k_\ell}$ and $\tilde\Theta_{-k_\ell}$ that  would yield a minus sign due to $\text{sign}(k_{\ell 0})$ in \eqref{eq:dSn/dt}. We see that the gauge collision rate for $\Delta n_{\alpha\beta}$  would vanish up to the linear order of small chemical potentials, while that for  $\Sigma n_{\alpha\beta}$ becomes  the main damping source in suppressing the off-diagonal correlation\footnote{This can also be checked more explicitly by spelling out the collision rate; see also Sections~3.3 and 3.4 in \cite{Beneke:2010dz}, after imposing the condition of local thermal equilibrium for charged-lepton Yukawa interactions.}. 

For diagonal charged-lepton Yukawa interactions, we replace $f_N$ in \eqref{eq:theta-k}-\eqref{eq:tthetak} with the right-handed charged-lepton distribution function $ \delta_{\alpha\gamma}f_e$.  Then we obtain the $\delta f_{\alpha\beta}$ dependence as
\begin{align}\label{eq:y-rate_Deltan}
   \text{Charged-lepton Yukawa~rate~for}~\Delta n_{\alpha\beta} &:~ (f_e^{\rm eq}+f_\phi^{\rm eq}) (\delta f_{\alpha\beta}-\delta \bar f_{\alpha\beta})\,,
    \\[0.2cm]
    \text{Charged-lepton Yukawa~rate~for}~\Sigma n_{\alpha\beta} &:~(f_e^{\rm eq}+f_\phi^{\rm eq}) (\delta f_{\alpha\beta}+\delta \bar f_{\alpha\beta})\,.
\end{align}
Again, the sign difference between $\delta f_{}$ and $\delta \bar f$ in $\Delta n_{\alpha\beta}$ and $\Sigma n_{\alpha\beta}$ arises from the momentum flip in $\Theta_{-k_\ell}$ and $\tilde\Theta_{-k_\ell}$. The above $\delta f$ dependence shows that the charged-lepton Yukawa interactions will work as damping sources for both $\Delta n_{\alpha\beta}$ and $\Sigma n_{\alpha\beta}$. However, the main damping effect in the evolution of $\Sigma n_{\alpha\beta}$ comes from the stronger gauge interactions, leading to a fact that $\Sigma n_{\alpha\beta}$ will damp away in a speed much faster than $\Delta n_{\alpha\beta}$ and hence $\delta f_{\alpha\beta}+\delta \bar f_{\alpha\beta}\approx 0$ will be developed in a timescale determined by gauge interactions. 

Following the above discussions, we can  analyse the evolution of the off-diagonal correlations in \eqref{eq:dDn/dt} and \eqref{eq:dSn/dt} at one-loop order. Initially, we have $\Delta n_{\alpha\beta}=0, \Sigma_{\alpha\beta}=0$. Then, we should identify the source terms that can create non-zero $\Delta n_{\alpha\beta}, \Sigma_{\alpha\beta}$.   The one-loop source term for $\Sigma n_{\alpha\beta}$ in  \eqref{eq:dSn/dt} comes from nonthermal neutrino Yukawa interactions featuring $f_N\approx \bar f_N$ since neither kinetic equilibrium nor chemical equilibrium is established for right-handed neutrinos. In this regime, we can infer from \eqref{eq:dSn/dt} that 
\begin{align}\label{eq:dSigma/dt}
      \frac{{\rm d}\Sigma n_{\alpha \beta}}{{\rm d}t}-i (\tilde b_\alpha-\tilde b_\beta)\Delta n_{\alpha\beta}\sim C_{N_j} y'_{\beta i}y^{\prime *}_{\alpha i}-C_V (3g_2^2+g_1^2)\Sigma n_{\alpha\beta}\,,
\end{align}
where  $C_{N_j}$ and $C_V$\footnote{ Note that the gauge collision rate depends on the quartic gauge coupling, as also mentioned in Section~\ref{sec:mix-osc-cor}. Here,  we have absorbed the other $g^2$ dependence into $C_V$.} denote the one-loop temperature-dependent coefficients of the neutrino Yukawa and gauge interactions, respectively, and  $C_{N_j}$ vanishes when   $f_N=f_N^{\rm eq}$. The source term from nonthermal neutrino Yukawa interactions first create the off-diagonal correlations in $\Sigma n_{\alpha\beta}$ when the source term keeps larger than the gauge damping effect.  Note that  $\Sigma n_{\alpha\beta}$ can develop both real and imaginary components owing to the complex neutrino Yukawa couplings. Once a non-zero $\Sigma n_{\alpha\beta}$ appears, it can contribute as a source to creating $\Delta n_{\alpha\beta}$ via the coherent term in \eqref{eq:dDn/dt}, yielding 
\begin{align}\label{eq:Dn_ab-scale}
      \frac{{\rm d}\Delta n_{\alpha \beta}}{{\rm d}t}-i (\tilde b_\alpha-\tilde b_\beta)\text{Re}\Sigma n_{\alpha\beta}\sim -C_\ell (y_\alpha^2+y_\beta^2) \Delta n_{\alpha\beta}-(\tilde b_\alpha-\tilde b_\beta)\text{Im}\Sigma n_{\alpha\beta}\,,
\end{align}
with $C_\ell$  the one-loop temperature-dependent coefficient from charged-lepton Yukawa collision rate. 
We see that a source term can arise in the evolution of  $\Delta n_{\alpha\beta}$ if $(\tilde b_\alpha-\tilde b_\beta)\text{Im}\Sigma n_{\alpha\beta}$ does not vanish, which indicates that
the off-diagonal correlation in $\Delta n_{\alpha\beta}$ can   be created only if there is CP violation from neutrino Yukawa interactions. This is consistent with our expectation that the final lepton asymmetry would be proportional to the imaginary part of  neutrino Yukawa couplings (see e.g., \eqref{eq:CPV-l}). At this initial stage, the source term is larger than the charged-lepton Yukawa damping term, where $\Delta n_{\alpha\beta}$
is  increasing.

After some short time, the gauge damping effect becomes important to suppress $\Sigma n_{\alpha\beta}$ such that the off-diagonal correlation $\Sigma n_{\alpha\beta}$  damps away via the strong gauge interactions.  
We may then infer from \eqref{eq:dSigma/dt} a quasi-steady constraint equation,
\begin{align}\label{eq:Delta-asym}
-i (\tilde b_\alpha-\tilde b_\beta)\Delta n_{\alpha\beta}= C_{N_j} y'_{\beta i}y^{\prime *}_{\alpha i}\,,
\end{align}
which will produce \eqref{eq:Dn_ab} as the asymptotic solution for $\Delta n_{\alpha\beta}$. Nevertheless, during the decrease of $\Sigma n_{\alpha\beta}$, it will also suppress the source term for $\Delta n_{\alpha\beta}$ in \eqref{eq:Dn_ab-scale}, leading to the decreasing stage of $\Delta n_{\alpha \beta}$. The above quasi-steady state would predict $\Delta n_{\alpha\beta}\sim \text{Im}\Sigma n_{\alpha\beta}$ from \eqref{eq:Dn_ab-scale} when ${\rm d}\Delta n_{\alpha \beta}/{\rm d}t\approx 0$. However, this is not valid as $\Sigma n_{\alpha\beta}$ has essentially damped away. 
Therefore, the constraint equation necessitates some source term for $\Delta n_{\alpha\beta}$ beyond the one-loop order. 

\begin{figure}[t]
	\centering
\includegraphics[scale=0.8]{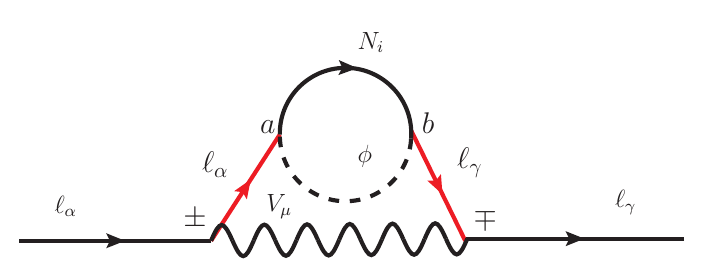} 
	\caption{\label{fig:lab-2loop} A two-loop diagram that contributes as a source to creating the off-diagonal correlation $\Delta n_{\alpha\beta}$, where $V_\mu=W_\mu, B_\mu$ denotes the gauge bosons,   $\pm$ and $a,b=\pm$ are the CTP indices, and the red lines denote thermal resummation of soft  lepton propagators.  }
\end{figure}

The discussions of the evolution for the off-diagonal correlations were also made in \cite{Beneke:2010dz,Garbrecht:2012pq}. However, the analysis was confined at  one-loop order. As such, similarly to what we observed above, it was also found in \cite{Beneke:2010dz,Garbrecht:2012pq} that the off-diagonal correlations will damp away. 
When going to  two-loop order,  we found that there is a source term for $\Delta n_{\alpha\beta}$ arising from Fig.~\ref{fig:lab-2loop}. Recall that the corresponding one-loop gauge rate, i.e., removing the inner loop from Fig.~\ref{fig:lab-2loop}, vanishes, as has been identified in \eqref{eq:gauge-Delta_nab}. At two-loop order, however, the nonthermal neutrino Yukawa interactions will induce flavour-changing corrections for $\alpha\neq \gamma$. It was found in \cite{Kanemura:2024dqv} that by taking thermal corrections to the soft lepton propagators (the red lines shown in Fig.~\ref{fig:lab-2loop}), there will be a significant enhancement factor $(y_\alpha^2-y_\gamma^2)^{-1}$ when one of the resummed lepton propagators goes on-shell. This can also be learned from RL via quasi-degenerate sterile neutrinos,  as presented in Appendix~\ref{app:RFL-N} under the CTP formalism. In particular, a resonant enhancement factor $(M_i^2-M_j^2)^{-1}$ will arise from Fig.~\ref{fig:N-self} when one of the sterile neutrinos goes on-shell, which contributes as a CP-violating source to the generation of lepton asymmetries.  Alternatively, one can infer from Fig.~\ref{fig:2+1loop-Nself} and the corresponding \eqref{eq:CPV-l} that the source term in Fig.~\ref{fig:lab-2loop} would scale as
\begin{align}
       \text{Two-loop~source~rate~for}~\Delta n_{\alpha\beta} &:~C'_{N_j}(3g_2^2+g_1^2)\frac{\text{Im}(y'_{\gamma i}y^{\prime *}_{\alpha i})}{y_\alpha^2-y_\gamma^2}\delta_{\gamma\beta}\,,
\end{align}
where $C'_{N_j}$ denotes the  temperature-dependent coefficient of the two-loop amplitude $[i\Sigma_{<,>}]_{\alpha\gamma}$ from Fig.~\ref{fig:lab-2loop}, and $\delta_{\gamma\beta}$ arises from taking the diagonal distribution functions in $[iS_{<,>}]_{\gamma\beta}$; see \eqref{eq:S<-approx}-\eqref{eq:S>-approx}.  Including this two-loop source term, we can reformulate \eqref{eq:Dn_ab-scale} in the asymptotic regime $\Sigma n_{\alpha\beta}\to 0$  as
\begin{align}\label{eq:Dn_ab-scale-2}
      \frac{{\rm d}\Delta n_{\alpha \beta}}{{\rm d}t}\sim -C_\ell (y_\alpha^2+y_\beta^2) \Delta n_{\alpha\beta}+C'_{N_j}(3g_2^2+g_1^2)\frac{\text{Im}(y'_{\gamma i}y^{\prime *}_{\alpha i})}{y_\alpha^2-y_\gamma^2}\delta_{\gamma\beta} \,,
\end{align}
from which we see both damping and source terms for the evolution of $\Delta n_{\alpha\beta}$. While there is loop suppression  encoded in $C'_{N_j}$, the enhancement factor $(3g_2^2+g_1^2)/(y_\alpha^2-y_\beta^2)$ in the source term can increase $\Delta n_{\alpha\beta}$ when $\Sigma n_{\alpha\beta}\to 0$ until the damping term becomes comparable in magnitude to the source term, eventually leading to a quasi-steady state ${\rm d}\Delta n_{\alpha \beta}/{\rm d}t\approx 0$. Note that whether a two-loop source term appears in $\Sigma n_{\alpha\beta}$ does not significantly change the above pattern, since on the one hand the two-loop source rate should be  sub-dominant corrections to the one-loop source rate, and on the other hand, $\Sigma n_{\alpha\beta}$ undergoes much stronger damping effects from gauge interactions, which is especially the case in the muon-electron correlation.

\subsection{Source and washout effects for lepton-number asymmetry}\label{app:collision-rate-2}

The collision rate for the lepton-number asymmetry of lepton doublets can be derived by using the same expressions of $\mathcal{C}_{\alpha\beta}$ and $\mathcal{C}^\dagger_{\alpha\beta}$ given in the previous appendix, but this time we should keep the full dependence of $f_{\alpha\beta}$ both in the diagonal and off-diagonal entries, and also keep the differences of  $f_N-\bar f_N$, $f_\phi-\bar f_\phi$, and $f_{\alpha\beta}-\bar f_{\alpha\beta}$.  Expanding $\text{Tr}(\mathcal{C}_{\alpha\alpha})$ from \eqref{eq:Tr1}-\eqref{eq:Tr2}, we arrive at
\begin{align}\label{eq:TrC}
   \sum_\alpha \text{Tr}(\mathcal{C}_{\alpha\alpha})=&-4\pi    \sum_{\alpha,\gamma,j} y'_{\gamma j}y^{\prime *}_{\alpha j}\Delta m_j^2\delta(k_\ell^2-2\tilde m^2_\gamma)
\\[0.2cm]
   &\times \int \text{d}\Pi_{\phi} \text{d}\Pi_{N} (2\pi)^4 \tilde{\delta}^{(4)}(k_\ell+k_\phi-k_N)\theta(k_{\ell 0})\left[\delta_{\gamma\alpha} (I^\phi_j-I^N_j)+I^{\ell}_{\gamma\alpha j}\right],   \nonumber 
\end{align}
where 
\begin{align}
    I^\phi_j&\equiv \delta f_\phi (E_\phi) \left[1-f_\ell^{\rm eq}(E_\ell)-f_{N_j}^{\rm eq}(E_N)-\frac{1}{2}\left(\delta f_{N_j}(E_N)+\delta \bar f_{N_j}(E_N)\right)\right],
    \\[0.2cm]
   I^N_j &\equiv \frac{1}{2}\left(\delta f_{N_j}(E_N)-\delta \bar f_{N_j}(E_N)\right)\left(f_\ell^{\rm eq}(E_\ell)+f_\phi^{\rm eq}(E_\phi)\right),
   \\[0.2cm]
   I^{\ell}_{\gamma\alpha j}&\equiv \delta f_{\gamma\alpha}(E_\ell)\left[f_\phi^{\rm eq}(E_\phi)+f_{N_j}^{\rm eq}(E_N)+ \frac{1}{2}\left(\delta f_{N_j}(E_N)+\delta \bar f_{N_j}(E_N)\right)\right].
\end{align}

Similarly, we can write down the Hermitian-conjugate rate by changing the indices $\alpha\gamma\to \gamma\beta$ in $i\Sigma_{<,>}$ and $\gamma\beta\to \alpha \gamma$ in $iS_{<,>}$ from \eqref{eq:Tr1}-\eqref{eq:Tr2}, and by taking $\alpha=\beta$. We then arrive at 
\begin{align}\label{eq:TrCdag}
     \sum_\alpha  \text{Tr}(\mathcal{C}^\dagger_{\alpha\alpha})=&-2\pi\sum_{\alpha,\gamma,j} y'_{\alpha j}y^{\prime *}_{\gamma j}\Delta m_j^2\delta(k_\ell^2-2\tilde m^2_\alpha)
 \\[0.2cm]
&\times    \int \text{d}\Pi_{\phi}\Pi_{N} (2\pi)^4 \tilde{\delta}^{(4)}(k_\ell+k_\phi-k_N)
     \theta(k_{\ell 0})\left[\delta_{\alpha\gamma} (I^\phi_j-I^N_j)+I^{\ell}_{\alpha \gamma j}\right]. \nonumber
\end{align}
Relabelling  the indices $\alpha,\gamma$ on the rhs of \eqref{eq:TrCdag} by $\alpha \leftrightharpoons \gamma$, we found that $\sum_\alpha \text{Tr}(\mathcal{C}^\dagger_{\alpha\alpha})=\sum_\alpha\text{Tr}(\mathcal{C}_{\alpha\alpha})$. Substituting these terms into \eqref{eq:dDn/dt}, we can reproduce \eqref{eq:dDn/dt-2}.

\section{Vertex corrections to Thermal Resonant Leptogenesis}\label{app:vertex}

\begin{figure}[t]
	\centering
\includegraphics[scale=0.6]{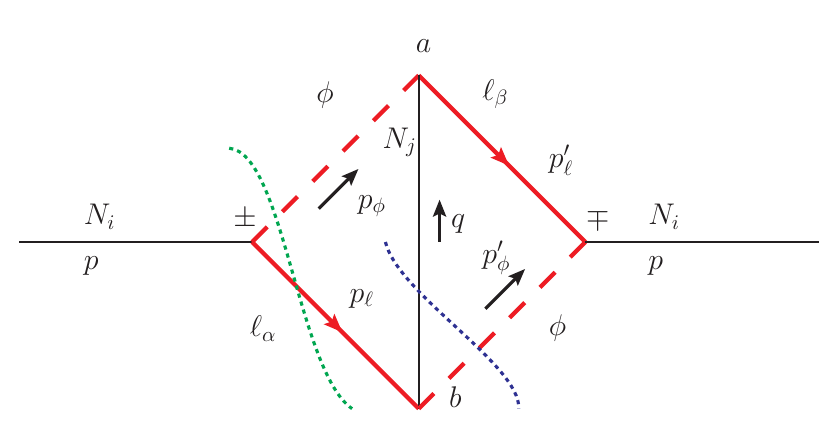} 
	\caption{\label{fig:vertex} The two-loop vertex diagrams of relativistic singlet neutrinos contributing to CP asymmetry, with $\pm$ and $a,b=\pm$ being the CTP indices. A green-dashed thermal cut decomposes the two-loop diagram to a tree-level  decay diagram and the associated one-loop diagram. The blue-dashed thermal cut generates a purely thermal-induced absorptive part in the one-loop Higgs decay. The red, thick lines for lepton and Higgs doublets reflect thermal corrections. The fermion flow  indicates the direction of momenta, while the momentum directions of scalar   and singlet neutrinos are denoted by the black arrows, with four-momenta  attached. }
\end{figure}

Let us discuss the contribution from vertex corrections. The CP-violating source from Fig.~\ref{fig:Higgsdecay} or Fig.~\ref{fig:2+1loop-Nself} can be attributed to wave-function corrections. Like the traditional leptogenesis scenarios, we may expect a CP-violating source from vertex corrections, as shown in Fig.~\ref{fig:vertex}  in terms of the two-loop neutrino self-energy diagrams. Nevertheless, the resulting CP-violating source does not exhibit a thermal resonant enhancement, and the appearance of a CP-violating phase requires a Majorana mass insertion (neutrino chirality flip). 
The simplest way to confirm  these expectations is to check the kinetic relations from the two thermal cuts (green-dashed and blue-dashed lines) shown in Fig.~\ref{fig:vertex}, as well as the Yukawa coupling dependence of the two-loop amplitude. 

Based on the thermal cuts,  the thermal resonance would arise if the off-shell propagation of the resummed $\ell_\beta$ satisfies $p^{\prime 2}_\ell=2\tilde m_\alpha^2$ such that the commutation term $\tilde b_\alpha-\tilde b_\beta$ can appear in the denominator of the loop amplitude. From the on-shell conditions of $\phi$, $N_i, N_j$ and $\ell_\alpha$, we have
\begin{align}\label{eq:p.pl}
    p_\phi^2=(p-p_\ell)^2=m_\phi^2=M_i^2+2\tilde m_\alpha^2-2p\cdot p_\ell\,,
    \\[0.2cm]
       p_\phi^{\prime 2}=(p_\ell-q)^2=m_\phi^2=2\tilde m_\alpha^2+M_j^2-2p_\ell \cdot q\,,\label{eq:pl.q}
\end{align}
from which we can get the 4-momentum products $p\cdot p_\ell$ and $p_\ell\cdot q$.  On the other hand, we have
\begin{align}\label{eq:p.q}
    p^{\prime 2}_\ell=(p_\phi+q)^2=m_\phi^2+M_j^2-2(p-p_\ell)\cdot q\,.
\end{align}
Substituting the solution $p_\ell\cdot q$ from~\eqref{eq:pl.q} into~\eqref{eq:p.q}, we can get the 4-momentum product $p\cdot q$:
\begin{align}
    2p\cdot q= 2M_j^2+2\tilde m_\alpha^2-p^{\prime 2}_\ell\,.
\end{align}
Then using 
\begin{align}
   p^{\prime 2}_\ell=(p-p_\ell+q)^2=M_i^2+2\tilde m_\alpha^2+M_j^2-2p\cdot p_\ell+2p\cdot q-2p_\ell\cdot q\,,
\end{align}
together with the three 4-momentum products, we arrive at
\begin{align}
       p^{\prime 2}_\ell=m_\phi^2+M_j^2\neq 2\tilde m_\alpha^2\,.
\end{align}
It indicates that the factor appearing in the denominator of the CP-violating source will be dominated by $m_\phi^2$ rather than proportional to  the difference $\tilde m_\alpha^2-\tilde m_\beta^2$, and hence no thermal resonant enhancement would appear. 

Furthermore, the $m_\phi^2$-dominance in the denominator implies that we can practically do the summation of lepton flavours trivially in the Yukawa product. To see this, let us first consider the situation where $N_j$ does not flip chirality. It is easy to  write down the dependence of the amplitude on the Yukawa couplings,
\begin{align}
\sum_{\alpha,\beta}y^{\prime}_{\alpha i} y'_{\beta j}y^{\prime *}_{\alpha j} y^{\prime *}_{\beta i}=(y^{\prime \dagger}y)_{ji}(y^{\prime \dagger}y)_{ij}\,,
\end{align}
which is a real value. It implies that, in the approximation of $m_\phi^2\gg \tilde m^2$, a CP-violating Yukawa phase cannot appear in the $L_{\rm tot}$-conserving vertex diagrams even if there is an absorptive component from the thermal cuts, and hence we should not expect CP violation from this kind of vertex corrections. 

Next, let us consider flipping the $N_j$ chirality in Fig.~\ref{fig:vertex}. In this case, the dependence of the amplitude on the Yukawa couplings reads 
\begin{align}
\sum_{\alpha,\beta}y^{\prime}_{\alpha i} y^{\prime *}_{\beta j}y^{\prime *}_{\alpha j} y^{\prime }_{\beta i}=(y^{\prime \dagger}y)^2_{ji}\,,
\end{align}
 which can be complex-valued.  Therefore, in addition to the absence of  thermal resonance, we found that a CP-violating phase arising from vertex corrections requires an $L$-violating chirality flip of singlet neutrinos, i.e. the singlet neutrinos must be Majorana. Hence, the amplitude will be further suppressed by the small ratio $M_j/T$, especially for right-handed neutrinos much lighter than the electroweak scale.

Therefore, thermal resonance only appears in the wave-function corrected diagrams, and  vertex corrections are sub-dominant effects in TRL, which works both for Dirac neutrinos and GeV-scale Majorana neutrinos.

\section{Resonant Leptogenesis in the Closed-Time-Path formalism}\label{app:RFL-N}

In this appendix, we first discuss the usual scenario of RL through an $L_{\rm tot}$-violating  source and then calculate  the CP-violating source  in the CTP formalism.  Here, we build the KB equation for right-handed neutrinos and evaluate the lepton-number asymmetry stored in the neutrino sector. For scenarios of interest to us, the lepton-number asymmetry in the SM lepton sector is simply given by $\Delta n_L=-\Delta n_N$, under the valid approximation of lepton-number conservation and for weak washout effects. This method is different from that presented in Section~\ref{sec:FC-leptons} where we built the KB equation of leptons without evaluating the  two-loop self-energy diagrams directly. Instead, for KB equations of right-handed neutrinos, we should calculate the two-loop self-energy diagram of neutrinos with resummed lepton propagators.

\begin{figure}[t]
	\centering
\includegraphics[scale=0.8]{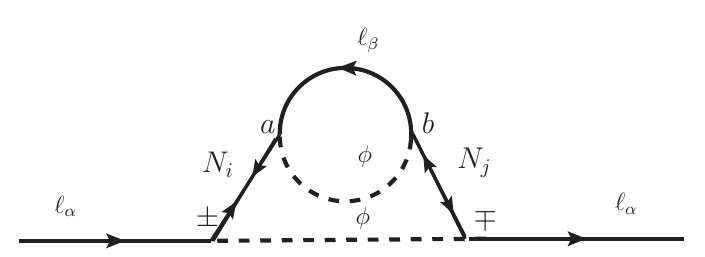} 
	\caption{\label{fig:N-self} The two-loop  self-energy diagrams of leptons contributing to CP asymmetry, with $\pm$ and $a,b=\pm$ being the CTP indices.  The CP-violating source is  induced by inserting Majorana masses (the flipping of chirality denoted by the arrows) and hence $L_{\rm tot}$ is violated.  }
\end{figure}

RL from Higgs decay with quasi-degenerate GeV-scale sterile neutrinos was calculated in~\cite{Hambye:2016sby} by applying the finite-temperature cutting rules~\cite{Frossard:2012pc}. It was later confirmed in~\cite{Hambye:2017elz} that using the density matrix formalism can also provide a consistent description. Here, we will follow the CTP formalism to calculate the CP-violating source originating from the two-loop self-energy diagrams of SM leptons, as shown in Fig.~\ref{fig:N-self}.  Instead of a full numerical analysis, the main purpose of this appendix is to provide the expression of the CP-violating source.  In particular, it allows to determine the regime where the $L_{\rm tot}$-conserving  source dominates over the $L_{\rm tot}$-violating  source, as presented in Section~\ref{sec:nu-mixing}.

The  evolution of the SM lepton-doublet asymmetries, $\Delta n_\ell\equiv \sum_\alpha \Delta n_{\ell_\alpha}$,  is determined by the following KB equation~\cite{Prokopec:2003pj,Prokopec:2004ic}
\begin{align}\label{eq:dnell/dt}
	\frac{\text{d}\Delta n_{\ell_\alpha}}{\text{d}t}=\frac{1}{2}\int \frac{{\rm d}^4p}{(2\pi)^4}\text{Tr}\left(i {\Sigma}_{\ell_\alpha}^>i{S}^<_{\ell_\alpha}-i{\Sigma}_{\ell_\alpha}^< i{S}^>_{\ell_\alpha}\right).
\end{align}
The trace acts on the Dirac spinor space, and ${S}_{\ell_\alpha}^{>,<}$ denote the lepton Wightman propagators.  ${\Sigma}_{\ell_\alpha}^{>,<}$  include one-loop self-energy diagrams that contribute to the washout rate and  two-loop self-energy diagrams that induce CP violation. We can formally write\footnote{Bear in mind that our convention for loop amplitudes is $-i\Sigma$.}
\begin{align}\label{eq:Sigma-Nij}
	i{\Sigma}_{\ell_\alpha}^<(p)&=2y_{\alpha i}^{\prime *} y'_{\alpha j}y^{\prime *}_{\beta i}y'_{\beta j}\int \frac{{\rm d}^4 p_N}{(2\pi)^4}\frac{{\rm d}^4 p_\phi}{(2\pi)^4}(2\pi)^4 \delta^4(p+p_\phi-p_N)P_R i{S}_{N_{ij}}^<P_L iG_\phi^>\,,
	\\[0.2cm]
		i{\Sigma}_{\ell_\alpha}^>(p)&=2y'_{\alpha i}y_{\alpha j}^{\prime *} y'_{\beta i}y^{\prime *}_{\beta j}\int \frac{{\rm d}^4 p_N}{(2\pi)^4}\frac{{\rm d}^4 p_\phi}{(2\pi)^4}(2\pi)^4 \delta^4(p+p_\phi-p_N)P_R i{S}_{N_{ji}}^>P_L iG_\phi^<\,,\label{eq:Sigma-Nji}
\end{align} 
where we have made an interchange for the dummy indices $i,j$  in $i{\Sigma}_{\ell_\alpha}^>$, explaining the appearance of complex conjugate in the Yukawa couplings and the flavour transpose of $i{S}_{N}^>$. Since we are not  working in flavoured leptogenesis where SM lepton flavours evolve differently, we will sum over all SM lepton flavours $\alpha, \beta$. In doing this, the Yukawa product becomes $y_{\alpha i}^{\prime *} y'_{\alpha j}y^{\prime *}_{\beta i}y'_{\beta j}=(y^{\prime \dagger} y')_{ij}^2$, which is in general complex for $i\neq j$ and provides  the CP-violating phase.

Following the decomposition approach in~\cite{Kanemura:2024dqv}, we can write down  the flavour-changing self-energy of sterile neutrinos ${S}^{<}_{N_{ij}},{S}^{>}_{N_{ji}}$ as 
\begin{align}\label{eq:SNij}
	i{S}_{N_{ij}}^<&=\left(i{S}_{N_i}^R\right)\left(-i{\Sigma}_{N}^T\right)\left(i{S}_{N_j}^<\right)+
	\left(i{S}_{N_i}^R\right)\left(-i{\Sigma}_{N}^<\right)\left(i{S}_{N_j}^R\right)+
	\left(i{S}_{N_i}^<\right)\left(-i{\Sigma}_{N}^<\right)\left(i{S}_{N_j}^R\right)
	\nonumber\\[0.2cm]
	&-
	\left(i{S}_{N_i}^R\right)\left(-i{\Sigma}_{N}^<\right)\left(i{S}_{N_j}^>\right)-
	\left(i{S}_{N_i}^<\right)\left(-i{\Sigma}_{N}^{\bar T}\right)\left(i{S}_{N_j}^R\right),
	\\[0.2cm]
		i{S}_{N_{ji}}^>&=\left(i{S}_{N_i}^R\right)\left(-i{\Sigma}_{N}^T\right)\left(i{S}_{N_j}^>\right)+
	\left(i{S}_{N_i}^R\right)\left(-i{\Sigma}_{N}^>\right)\left(i{S}_{N_j}^R\right)+
	\left(i{S}_{N_i}^<\right)\left(-i{\Sigma}_{N}^>\right)\left(i{S}_{N_j}^R\right)
	\nonumber\\[0.2cm]
	&-
	\left(i{S}_{N_i}^R\right)\left(-i{\Sigma}_{N}^>\right)\left(i{S}_{N_j}^>\right)-
	\left(i{S}_{N_i}^>\right)\left(-i{\Sigma}_{N}^{\bar T}\right)\left(i{S}_{N_j}^R\right).\label{eq:SNji}
\end{align}
The retarded propagator $i{S}_{N}^R$ can be obtained from the four components of sterile neutrino propagators in the CTP formalism, namely $i{S}_{N}^{T}, i{S}_{N}^{\bar T},i{S}_{N}^{<},i{S}_{N}^{>}$, via
\begin{align}\label{eq:SNR}
i{S}_{N_i}^R(p)&=\frac{1}{2}\left(i{S}_{N_i}^{T}-i{S}_{N_i}^{\bar T}+i{S}_{N_i}^{>}-i{S}_{N_i}^{<}\right)=\frac{i(\slashed{p}+M_i)}{p^2-M_i^2+i\text{sign}(p_0)\epsilon}\,,
\end{align}
where $\text{sign}(p_0)$ denotes the sign of frequency $p_0$ attached to the positive infinitesimal $\epsilon$,  and the propagators read 
\begin{align}\label{eq:SN<}
&i {S}_{N_i}^<(p)=-2\pi \delta(p^2-M_i^2)(\slashed{p}+M_i)\left[\theta(p_0)f_{N_i}(p_0)-\theta(-p_0)(1-\bar f_{N_i}(-p_0))\right],
	\\[0.2cm]
&	i {S}_{N_i}^>(p)=-2\pi \delta(p^2-M_i^2)(\slashed{p}+M_i)\left[-\theta(p_0)(1-f_{N_i}(p_0))+\theta(-p_0) \bar f_{N_i}(-p_0)\right],\label{eq:SN>}
\\[0.2cm]	
&i{S}_{N_i}^T(p)=\frac{i(\slashed{p}+M_i)}{p^2-M_i^2+i\epsilon}-2\pi \delta(p^2-M_i^2)(\slashed{p}+M_i)\left(\theta(p_0)f_{N_i}(p_0)+\theta(-p_0) \bar f_{N_i}(-p_0)\right),\label{eq:SNT}
\\[0.2cm]
&i{S}_{N_i}^{\bar T}(p)=\frac{-i(\slashed{p}+M_i)}{p^2-M_i^2-i\epsilon}-2\pi \delta(p^2-M_i^2)(\slashed{p}+M_i)\left(\theta(p_0)f_{N_i}(p_0)+\theta(-p_0) \bar f_{N_i}(-p_0)\right),\label{eq:SNTbar}
\end{align}
with $\theta(x)$ the Heaviside step-function. The distribution function $\bar f_N$ denotes the \textit{anti-neutrino} state defined as having the opposite helicity number to neutrinos.  

In thermal equilibrium, the distribution function gives
\begin{align}
    f^{\rm eq}_{N_i}(p_0)=\frac{1}{e^{p_0/T}+1}\,, 
\end{align}
with $p_0=\sqrt{|{\bf p}|^2+M_i^2}$, and the Wightman functions satisfy the KMS relation 
\begin{align}
    i{S}_{N_i}^>(p)=-e^{p_0/T}i{S}_{N_i}^{<}(p)\,.
    \end{align}

It it worthwhile to mention that the neutrino CTP propagators we used correspond to the KB ansatz and the on-shell approximation. It was shown  in~\cite{BhupalDev:2014oar} that such approximations based on the KB ansatz and the on-shell condition may not be sufficient to capture the full flavour mixing and oscillation effects, when the neutrinos are in quasi-degenerate and when the thermal corrections to neutrinos are important. Since the leptogenesis mechanism presented in this work does not rely on the RL from quasi-degenerate neutrinos, and since we are to  make a simple comparison between the two sources of resonant enhancements, i.e., the one from heavy  neutrino mixing  and the other from thermal lepton mixing, we will neglect the corrections potentially missed by using \eqref{eq:SN<}-\eqref{eq:SNTbar}. Nevertheless, we expect that such corrections should be small as thermal corrections are sub-dominant for nonthermal neutrinos with small Yukawa couplings.

\begin{figure}[t]
	\centering
\includegraphics[scale=0.8]{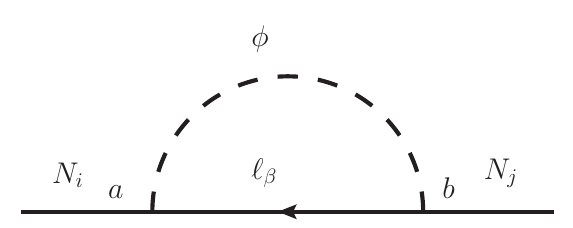} 
	\caption{\label{fig:4self-N} Flavour-changing self-energy diagrams of singlet neutrinos as the inner loop in  Fig.~\ref{fig:N-self}.}
\end{figure}

The self-energy amplitudes $-i{\Sigma}_N^{ab}$ with $a,b=++,--, +-,-+$ can be obtained straightforwardly  via  the Feynman diagram shown in  Fig.~\ref{fig:4self-N}, after the application of the thermal lepton propagators
\begin{align}\label{eq:Sl<}
 i {S}_\ell^<(p)&=-2\pi \delta(p^2)P_L\slashed{p}P_R\left[\theta(p_0)f_{\ell}(p_0)-\theta(-p_0)(1-\bar f_\ell(-p_0))\right],
	\\[0.2cm]
i {S}_\ell^>(p)&=-2\pi \delta(p^2)P_L\slashed{p}P_R\left[-\theta(p_0)(1-f_\ell(p_0))+\theta(-p_0)\bar f_\ell(-p_0)\right],\label{eq:Sl>}
	\\[0.2cm]
 i{S}_\ell^T(p)&=\frac{iP_L\slashed{p}P_R}{p^2+i\epsilon}-2\pi \delta(p^2)P_L\slashed{p}P_R\left(\theta(p_0)f(p_0)+\theta(-p_0)\bar f(-p_0)\right),\label{eq:SlT}
	\\[0.2cm]
	i{S}_\ell^{\bar T}(p)&=\frac{-i P_L\slashed{p}P_R}{p^2-i\epsilon}-2\pi \delta(p^2)P_L\slashed{p}P_R\left(\theta(p_0)f_\ell(p_0)+\theta(-p_0)\bar f_\ell(-p_0)\right),\label{eq:SlTbar}
\end{align}
and the Higgs doublet,
\begin{align}\label{eq:G<}
i G_\phi^<(p)&=2\pi \delta(p^2-m_\phi^2)\left[\theta(p_0)f_\phi(p_0)+\theta(-p_0)(1+\bar f_\phi(-p_0))\right],
\\[0.2cm]
i G_\phi^>(p)&=2\pi \delta(p^2-m_\phi^2)\left[\theta(p_0)(1+f_\phi(p_0))+\theta(-p_0)\bar f_\phi(-p_0)\right],
\\[0.2cm]
iG_\phi^T(p)&=\frac{i}{p^2-m_\phi^2+i\epsilon}+2\pi \delta(p^2-m_\phi^2)\left(\theta(p_0)f_\phi(p_0)+\theta(-p_0)\bar f_\phi(-p_0)\right),
\\[0.2cm]
iG_\phi^{\bar T}(p)&=\frac{-i}{p^2-m_\phi^2-i\epsilon}+2\pi \delta(p^2-m_\phi^2)\left(\theta(p_0)f_\phi(p_0)+\theta(-p_0)\bar f_\phi(-p_0)\right).\label{eq:GTbar}
\end{align}
Note that $m_\phi$ in the Higgs propagators contains the thermal mass effect. However,  we do not include the thermal mass correction to lepton propagators, which is a sub-dominant effect in evaluating the CP-violating source from Fig.~\ref{fig:N-self}, but is essential for the CP-violating source from Fig.~\ref{fig:2+1loop-Nself}, as detailed in Section~\ref{sec:FC-leptons}. 

Defining the CP-violating source on the rhs of \eqref{eq:dnell/dt} as $\mathcal{S}_{}$, we can write down 
\begin{align}
\mathcal{S}_{}=-4\pi \int\frac{{\rm d}^4 p}{(2\pi)^4}\frac{{\rm d}^4 p_N}{(2\pi)^4}\frac{{\rm d}^4 p_\phi}{(2\pi)^4}(2\pi)^4 \delta^4(p+p_\phi-p_N)\delta(p^2)\sum_{s=1}^5\mathcal{I}_s\,,
\end{align}
where $\mathcal{I}_s$ is defined as
\begin{align}
\sum_{s=1}^5\mathcal{I}_s&=y_4^*\left(f_\ell(|p_0|)-\theta(-p_0)\right)iG_\phi^<(p_\phi)\text{Tr}\left(P_R i{S}_{N_{ji}}^>P_L\slashed{p}\right)
\nonumber \\[0.2cm]
&-y_4\left(f_\ell(|p_0|)-\theta(p_0)\right)iG_\phi^>(p_\phi)\text{Tr}\left(P_R i{S}_{N_{ij}}^<P_L\slashed{p}\right),
\end{align}
with summation over the five terms given by  the ordered decomposition shown in \eqref{eq:SNij}-\eqref{eq:SNji}, and 
\begin{align}
	y_4\equiv y_{\alpha i}^{\prime *} y'_{\alpha j}y^{\prime *}_{\beta i}y'_{\beta j}\,.
\end{align}
 Let us proceed with these terms. The first term yields 
\begin{align}\label{eq:I1}
	\mathcal{I}_1=\frac{-2i\pi}{p_N^2-M_i^2}&\text{Tr}\left[P_R (\slashed{p}_N+M_j)P_R\slashed{q}P_L (\slashed{P}_N+M_i)P_L \slashed{p}\right]\delta(p_N^2-M_j^2)
\\[0.2cm]
	&\times \Big[y_4^*\left(f_\ell(|p_0|)-\theta(-p_0)\right)iG_\phi^<(p_\phi)(-i\Sigma_N^T)\left(-\theta(p_{N0})+f_{N_j}(|p_{N0}|)\right)
		\nonumber\\[0.2cm]
	&\phantom{\times} -y_4\left(f_\ell(|p_0|)-\theta(p_0)\right)iG_\phi^>(p_\phi)(-i\Sigma_N^T)\left(-\theta(-p_{N0})+f_{N_j}(|p_{N0}|)\right)
	\Big],	\nonumber 
\end{align}
where we have extracted the spinor structures out of the self-energy amplitude via 
\begin{align}
    -i{\Sigma}_N(q)\to P_L \slashed{q}P_R  (-i \Sigma_N)\,. 
\end{align}
By flipping all the momenta in the first term of the curly bracket in \eqref{eq:I1}, and using the following relations
\begin{align}
G_\phi^<(-p_\phi)=G_\phi^>(p_\phi)\,,\quad  \Sigma_N^T(-p)=\Sigma_N^T(p)\,,
\end{align} 
we can simplify $\mathcal{I}_1$ as
\begin{align}
	\mathcal{I}_1=\frac{-\pi M_i M_j }{M_j^2-M_i^2}\text{Im}y_4&\, p\cdot q\,\delta(p_N^2-M_j^2)iG_\phi^>(p_\phi)(-i\Sigma_N^T)
\nonumber    \\[0.2cm]
    &\times \left(f_\ell(|p_{0}|)-\theta(p_0)\right)\left(f_{N_j}(|p_{N 0}|)-\theta(-p_{N 0})\right),
\end{align}
where we have evaluated the Dirac trace
\begin{align}
\text{Tr}\left[P_R (\slashed{p}_N+M_j)P_R\slashed{q}P_L (\slashed{P}_N+M_i)P_L \slashed{p}\right]=2M_i M_j p\cdot q\,.
\end{align} 

The second function $\mathcal{I}_2$ reads
\begin{align}
	\mathcal{I}_2&=2i\pi \text{sign}(p_{N0})M_i M_j p\cdot q \left(\frac{\delta(p_N^2-M_j^2)}{p_N^2-M_i^2}+\frac{\delta(p_N^2-M_i^2)}{p_N^2-M_j^2}\right)
\\[0.2cm]
	&\times \Big[ y_4^* \left(f_\ell(|p_0|)-\theta(-p_0)\right)iG_\phi^<(p_\phi)(-i\Sigma_N^>)
-y_4 \left(f_\ell(|p_0|)-\theta(p_0)\right)iG_\phi^>(p_\phi)(-i\Sigma_N^<)\Big].	\nonumber
\end{align}
Flipping the momenta in the first term of the curly bracket with $\Sigma_N^>(-p)=\Sigma_N^<(p)$, and   interchanging the dummy indices $i,j$ in the second term of the parentheses, we arrive at 
\begin{align}
	\mathcal{I}_2=\frac{-4i\pi\text{sign}(p_{N0})(y_4^*+y_4)}{M_j^2-M_i^2}M_i M_j & p\cdot q\,\delta(p_N^2-M_j^2)
    \nonumber \\[0.2cm]
    &\times \left(f_\ell(|p_0|)-\theta(p_0)\right) iG_\phi^>(p_\phi)(-i\Sigma_N^<)\,.
\end{align}

The function $\mathcal{I}_3$ yields
\begin{align}
	\mathcal{I}_3&=\frac{-4i\pi}{p_N^2-M_j^2} M_i M_j p\cdot q\delta(p_N^2-M_i^2)\left(f_{N_i}(|p_{N0}|)-\theta(-p_{N0})\right)
\\[0.2cm]
	&\times\Big[y_4^*\left(f_\ell(|p_0|)-\theta(-p_0)\right)iG_\phi^<(p_\phi)(-i\Sigma_N^>)-y_4\left(f_\ell(|p_0|)-\theta(p_0)\right)iG_\phi^>(p_\phi)(-i\Sigma_N^<)\Big],	\nonumber
\end{align}
which, after flipping the momenta in the first term of the curly bracket, reduces to 
\begin{align}
	\mathcal{I}_3&=\frac{-8\pi \text{Im}y_4}{M_i^2-M_j^2}M_i M_j p\cdot q\, \delta(p_N^2-M_i^2)
    	\\[0.2cm]
	&\times \left(f_{N_i}(|p_{N0}|)-\theta(-p_{N0})\right)\left(f_\ell(|p_0|)-\theta(p_0)\right)iG_\phi^>(p_\phi)(-i\Sigma_N^<)
	\nonumber\\[0.2cm]
	&+\frac{4i\pi\text{sign}(p_{N0})y_4^*}{M_i^2-M_j^2}M_i M_j p\cdot q\, \delta(p_N^2-M_i^2)\left(f_\ell(|p_0|)-\theta(p_0)\right)iG_\phi^>(p_\phi)(-i\Sigma_N^<)\,.\nonumber
\end{align}

The function $\mathcal{I}_4$ reads
\begin{align}
	\mathcal{I}_4&=\frac{4i\pi}{p_N^2-M_i^2}M_i M_j p\cdot q \delta(p_N^2-M_j^2)\left(f_{N_j}(|p_{N0}|)-\theta(p_{N0})\right)
\\[0.2cm]
&\times 	\Big[y_4^*\left(f_\ell(|p_0|)-\theta(-p_0)\right)iG_\phi^<(p_\phi)(-i\Sigma_N^>)-y_4\left(f_\ell(|p_0|)-\theta(p_0)\right)iG_\phi^>(p_\phi)(-i\Sigma_N^<)\Big],
\nonumber
\end{align} 
which, after the same flipping of the momenta and the interchanging of the dummy indices $i,j$, becomes identical to  $\mathcal{I}_3$. 

Finally, the function $\mathcal{I}_5$ is given by 
\begin{align}
	\mathcal{I}_5&=\frac{4i\pi}{p_N^2-M_j^2} M_i M_j p\cdot q \, \delta(p_N^2-M_i^2)
\\[0.2cm]
	&\times \Big[y_4^*\left(f_\ell(|p_0|)-\theta(-p_0)\right)iG_\phi^<(p_\phi)(-i\Sigma_N^{\bar T})\left(f_{N_i}(|p_{N0}|)-\theta(p_{N0})\right)
	\nonumber\\[0.2cm]
&\phantom{\times}-y_4\left(f_\ell(|p_0|)-\theta(p_0)\right)iG_\phi^>(p_\phi)(-i\Sigma_N^{\bar T})\left(f_{N_i}(|p_{N0}|)-\theta(-p_{N0})\right)
	\Big],\nonumber 
\end{align}
which can  be simplified to
\begin{align}
	\mathcal{I}_5=\frac{-8\pi \text{Im}y_4}{M_j^2-M_i^2}& M_i M_j p\cdot q \,\delta(p_N^2-M_j^2)
	\\[0.2cm]
	&\times \left(f_\ell(|p_0|)-\theta(p_0)\right)iG_\phi^>(p_\phi)(-i\Sigma_N^{\bar T})\left(f_{N_j}(|p_{N0}|)-\theta(-p_{N0})\right).\nonumber
\end{align}

Assembling the $\mathcal{I}_s$ functions, we arrive at
\begin{align}\label{eq:Is-sum-fin}
	\sum_{i=s}^{5}\mathcal{I}_s=&\frac{8\pi \text{Im}y_4}{M_j^2-M_i^2} M_i M_j p\cdot q \,\delta(p_N^2-M_j^2)\left(f_\ell(|p_0|)-\theta(p_0)\right)iG_\phi^>(p_\phi)
	 \\[0.2cm]
	&\times\Big\{\left[(-i\Sigma_N^<)-(-i\Sigma_N^>)\right]\left(f_{N_j}(|p_{N0}|)-\theta(-p_{N0})\right)-\text{sign}(p_{N0})(-i\Sigma_N^<)\Big\},\nonumber
\end{align}
where we have used the relation $\Sigma_N^T+\Sigma_N^{\bar T}=\Sigma_N^<+\Sigma_N^>$. In the thermal limit $f_{N_j}=f^{\rm eq}_{N_j}$, we can check that the above curly bracket vanishes under the KMS relation 
\begin{align}
    \Sigma_N^>(p_N)=-e^{p_{N0}/T}\Sigma_N^<(p_N)\,,
\end{align} 
which guarantees that no CP asymmetry can be generated in thermal equilibrium. Using this fact, the curly bracket in the approximation of $f_\ell=f_\ell^{\rm eq}$ and $f_\phi=f_{\phi}^{\rm eq}$ can be reduced to 
\begin{align}
    \{..\}~\text{of}~\eqref{eq:Is-sum-fin}=\left[(-i\Sigma_N^<)-(-i\Sigma_N^>)\right]\delta f_{N_j}(|p_{N0}|)\,. 
\end{align}
With the Feynman diagram shown in Fig.~\ref{fig:4self-N}, the difference $i\Sigma_N^<-i\Sigma_N^>$ from the above equation yields
\begin{align}
	i\Sigma_N^<-i\Sigma_N^>=\frac{1}{(2\pi)^2}\int {\rm d}^4 q \delta(q^2)&\delta(M_j^2-m_\phi^2+2q \cdot p_{})
    \nonumber \\[0.2cm]
    &\times\left(\text{sign}(q_0)f_\phi(|q_{\phi 0}|)+\text{sign}(q_{\phi 0})f_\ell(|q_0|)\right).
\end{align}
Therefore, the CP-violating source can be rewritten as
\begin{align}
	\mathcal{S}_{}=\frac{\text{Im}y_4 M_i M_j}{4\pi^5 (M_j^2-M_i^2)}&\int {\rm d}p_0 \,{\rm d}\cos\theta_p\, |{\bf p}|^2 {\rm d}|{\bf p}|\int p_{N0}\,{\rm d}\cos\theta_{N}\, |{\bf p}_N|^2 {\rm d}|{\bf p}_N|
\nonumber \\[0.2cm]
&\times\int q_{0}\,{\rm d}\varphi_q\, {\rm d}\cos\theta_{q} \,|{\bf q}|^2 {\rm d}|{\bf q}| \left(q_0 p_0-{\bf q}\cdot {\bf  p}\right)
\nonumber \\[0.2cm]
	&\times  \delta(p_0^2-|{\bf p}|^2)\delta(p_{N0}^2-E_{N_j}^2)\delta(q_0^2-|{\bf q}|^2)
    \nonumber \\[0.2cm]
    &\times\delta(M_j^2-m_\phi^2-2p_{N0}p+2 {\bf p}_N \cdot {\bf p}) \delta(M_j^2-m_\phi^2+2q_0 p_{N0}-2{\bf q}\cdot {\bf p}_N)
\nonumber\\[0.2cm]
	&\times\left(f_\ell(|p_0|)-\theta(p_0)\right)\left(f_\phi(|p_{N0}-p_0|)-\theta(p_{N0}-p_0)\right)
    \nonumber\\[0.2cm]
	&\times \left(\text{sign}(q_0)f_\phi(|q_{\phi 0}|)+\text{sign}(q_{\phi 0})f_\ell(|q_0|)\right)\delta f_{N_j}(|p_{N0}|)\,,	 
\end{align}
where $E_{N_j}=\sqrt{|{\bf p}_N|^2+M_j^2}$, and  we have taken ${\bf p}_N=|{\bf p}_N|(0,0, \vec e_z)$ to fix the $z$-axis.  Then the spatial momentum product ${\bf p}_N \cdot {\bf p}$ within the Dirac $\delta$-function can be used to fix a plane where the projection of ${\bf p}$ on, e.g., the $x$-axis is zero. In this setup,  the momentum components and products read
\begin{align}
&{\bf p}=|{\bf p}|(0,\sin\theta_p,\cos\theta_p)\,, \quad {\bf q}=|{\bf q}|(\cos\varphi_q\sin\theta_q, \sin\varphi_q \sin\theta_q, \cos\theta_q)\,,
\\[0.2cm]
& {\bf q}\cdot {\bf p}_N=|{\bf q}||{\bf p}_N|\cos\theta_q\,,\quad  {\bf p}\cdot {\bf p}_N=| {\bf p}| |{\bf p}_N| \cos\theta_p\,,
\\[0.2cm]
& {\bf p} \cdot {\bf q}=|{\bf p}||{\bf q}|(\sin\varphi_q\sin\theta_q\sin\theta_p+\cos\theta_p \cos\theta_q)\,.
\end{align} 
We can perform the angular integration of  $\theta_q$  via $\delta(M_j^2-m_\phi^2+2p_{N0}q_0-2 {\bf p}_N \cdot {\bf  q})$, and of $\theta_p$ via $\delta(M_j^2-m_\phi^2-2p_{N0}p_0+2 {\bf p}_N \cdot {\bf p})$. In doing this, we will  obtain the integration  limits of $|{\bf q}|$ and $|{\bf p}|$
\begin{align}
	Q_-\equiv \frac{m_\phi^2-M_j^2}{2|{\bf p}_N|+2E_{N_j}}\leqslant |{\bf q}|, |{\bf p}|\leqslant	\frac{m_\phi^2-M_j^2}{2E_{N_j}-2|{\bf p}_N|}\equiv Q_+\,.
\end{align}
Finally, we arrive at the CP-violating source as
\begin{align}
	\mathcal{S}_{}=\frac{M_i M_j \text{Im}y_4}{128\pi^4(M_j^2-M_i^2)}&\int_{M_j}^\infty {\rm d}E_{N_j}\int_{Q_-}^{Q_+} {\rm d}|{\bf p}|\int_{Q_-}^{Q_+} {\rm d}|{\bf q}|
    \\[0.2cm]
&\times\frac{\left(m_\phi^2-M_j^2-2E_{N_j}|{\bf p}|\right)\left(m_\phi^2-M_j^2-2E_{N_j}|{\bf q}|\right)-4|{\bf p}| |{\bf q}| |{\bf p}_N|^2}{|{\bf p}_N|^3}
\nonumber\\[0.2cm]
	&\times  \left(f_\ell(|{\bf p}|)+f_\phi(E_{N_j}+|{\bf p}|)\right)\left(f_\ell(|{\bf q}|)+f_\phi(E_{N_j}+|{\bf q}|)\right)\delta f_{N_j}(E_{N_j})\,. \nonumber
\end{align}

After straightforward calculations, we can write down a compact result for the evolution of $\Delta n_{\ell}$ in the approximation of weak washout and in the limit of $M_j\ll m_\phi$, which reads
 \begin{align}\label{eq:dDelta-nell/dt}
 	\frac{{\rm d}\Delta n_{\ell}}{{\rm d}t}=\sum_{\alpha,\beta}\,\sum_{\substack{i,j
    \\ i\neq j}}\frac{\text{Im}\left(y_{\alpha i}^{\prime *} y'_{\alpha j}y^{\prime *}_{\beta i}y'_{\beta j}\right)}{128\pi^4 } \frac{M_i M_j}{M_j^2-M_i^2}\int_0^\infty  {\rm d}|{\bf p}_N| \int_{\bar Q_-}^{\infty}{\rm d}|{\bf p}|\int_{\bar Q_-}^{\infty}{\rm d}|{\bf q}|\,\mathcal{Q}\,\mathcal{F}_j\,,
 \end{align}
where summation over $i,j$ takes into account the   singlet neutrino flavours contributing to leptogenesis,  $Q_+(M_j= 0)\to\infty$, and 
  \begin{align}\label{eq:pq-low}
 \bar Q_-=Q_- (M_j=0)=\frac{m_\phi^2}{4|{\bf p}_N|}\,.
 \end{align}
The momentum function $\mathcal{Q}$ given in \eqref{eq:dDelta-nell/dt} reads
 \begin{align}\label{eq:Q-ell}
 \mathcal{Q}=\frac{m_\phi^2\left[m_\phi^2-2|{\bf p}_N|(|{\bf p}|+|{\bf q}|)\right]}{|{\bf p}_N|^3}\,,
 \end{align}
while the  statistics function $\mathcal{F}_j$ gives
 \begin{align}
 	\mathcal{F}_j=\left(f^{\rm eq}_\ell(|{\bf p}|)+f^{\rm eq}_\phi(|{\bf p}_N|+|{\bf p}|)\right)\left(f^{\rm eq}_\ell(|{\bf q}|)+f^{\rm eq}_\phi(|{\bf p}_N|+|{\bf q}|)\right)\delta f_{N_j}(|{\bf p}_N|)\,,
 \end{align}
where  $\delta f_{N_j}\equiv f_{N_j}-f_{N_j}^{\rm eq}$, and we have used thermal distribution functions for leptons and the Higgs doublet,
 \begin{align}
 	f^{\rm eq}_{\ell}(|{\bf p}|)=\frac{1}{e^{|{\bf p}|/T}+1}\,,\quad  	f^{\rm eq}_{\phi}(E)=\frac{1}{e^{E/T}-1}\,.
 \end{align}

In the above derivations,  we have taken the relativistic limit for singlet neutrinos $E_N\approx |{\bf p}_N|$ and $M_j\ll m_\phi$. The Higgs mass $m_\phi$ includes the vacuum mass $m_h\approx 125$~GeV after electroweak gauge symmetry breaking and a thermal mass. For simplicity, we consider the Heaviside step-function to simulate the electroweak cross-over
 \begin{align}\label{eq:Higgs-thermal-mass}
 	m_\phi^2\approx m_h^2\theta(T_c-T)+m_T^2\,,\quad m_T^2=\left(\frac{3g_2^2+g_1^2}{16}+\frac{\lambda}{2}+\frac{y_t^2}{4}\right)T^2\equiv c_hT^2\,,
 \end{align}
where $T_c\approx 160$~GeV denotes the cross-over temperature~\cite{DOnofrio:2014rug} and $c_h\approx 0.4$. 

As a final remark, we should mention that we did not include vacuum and thermal width effects in the denominator of  \eqref{eq:dDelta-nell/dt}, which in some parameter regimes can suppress the resonant enhancement in the nearly degenerate limit $M_j\to M_i$. Since tuning the mass degeneracy was not the focus of this work, we always assumed that the degeneracy under consideration is not affected by vacuum and thermal width effects. 

\vfill\eject

\bibliographystyle{JHEP}
\bibliography{Refs}

\end{document}